\documentclass[journal, comsoc]{IEEEtran}

\usepackage{cite}
\usepackage{amsbsy,amsmath,amssymb,epsfig,bbm,mathrsfs,multirow,amsthm,amsfonts}
\usepackage{algorithm,algorithmic}
\usepackage{graphicx}
\usepackage{textcomp}
\usepackage{subfigure}
\usepackage[table,xcdraw]{xcolor}
\usepackage{multicol}
\usepackage{url}
\usepackage{multirow}
\usepackage{xcolor}
\usepackage{chngpage}
\usepackage{array}
\PassOptionsToPackage{hyphens}{url}
\usepackage{hyperref}
\hypersetup{colorlinks, citecolor=green, filecolor=pink, linkcolor=blue, urlcolor=blue }
\graphicspath{{./figures/}}

\def\BibTeX{{\rm B\kern-.05em{\sc i\kern-.025em b}\kern-.08em
		T\kern-.1667em\lower.7ex\hbox{E}\kern-.125emX}}

\usepackage{comment}
\usepackage{nicematrix}
\usepackage[T1]{fontenc}
\usepackage{makecell}
\usepackage{amstext} 
\usepackage{caption}
\definecolor{mygray}{gray}{0.6}
\definecolor{myblue}{rgb}{0.8,0.85,1}

\usepackage{array, tabularx, boldline}
\newcolumntype{L}[1]{>{\raggedright\let\newline\\\arraybackslash\hspace{0pt}}m{#1}}
\newcolumntype{C}[1]{>{\centering\let\newline\\\arraybackslash\hspace{0pt}}m{#1}}
\newcolumntype{R}[1]{>{\raggedleft\let\newline\\\arraybackslash\hspace{0pt}}m{#1}}

\usepackage{booktabs}
\usepackage{cellspace}
\usepackage{amssymb}% http://ctan.org/pkg/amssymb
\usepackage{pifont}% http://ctan.org/pkg/pifont
\newcommand{\cmark}{\ding{51}}%
\newcommand{\xmark}{\ding{55}}%
\usepackage{xcolor}
\begin{document}
	
	\title{A Full Dive into Realizing the Edge-enabled Metaverse: Visions, Enabling Technologies, and Challenges}
	
	\author{Minrui Xu, Wei Chong Ng, Wei Yang Bryan Lim, Jiawen Kang, Zehui Xiong, Dusit Niyato, \\ Qiang Yang, Xuemin Sherman Shen, and Chunyan Miao
		\thanks{Minrui~Xu, Dusit~Niyato, and Chunyan~Miao are with the School of Computer Science and Engineering, Nanyang Technological University, Singapore (e-mail: minrui001@e.ntu.edu.sg; dniyato@ntu.edu.sg; ascymiao@ntu.edu.sg).}
		\thanks{Wei Chong Ng and Wei Yang Bryan Lim are with Alibaba Group and Alibaba-NTU Joint Research Institute, Nanyang Technological University, Singapore (e-mail: weichong001@e.ntu.edu.sg; limw0201@e.ntu.edu.sg).}
		\thanks{Jiawen~Kang is with the School of Automation, Guangdong University of Technology, China (e-mail: kavinkang@gdut.edu.cn).}
		\thanks{Zehui~Xiong is with the Pillar of Information Systems Technology and Design, Singapore University of Technology and Design, Singapore 487372, Singapore (e-mail: zehui\_xiong@sutd.edu.sg).}
		\thanks{Qiang~Yang is with the Department of Computer Science and Engineering, Hong Kong University of Science and Technology, Hong Kong, and Webank (e-mail: qyang@cse.ust.hk).}
		\thanks{Xuemin~Sherman~Shen is with the Department of Electrical and Computer Engineering, University of Waterloo, Waterloo, ON, Canada, N2L 3G1 (e-mail: sshen@uwaterloo.ca).}
	}
	\maketitle

	\begin{abstract}
		Dubbed ``the successor to the mobile Internet", the concept of the Metaverse has grown in popularity. While there exist lite versions of the Metaverse today, they are still far from realizing the full vision of an immersive, embodied, and interoperable Metaverse. Without addressing the issues of implementation from the communication and networking, as well as computation perspectives, the Metaverse is difficult to succeed the Internet, especially in terms of its accessibility to billions of users today. In this survey, we focus on the edge-enabled Metaverse to realize its ultimate vision. We first provide readers with a succinct tutorial of the Metaverse, an introduction to the architecture, as well as current developments. To enable ubiquitous, seamless, and embodied access to the Metaverse, we discuss the communication and networking challenges and survey cutting-edge solutions and concepts that leverage next-generation communication systems for users to immerse as and interact with embodied avatars in the Metaverse. Moreover, given the high computation costs required, e.g., to render 3D virtual worlds and run data-hungry artificial intelligence-driven avatars, we discuss the computation challenges and cloud-edge-end computation framework-driven solutions to realize the Metaverse on resource-constrained edge devices. Next, we explore how blockchain technologies can aid in the interoperable development of the Metaverse, not just in terms of empowering the economic circulation of virtual user-generated content but also to manage physical edge resources in a decentralized, transparent, and immutable manner. Finally, we discuss the future research directions towards realizing the true vision of the edge-enabled Metaverse.
	\end{abstract}
	
	\begin{IEEEkeywords}
		Metaverse, Edge Networks, Communication and Networking, Computation, Blockchain, Internet Technology 
	\end{IEEEkeywords}

	\section{Introduction}
	\label{sec:intro}
	
	\subsection{Birth of the Metaverse}

	The concept of the \textit{Metaverse} first appeared in the science fiction novel \textit{Snow Crash} written by Neal Stephenson in 1992. More than twenty years later, the Metaverse has re-emerged as a buzzword. In short, the Metaverse is commonly described as an embodied version of the Internet. Just as how we navigate today's web pages with a mouse cursor, users will explore the virtual worlds within the Metaverse with the aid of augmented reality (AR), virtual reality (VR), and the tactile Internet. 
	
	% Moreover, powered by Artificial Intelligence (AI), blockchain technology, and 5G and Beyond (B5G),  the Metaverse is envisioned to facilitate peer-to-peer interactions and support novel, decentralized ecosystems of service provisions that will blur the lines between the physical and virtual worlds. 
	
	To date, tech giants have invested heavily towards realizing the Metaverse as ``the successor to the mobile Internet".  In the future, the Metaverse will succeed the Internet towards revolutionizing novel ecosystems of service provisions in all walks of life, e.g., in healthcare \cite{yu2012building}, education \cite{ng2021optimal}, entertainment, e-commerce \cite{jeong2022innovative}, and smart industries \cite{kwon2021smart}. In 2021, Facebook was rebranded as ``Meta"\footnote{\url{https://about.fb.com/news/2021/10/facebook-company-is-now-meta/}}, as it reinvents itself to be a  ``Metaverse company" from a ``social media company", to reinforce its commitment towards the development of the Metaverse. 
	
	There are two fundamental driving forces behind the excitement surrounding the Metaverse. First, the Covid-19 pandemic has resulted in a paradigm shift in how work, entertainment, and socializing are experienced today \cite{bick2021work,pandi2021distant, bates2021global}. As more users grow accustomed to carrying out these conventionally physical activities in the virtual domain, the Metaverse has been positioned as a \textit{necessity} in the near future. Second, emerging technological enablers have made the Metaverse a growing \textit{possibility}. For example, beyond 5G/6G (B5G/6G) communication systems promise eMBB (enhanced Mobile Broadband) and URLLC (Ultra Reliable Low Latency Communication)~\cite{you2021towards,thompson20145g,shen2021holistic, xiang20165g, tang2022roadmap}, both of which enable AR/VR and haptic technologies~\cite{sharma2020toward} that will enable users to be visually and physically immersed in the virtual worlds.  
	
	  The Metaverse is regarded as an advanced stage and the long-term vision of digital transformation~\cite{metaverse}. As an exceptional multi-dimensional and multi-sensory communication medium~\cite{mann2018all}, the Metaverse overcomes the tyranny of distance by enabling participants in different physical locations to immerse in and interact with each other in a shared 3D virtual world~\cite{duan2021metaverse}. To date, there exist ``lite" versions of the Metaverse that have evolved mainly from Massive Multiplayer Online (MMO) games. Among others, Roblox\footnote{\url{https://www.roblox.com/}} and Fortnite\footnote{\url{https://www.epicgames.com/fortnite/en-US/home}} started as online gaming platforms. Recently, virtual concerts were held on Roblox and Fortnite and attracted millions of views. Beyond the gaming industry, cities around the world have begun ambitious projects to build their meta-cities~\cite{zhou2022self} in the Metaverse.
	
	However, we are still far from realizing the full vision of the Metaverse~\cite{lim2022realizing}. Firstly, the aforementioned ``lite" versions are distinct virtual worlds operated by separate entities. In other words, one's Fortnite avatar and virtual items cannot be used in the Roblox world. In contrast, the Metaverse is envisioned to be a \textit{seamless} integration of various virtual worlds, with each virtual world developed by separate service providers. Similar to life in the physical world, one's belongings and assets should preserve their values when brought over from one virtual world to another. Secondly, while MMO games can host more than a hundred players at once, albeit with high-specification system requirements, an open-world Virtual Reality MMO (VRMMO) application is still a relatively nascent concept even in the gaming industry. As in the physical world, millions of users are expected to co-exist within the virtual worlds of the Metaverse. It will be a challenge to develop a shared Metaverse, i.e., one that does not have to separate players into different server sessions. Thirdly, the Metaverse is expected to integrate the physical and virtual worlds, e.g., by creating a virtual copy of the physical world using the digital twin (DT) \cite{wu2021digital}. The stringent sensing, communication, and computation requirements impede the real-time and scalable implementation of the Metaverse. Finally, the birth of the Metaverse comes amid increasingly stringent privacy regulations \cite{cheng2020federated}. A data-driven realization of the Metaverse brings forth new concerns, as new modes of accessing the Internet using AR/VR imply that new modalities of data such as eye-tracking data can be captured and leveraged by businesses.

		Though the Metaverse is considered to be the successor to the Internet, edge devices, i.e., devices connected to the Metaverse via radio access networks~\cite{guevara2020classification}, may lack the capabilities to support interactive and resource-intensive applications and services, such as interacting with avatars and rendering 3D worlds. To truly succeed as the next-generation Internet, it is of paramount importance that users can ubiquitously access the Metaverse, just as the Internet entertains billions of users daily. Fortunately, as AR/VR, the tactile Internet, and hologram streaming are key driving applications of 6G \cite{saad2019vision,yang20226g}, it has become apparent that future communication systems will be developed to support the Metaverse. Moreover, the timely shift from the conventional focus on classical communication metrics such as data rates towards the co-design of computation and communication systems~\cite{letaief2021edge}, implies that the design of next-generation mobile edge networks will also attempt to solve the problem of bringing the Metaverse to computationally constrained mobile users. Finally, the paradigm shift from centralized big data to distributed or decentralized small data \cite{saad2019vision} across the ``Internet of Everything"  implies that the blockchain~\cite{yang2022fusing,gadekallu2022blockchain} will play a fundamental role in realizing the Metaverse at mobile edge networks. With the above considerations in mind, this survey focuses on discussing the edge-enabled Metaverse. We will discuss the key communication and networking, computation, and blockchain/distributed ledger technologies (DLT) that will be instrumental in solving the aforementioned challenges and realizing the edge-enabled Metaverse.

	\subsection{Related Works and Contributions}
	
	% \usepackage{multirow}
	% \usepackage{graphicx}
	\begin{table*}[]
		\centering
		\resizebox{\textwidth}{!}{%
			\begin{tabular}{|l|l|l|}
				\hline
				\textbf{Reference} &
				\textbf{Key focus of survey} &
				\textbf{How our survey differs} \\ \hline
				\cite{duan2021metaverse}$\star$ &
				\begin{tabular}[c]{@{}l@{}}Discusses the architecture of the Metaverse, current examples, and its role in promoting \\social good, and the development of a campus prototype\end{tabular} &
				\begin{tabular}[c]{@{}l@{}}Our survey discusses how enablers of Metaverse can be implemented at a scale\\ at mobile edge networks\end{tabular} \\ \hline
				\cite{lee2021all}$\star$ &
				Discusses technology enablers of the Metaverse, e.g., VR, AR, and IoT &
				\multirow{4}{*}{\begin{tabular}[c]{@{}l@{}}Instead of focusing on \textit{what} are the enabling technologies, we focus on \textit{how} \\ the enablers of the Metaverse can be implemented at mobile edge networks. We\\ discuss the implementation challenges and solutions from the communication and networking, \\ computation, and blockchain perspectives\end{tabular}} \\ \cline{1-2}
				\cite{huynh2022artificial}$\star$ &
				Discusses how AI can aid in the development of the Metaverse &
				\\ \cline{1-2}
				\cite{yang2022fusing}$\star$ &
				Discusses how AI and blockchain can enrich applications in the Metaverse &
				\\ \cline{1-2}
				\cite{ning2021survey}$\star$ &
				Discusses the industrial examples, applications, and forecasts of the Metaverse &
				\\ \hline
				\cite{siriwardhana2021survey,erol2018caching,ahmed2017mobile} &
				\begin{tabular}[c]{@{}l@{}}Discusses how VR/mobile AR can be realized with 5G mobile edge computing, the \\ computing architectures and applications\end{tabular} &
				\begin{tabular}[c]{@{}l@{}}AR/VR are just ways to access the Metaverse. Our survey adopts a holistic view of \\ implementation challenges, e.g., of distributed computation and blockchain. Our survey also \\ focuses on the challenges of adopting a Metaverse that will host a massive number of users \\ simultaneously, similar to VRMMO games\end{tabular} \\ \hline
				\cite{wang2020convergence,lim2020federated} &
				\begin{tabular}[c]{@{}l@{}}Discusses how AI can be conducted at the edge and how AI can facilitate edge \\ resource management\end{tabular} &
				\multirow{2}{*}{\begin{tabular}[c]{@{}l@{}}As the optimization or resource allocation objectives are not Metaverse specific, several studies \\ look at network optimization from the conventional perspectives. Our survey introduces the \\ challenge of multi-sensory optimization objectives such as QoA.\end{tabular}} \\ \cline{1-2}
				\cite{yu2017survey,abbas2017mobile,lin2017survey} &
				\begin{tabular}[c]{@{}l@{}}Discusses how edge computing can be used to improve IoT through computation \\ and communication support\end{tabular} &
				\\ \hline
			\end{tabular}%
		}
		\caption{Summary of related works vs. our survey. The references marked with the asterisk symbol are Metaverse-specific.}
		\label{tab:summary}
	\end{table*}
	
	Given the popularity of the Metaverse, several surveys on related topics have appeared recently. A summary of these surveys and how our survey adds value is shown in Table \ref{tab:summary}. 
	
	The study in \cite{duan2021metaverse} is among the first to provide a tutorial on the Metaverse. In \cite{duan2021metaverse}, the authors discuss the Metaverse's role in promoting social good. Then, the architecture of the Metaverse, as well as examples of the emerging developments in the industry are described. The survey of \cite{lee2021all} discusses the technological enablers of the Metaverse in greater detail. The topics of eXtended reality (XR), IoT and robotics, user interface (UI) design, AI, and blockchain are introduced. 
	
	Following \cite{duan2021metaverse,lee2021all}, other surveys subsequently discuss more specific subsets of topics affiliated with the Metaverse. Given the recent advancements of AI, the study in \cite{huynh2022artificial} discusses how AI can play a part in developing the Metaverse, e.g., in terms of natural language processing to create intelligent chatbots or machine vision to allow AR/VR devices to effectively analyze and understand the user environment. The study in \cite{yang2022fusing} discusses how convergence of AI and blockchain can facilitate service delivery in the Metaverse. For example, AI can be used to train virtual characters that populate the Metaverse, whereas the blockchain can be used to facilitate transactions within the Metaverse. The study in \cite{ning2021survey} focuses its discussion on the present applications and industrial developments of the Metaverse, as well as provides forecasts on the future prospects from an industrial perspective.
	
	However, these surveys do not focus on the \textit{implementation} challenges of the Metaverse at mobile edge networks from a resource allocation~\cite{chu2022metaslicing}, communication, and networking, or computation perspective. While it is important to understand \textit{what} are the enabling technologies of the Metaverse, it is also important to discuss \textit{how} they can be implemented on mobile edge networks. For example, while \cite{lee2021all} surveys applications of AR/VR that can enhance the immersion experience of a user in the Metaverse, an equivalently important question to ask is: how can the communication and networking hurdles be crossed to bring AR/VR to potentially millions of edge devices while respecting the stringent latency and data rate requirements? Surveys on AR/VR service delivery \cite{siriwardhana2021survey,erol2018caching,ahmed2017mobile} do focus on how AR/VR can be implemented over 5G mobile edge networks, as well as discuss the computing and offloading architectures involved. However, these surveys are not Metaverse-specific and therefore do not address some of the key challenges discussed in our survey. For example, the Metaverse will have massive users accessing and interacting with the virtual worlds with AR/VR \textit{simultaneously}. These users will also interact with each other, thereby setting a stringent constraint on the round-trip communication latency. The challenges of interoperability, heterogeneity, and massive simultaneous stochastic network demands are important topics that we will discuss in this survey. In addition, the primary focus of \cite{siriwardhana2021survey,erol2018caching,ahmed2017mobile} is on AR/VR service delivery. In contrast, our survey adopts a holistic view of how other enablers of the Metaverse such as AI model training or blockchain can be implemented at the edge. Our approach therefore opens up further discussions and challenges as compared to that of conventional AR/VR service delivery.
	
	Similarly, while the aforementioned surveys \cite{yang2022fusing, lee2021all,huynh2022artificial} discuss the AI algorithms that can empower autonomous applications in the Metaverse, it is equally important to explore how these data and computing hungry algorithms can be executed on resource-constrained edge devices. Currently, surveys on how AI can be implemented at the edge conduct their discussions from the perspectives of edge intelligence \cite{wang2020convergence,lim2020federated}. Surveys on the convergence of the Internet of Things (IoT) and edge computing \cite{abbas2017mobile,lin2017survey,yu2017survey} also studies challenges and resource allocation solutions from the conventional IoT perspective. These surveys review and discuss studies based on conventional metrics such as data rate or energy efficiency. On the other hand, as the Metaverse involves 3D immersion and physical-virtual world synchronization, our survey aims to discuss how the challenges of multi-sensory optimization objectives such as quality of augmentation (QoA) can be solved, e.g., through leveraging next-generation communication paradigms such as semantic/goal-aware communication.
	
	Moreover, beyond discussing the blockchain's function of enabling decentralized and sustainable economic ecosystems, the importance of blockchain in managing edge communication and networking resources is under-explored in the aforementioned surveys\cite{duan2021metaverse,lee2021all,yang2022fusing}. Without addressing these fundamental implementation issues and proposing viable solutions, it is practically challenging for the Metaverse to be the ``successor to the mobile Internet" beyond the proof of concept. This motivates us to write a survey that discusses challenges from the perspectives of communication and networking, computation, and blockchain at mobile edge networks. Through discussing conceivable solutions for the challenges, our survey provides critical insights and useful guidelines for the readers to better understand how these implementation issues could be resolved towards realizing an immersive, scalable, and ubiquitous Metaverse.
	
	The contributions of our survey are as follows:
	\begin{itemize}
		
		\item We first provide a tutorial that motivates a definition and introduction to the architecture of the Metaverse. In our tutorial, we highlight some examples of current implementations and development frameworks for the Metaverse. Through this succinct tutorial, the readers can catch up with the forefront developments in this topic.
		
		\item We identify key communication and networking challenges towards realizing the \textit{immersive} Metaverse. Different from conventional AR/VR, the Metaverse involves massive volumes of user interactions and differentiated services. We discuss desirable characteristics of the Metaverse from the communication and networking perspective. Then, we survey key communication and networking solutions that will be fundamental towards realizing these characteristics. We particularly focus on next-generation communication architectures such as semantic communication and real-time physical-virtual synchronization. This way, readers are able to understand how the development of future communication systems can potentially play their roles in the edge-enabled Metaverse.
		
		\item We discuss key computation challenges towards realizing the \textit{ubiquitous} Metaverse for resource-constrained users at the network edge. We survey methods from two classifications of works. The first classification of our surveyed solutions leverages the cloud-edge-end computation paradigm to tap on cloud/edge computation resources for resource-constrained users to access the Metaverse.  The second classification of our work studies selected AI-driven techniques to improve the efficiency of computation, e.g., in terms of accelerating processing speed and reducing storage cost. Our survey in this section aims to provide general readers with a holistic view of how the Metaverse can be accessed at anytime and anywhere, especially on computation-constrained edge devices.
		
		\item  We explore how blockchain technologies can aid in the development of the \textit{interoperatable} Metaverse. Different from the aforementioned surveys that focus only on the application value of blockchain, e.g., in empowering economic ecosystems in the Metaverse, we also emphasize the infrastructure value of blockchain. Specifically, we expound on how the blockchain can support the edge communication and networking aspects of the Metaverse, e.g., through secure and decentralized edge data storing and sharing for reliable Metaverse, blockchain sharding, and cross-chain for scalability and interoperability of the Metaverse. This is in line with the B5G/6G vision \cite{saad2019vision} that DLT will play a fundamental role in enabling native secure and reliable communication and networking.
		
		\item We outline the research directions to pave the path towards future research attempts on realizing the Metaverse at mobile edge networks. Our survey serves as an initial step that precedes a comprehensive and insightful investigation of the Metaverse at mobile edge networks, helping researchers quickly grasp the overview of visions, enabling technologies, and challenges in the area.
		
	\end{itemize}
	
	The survey is organized as follows. Section \ref{sec:tutorial} discusses the architecture of the Metaverse and current developments. Section \ref{sec:communication} discusses communication and networking challenges and key solutions. Section \ref{sec:computation} discusses the computation challenges and solutions. Section \ref{sec:blockchain} discusses the blockchain technology and its functions in the edge-empowered Metaverse. Section \ref{sec:open} discusses the future research directions. Section \ref{sec:conclude} concludes.

	\section{Architecture, Developments, and Tools of the Metaverse}
	\label{sec:tutorial}
	
	In this section, we present the architecture, examples, and development tools for the Metaverse. 
	
	\begin{figure*}
		\centering
		\includegraphics[width=0.8\linewidth]{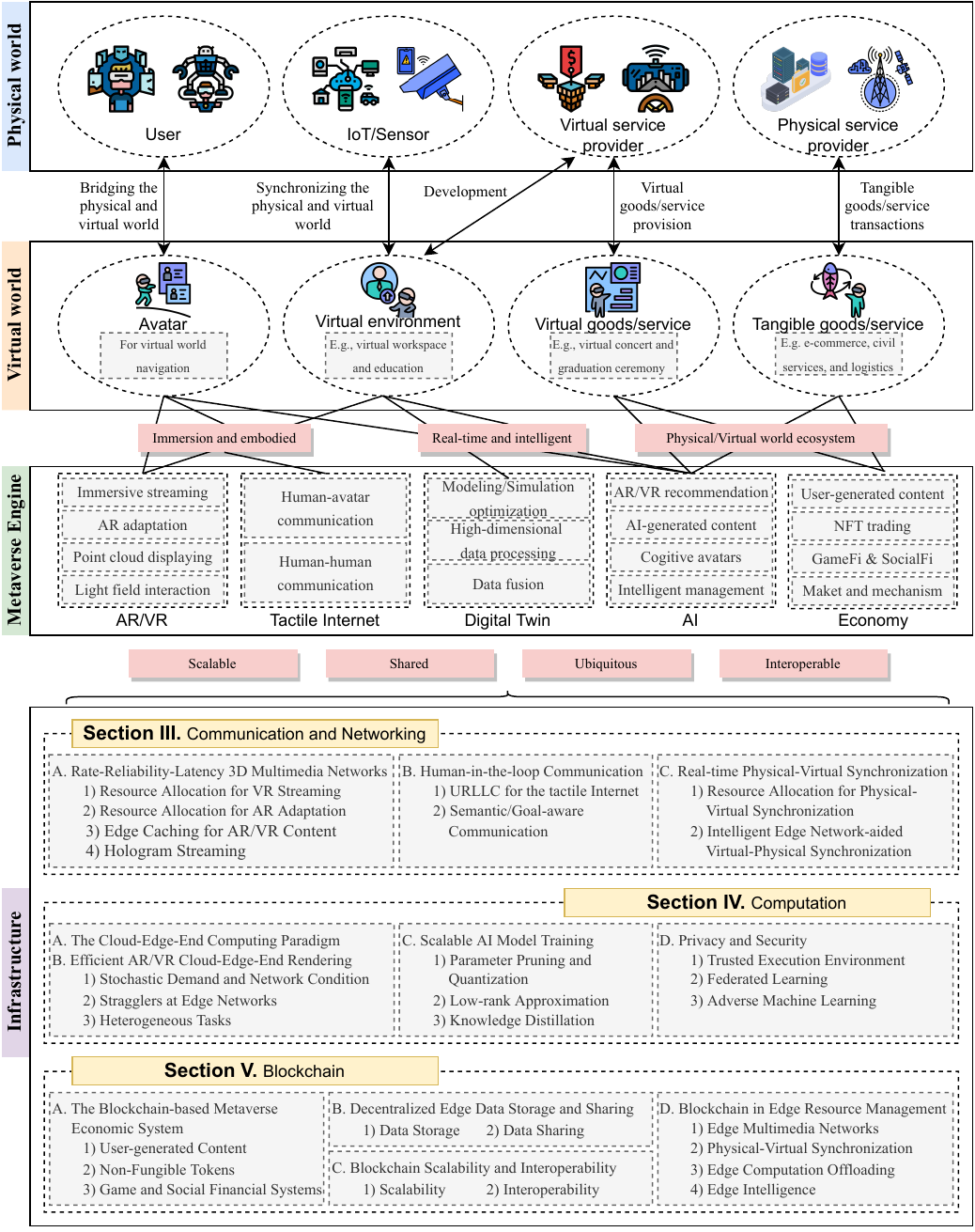}
		\caption{The Metaverse architecture features the immersive and real-time physical-virtual synchronization supported by the communication and networking, computation, and blockchain infrastructure subsequently discussed in the labeled sections.}
		\label{fig:archi}
	\end{figure*}
	
	\subsection{Definition and Architecture}
	\label{sec:archi}
	
	The Metaverse is an \textit{embodied} version of the Internet that comprises a \textit{seamless} integration of \textit{interoperable, immersive, and shared virtual ecosystems} navigable by user avatars. In the following, we explain each key word in this definition.
	
	\begin{itemize}
		
		\item \textit{Embodied:} The Metaverse blurs the boundary between the virtual and physical worlds \cite{abc}, where users can ``physically" interact with the virtual worlds with a tangible experience, e.g., by using 3D visual, auditory, kinesthetic, and haptic feedback.  The Metaverse can utilize AR to extend the virtual worlds into the physical world.
		
		\item \textit{Seamless/Interoperable:} Much like the physical world, belonging of users' avatars in one virtual world should not lose its value when brought to another seamlessly, even if the virtual worlds are developed by separate entities. In other words, no a single company ``owns” the Metaverse.
		
		\item \textit{Immersive:} The Metaverse can be ``experienced" beyond 2D interactions that allow users to interact with other users similar to that in the physical world.
		
		\item \textit{Shared:} Much like the physical world, thousands of users should be able to co-exist in a single server session, rather than having users separated into different virtual servers. With users accessing the Metaverse and immersing themselves anytime and anywhere, the life-like interaction of users is thus shared globally, i.e., an action may affect any other users just as in an open world, rather than only for users at a specific server.
		
		\item \textit{Ecosystem:} The Metaverse will support end-to-end service provisions for users with digital identities (DIDs), e.g., content creation, social entertainment, in-world value transfer regardless of nationalities of users, and physical services that will cross the boundary between the physical and virtual worlds. The Metaverse ecosystem empowered by the decentralized design of blockchains is expected to be sustainable as it has a closed-loop independent economic system with transparent operating rules.
	\end{itemize}
	
	Next, we discuss the architecture of the Metaverse with some foundemental enabling technologies (Fig. \ref{fig:archi}).
	
	\begin{enumerate}
		
		\item \textit{Physical-virtual synchronization}: Each non-mutually-exclusive stakeholder in the physical world controls components that influence the virtual worlds. The action of the stakeholder has consequences in the virtual worlds that may, in turn, give feedback to the physical world. The key stakeholders are:
		
		\begin{itemize}
			
			\item \textit{Users} can immerse the virtual worlds as avatars through various equipment, such as, Head Mounted Displays (HMDs) or AR goggles \cite{zhou2009virtual}. In the virtual societies The users can, in turn, execute actions to interact with other users or virtual objects.

			\item \textit{IoT and sensor networks} deployed in the physical world collect data from the physical world. The insights derived are used to update the virtual worlds, e.g., to maintain DTs for physical entities. 
			% The DT is a digital replica of the physical world which provides rich information to reflect the physical world and enable simulations to be conducted to explore how an intended action will affect the physical world \cite{tao2018digital}. 
			Sensor networks may be independently owned by sensing service providers (SSPs) that contribute live data feeds (i.e., sensing as a service \cite{sheng2013sensing}) to virtual service providers (VSPs) to generate and maintain the virtual worlds.
			
			\item \textit{Virtual service providers (VSPs)} develop and maintain the virtual worlds of the Metaverse. Similar to user-created videos today (e.g., YouTube), the Metaverse is envisioned to be enriched with user-generated content (UGC) that includes games, art, and social applications \cite{kim2012institutionalization}. These UGC can be created, traded, and consumed in the Metaverse.
			
			\item \textit{Physical service providers (PSPs)} operate the physical infrastructure that supports the Metaverse engine and respond to transaction requests that originate from the Metaverse. This includes the operations of communication and computation resources at edge networks or logistics services for the delivery of physical goods transacted in the Metaverse.

		\end{itemize}

		\item The \textit{Metaverse engine} obtains inputs such as data from stakeholder-controlled components that are generated, maintained, and enhanced by entities and their activities in the physical and virtual worlds. % are generated, maintained, and enhanced with these inputs.

		\begin{itemize}
			
			\item \textit{AR/VR} enables users to experience the Metaverse visually, whereas \textit{haptics} enables users to experience the Metaverse through the additional dimension of touch, e.g., using haptic gloves. This enhances user interactions, e.g., through transmitting a handshake across the world, and opens up the possibilities of providing physical services in the Metaverse, e.g., remote surgery. These technologies are developed by the standards that facilitate interoperability, e.g., Virtual Reality Modelling Language (VRML)\footnote{\url{https://www.web3d.org/documents/specifications/14772/V2.0/part1/javascript.html}}, that govern the properties, physics, animation, and rendering of virtual assets, so that users can traverse the Metaverse smoothly and seamlessly.
			
			\item \textit{Tactile Internet} enables users in the Metaverse to transmit/receive haptic and kinesthetic information over the network with a round-trip delay of approximately 1 ms~\cite{6755599}. For example, Meta's haptic glove project\footnote{\url{https://about.fb.com/news/2021/11/reality-labs-haptic-gloves-research/}} aims to provide a viable solution to one of Metaverse's essential problems: using the tactile Internet with haptic feedback to enable users in the physical world to experience the embodied haptic and kinesthetic feeling of their avatars during interacting with virtual objects or other avatars.

			\item \textit{Digital twin} enables some virtual worlds within the Metaverse to be modeled after the physical world in real-time. This is accomplished through modeling and data fusion \cite{liu2018role,khan2022federated}. DT adds to the realism of the Metaverse and facilitates new dimensions of services and social interaction~\cite{bellavista2022digital}. For example, Nvidia Omniverse\footnote{\url{https://blogs.nvidia.com/blog/2021/04/13/nvidia-bmw-factory-future/}} allows BMW to blend its physical automotive factories and VR, AI, and robotics to improve its industrial precision and flexibility, which finally increases BMW's planning efficiency by around 30 percent.
			\item \textit{Artificial Intelligence (AI)} can be leveraged to incorporate intelligence into the Metaverse for improved user experience, e.g., for efficient 3D object rendering, cognitive avatars, and AIGC. For example, the MetaHuman project\footnote{\url{https://www.unrealengine.com/en-US/digital-humans}} by EpicGames utilizes machine learning (ML) to generate life-like digital characters quickly. The generated characters may be deployed by VSPs as conversational virtual assistants to populate the Metaverse. A comprehensive review of AI training and inference in the Metaverse is presented in \cite{huynh2022artificial}.
			
			\item \textit{Economy} governs incentives and content creation, UGC trading, and service provisioning to support all aspects of the Metaverse ecosystem. For example, VSP can pay the price to SSP in exchange for data streams to synchronize the virtual and physical world \cite{han2022dynamic}. Metaverse service providers can also purchase computing resources from cloud services \cite{ng2022unified} to support resource-constrained users. Moreover, the economic system is the driving engine that incentivizes the sustainable development of digital assets and DIDs for the sustainable development of the Metaverse.
			
		\end{itemize}
		\item The \textit{infrastructure layer} enables the Metaverse to be accessed at the edge.
		\begin{itemize}
			
			\item \textit{Communication and Networking:} To prevent breaks in the presence (BIP), i.e., disruptions that cause a user to be aware of the real-world setting, AR/VR and haptic traffic have stringent rate, reliability, and latency requirements \cite{park2018urllc}. Due to the expected explosive data traffic growth, ultra-dense networks deployed in edge networks can potentially alleviate the constrained system capacity. Moreover, the B5G/6G communication infrastructure~\cite{shen2021holistic} will play an instrumental role towards enabling post-Shannon communication to alleviate bandwidth constraints in managing the explosive growth of communication costs. We discuss these concepts in greater detail in Section \ref{sec:communication}.
			
			\item \textit{Computation:} Today, MMO games can host more than a hundred players in a single game session and hence require high-specification GPU requirements. VRMMO games, which are the rudiments of the Metaverse system, are still scarce in the industry. The reason is that VRMMO games may require devices such as HMDs to be connected to powerful computers to render both the immersive virtual worlds and the interactions with hundreds of other players. To enable ubiquitous access to the Metaverse, a promising solution is the cloud-edge-end computation paradigm. Specifically, local computations can be performed on edge devices for the least resource-consuming task, e.g., computations required by the physics engine to determine the movement and positioning of an avatar. To reduce the burden on the cloud for scalability and further reduce end-to-end latency, edge servers can be leveraged to perform costly foreground rendering, which requires less graphical details but lower latency \cite{guo2020adaptive}. The more computation-intensive but less delay-sensitive tasks, e.g., background rendering, can, in turn, be executed on cloud servers. Moreover, AI techniques of model pruning and compression, as well as distributed learning, can reduce the burden on backbone networks. We further discuss these concepts in Section \ref{sec:computation}.
			
			\item \textit{Blockchain:}  The DLT provided by the blockchain will be the key to {provide proof-of-ownership of} virtual goods, as well as establishing the economic ecosystem within the Metaverse. It is difficult for current virtual goods to be of value outside the platforms on which they are traded or created. Blockchain technology will play an essential role in reducing the reliance on such centralization. For example, a Non-fungible token (NFT) serves as a mark of a virtual asset's uniqueness and authenticates one's ownership of the asset \cite{wang2021non}. This mechanism protects the value of virtual goods and facilitates peer-to-peer trading in a decentralized environment. As virtual worlds in the Metaverse are developed by different parties, the user data may also be managed separately. To enable seamless traversal across virtual worlds, multiple parties will need to access and operate on such user data. Due to value isolation among multiple blockchains, cross-chain is a crucial technology to enable secure data interoperability. In addition, blockchain technology has found recent successes in managing edge resources \cite{yang2019integrated}. We discuss these concepts in Section \ref{sec:blockchain}.
		\end{itemize}
		
	\end{enumerate}

	\begin{figure*}[t]
		\centering
		\includegraphics[width=0.8\linewidth]{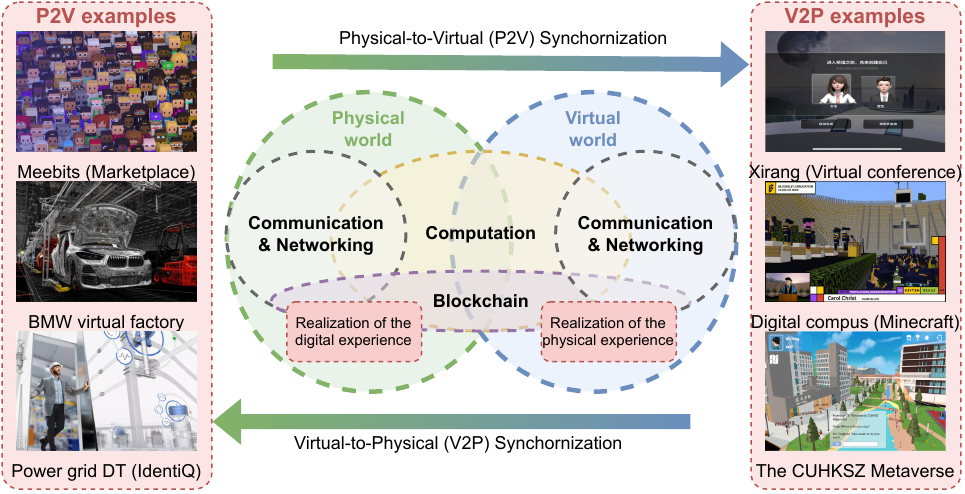}
		\caption{The development path in the Metaverse from the perspective of the physical and virtual worlds.}
		\label{fig:deve}
	\end{figure*}
	\subsection{Current Developments}
	{As shown in Fig.~\ref{fig:deve}, the development of the Metaverse is still in its infancy. We are only in a ``proto-verse" on the development side, with much more to be done to realize the vision of the Metaverse discussed in Section \ref{sec:archi}. Based on the communication and networking, computation, and blockchain infrastructure, the physical-virtual synchronization, including the physical-to-virtual (P2V) synchronization and the virtual-to-physical (V2P) synchronization, will help the Metaverse to realize the digital experience in the physical world while realize the physical experience in the virtual worlds.} For example, by designing more comfortable, cheaper, and lighter AR/VR devices and overcoming the challenges that would be encountered in bringing them to reality. The development of the Metaverse should be viewed from two perspectives. The first perspective is, how can actions in the virtual world affect the physical world? The digital goods in the Metaverse will have real (monetary) value. For example, the company Larva Labs built Meebits\footnote{\url{https://meebits.larvalabs.com/}} as a marketplace to trade avatars on the Ethereum blockchain. In addition, DTs have been utilized to empower smart manufacturing. The second perspective is, how can the actions in the physical world be reflected in the virtual world? This is driven by the digitization and intelligentization of physical objects. For example, a virtual 3D environment will reflect the status of the physical world in real-time to enable remote working, socializing, and service delivery such as education~\cite{microsoftmesh}.
	%Just like the Internet, the Metaverse should have the characteristic of interoperability. This means that the ultimate Metaverse should allow users to traverse smoothly from one sub-Metaverse platform to another.
	
	% Mark Zuckerberg, CEO of Facebook, has already started working on his own Metaverse and changed the company name to Meta. Meta believes that in the future, more people will spend time socializing in the Metaverse~\cite{fb2}. Meta has developed a meeting software, Horizon Workrooms, which allows the users to interact in a 3D conference room. With the help of VR headsets, the users are provided with “spatial sound,” and hand motion tracking~\cite{fb1}. 
	
	%The companies that are participating in the Metaverse are not limited to only technology companies. Instead, the Metaverse industry is also open to other fields such as entertainment, finance, healthcare, and education. 
	
	Companies across many sectors, such as entertainment, finance, healthcare, and education, can work closely to develop Metaverse hand in hand. For example, theme parks have become one of the ideal candidates to be situated in the Metaverse since they are expensive to construct in the physical world but can be done quickly and safely in the Metaverse. In 2020, Disney unveiled their plans to create a theme park in the Metaverse. The company reiterated this vision at the 2021 Q4 earnings call~\cite{dis}. This section looks at some existing and prospective developments that are ``early forms" of the Metaverse. These developments are mainly classified into two types of categories as follows.
	
	\begin{table*}
		\small\centering
		\caption{Features of representative Metaverse examples.}
		\begin{tabular}{|c|c|c|c|c|c|c|c|c|}
			\hline
			\multicolumn{3}{|c|}{\textbf{Metaverse Example}} & \textbf{Blockchain} & \textbf{AR/VR} & \makecell[c]{\textbf{DT}}& \makecell[c]{\textbf{UGC}} & \multicolumn{2}{c|}{\textbf{AI}} \\\cline{1-9}
			
			\multirow{12}{*}{\textbf{MMORPG}} &\multicolumn{2}{c|}{World of Warcraft}  & \color{red}\xmark &\color{red}\xmark & \color{red}\xmark &\color{green}\cmark &\color{green}\cmark & AI enemies\\\cline{2-9}
			&\multicolumn{2}{c|}{Grand Theft Auto Online} &  \color{red}\xmark &\color{red}\xmark&\color{green}\cmark &\color{green}\cmark&\color{green}\cmark&\makecell[c]{AI in rockstar \\advanced game engine}\\\cline{2-9}
			&\multicolumn{2}{c|}{Final Fantasy XIV} &\color{red}\xmark &\color{red}\xmark &\color{red}\xmark &\color{green}\cmark&\color{green}\cmark& AI enemies\\\cline{2-9}
			&\multicolumn{2}{c|}{Pokemon Go} &\color{red}\xmark &\color{green}\cmark &\color{green}\cmark &\color{red}\xmark&\color{green}\cmark & \makecell[c]{Plan route \\more strategically}\\\cline{2-9}
			&\multicolumn{2}{c|}{Animal Crossing: New Horizons} &\color{red}\xmark &\color{red}\xmark &\color{red}\xmark &\color{green}\cmark&\color{green}\cmark& \makecell[c]{AI NPCs}\\\cline{2-9}
			&\multicolumn{2}{c|}{Second Life} &\color{red}\xmark &\color{red}\xmark &\color{red}\xmark &\color{green}\cmark&\color{green}\cmark&\makecell[c]{AI NPCs}\\\cline{2-9}
			&\multicolumn{2}{c|}{Minecraft} &\color{red}\xmark &\color{green}\cmark &\color{red}\xmark &\color{green}\cmark&\color{green}\cmark&\makecell[c]{Platform to train AI}\\\cline{2-9}
			&\multicolumn{2}{c|}{DragonSB} &\color{green}\cmark &\color{red}\xmark &\color{red}\xmark &\color{green}\cmark&\color{green}\cmark&AI enemies\\\cline{1-9}
			
			\multirow{18}{*}{\makecell[c]{\textbf{Applications}\\ \textbf{of the} \\\textbf{Metaverse}}}  &\multirow{3}{*}{Smart Cities}  & Metaverse Seoul&\color{green}\cmark &\color{green}\cmark &\color{green}\cmark &\color{green}\cmark&\color{green}\cmark& \makecell[c]{AI NPCs}\\\cline{3-9}
			& & \makecell[c]{Barbados Metaverse\\ Embassy}&\color{green}\cmark &\color{green}\cmark &\color{green}\cmark &\color{green}\cmark&\color{green}\cmark& \makecell[c]{AI NPCs}\\\cline{2-9}
			& \multirow{3}{*}{Entertainment} & AltspaceVR&\color{green}\cmark &\color{green}\cmark &\color{green}\cmark &\color{green}\cmark&\color{green}\cmark& \makecell[c]{AI NPCs}\\\cline{3-9}
			& &Decentraland&\color{green}\cmark &\color{green}\cmark &\color{red}\xmark &\color{green}\cmark&\color{green}\cmark& \makecell[c]{AI NPCs}\\\cline{2-9}
			& \multirow{3}{*}{Education} &Xirang&\color{green}\cmark &\color{green}\cmark &\color{green}\cmark &\color{green}\cmark&\color{green}\cmark& \makecell[c]{AI NPCs}\\\cline{3-9}
			& &CUHKSZ&\color{green}\cmark &\color{red}\xmark &\color{green}\cmark  &\color{green}\cmark&\color{green}\cmark& \makecell[c]{AI NPCs}\\\cline{2-9}
			& \multirow{3}{*}{Work} & Horizon Workrooms &\color{green}\cmark &\color{green}\cmark&\color{green}\cmark &\color{green}\cmark&\color{green}\cmark&\makecell[c]{Tools for designing \\virtual content}\\\cline{3-9}
			& &Microsoft Mesh &\color{green}\cmark &\color{green}\cmark &\color{green}\cmark &\color{green}\cmark&\color{green}\cmark &\makecell[c]{Tools for avatars, holoportation,\\ and spatial rendering}\\\cline{2-9}
			& Healthcare &  Telemedicine &\color{green}\cmark &\color{green}\cmark &\color{green}\cmark &\color{green}\cmark&\color{green}\cmark&\makecell[c]{Data analysis \\and collaboration}\\
			\hline
		\end{tabular}
		\label{table:example}
	\end{table*}
	
	\subsubsection{Massive Multiplayer Online Role-playing Games} Many have argued that the Metaverse is not new since it has long been implemented in video games. While video games of today may not completely be classified as the Metaverse according to the technical definition provided in Section~\ref{sec:tutorial}, many video games have features that are rather similar to the Metaverse. For example, in massive multiplayer online role-playing games (MMORPG), players can enter the game and live in the digital world using their avatars. These avatars have roles tagged to the players, just as how one has a job in the physical world. MMORPGs can support a massive number of players to interact in a 3D rendered virtual world. Some key examples are as follows:
	\begin{itemize}
		\item \textbf{Second Life: }This is among the first attempt to create a complete virtual world in which players could live. In second life, players control their avatars and can do almost anything they can in the physical world, from job seeking to get married~\cite{secondlife}. Besides, players can customize the environment they live in. The economy in Second Life is supported by the Linden Dollar\footnote{\url{https://www.investopedia.com/terms/l/linden-dollar.asp}}, a virtual currency, which can be traded for real-world currency.
		\item \textbf{Minecraft: }The popular game allows anyone to create their own content by using 3D cubes in an infinite virtual world, and it supports user access using VR devices such as Oculus Rift. Digital copies of cities around the world, e.g., Kansas and Chicago, have been built using Minecraft and can be accessed by any user\footnote{\url{https://www.minecraftmaps.com/tags/real-cities-in-minecraft}}.
		\item \textbf{DragonSB: }DragonSB is the first Metaverse MMORPG built on Terra Protocol and Binance Smart Chain Platform~\cite{dSB}. The players will be rewarded with SB tokens, where the token will be listed on both centralized and decentralized exchanges.
	\end{itemize}
	Other popular MMORPGs are World of Warcraft, Grand Theft Auto Online, Final Fantasy XIV, Pokemon Go, and Animal Crossing: New Horizons. For a detailed discussion of these games, the reader may refer to \cite{duan2021metaverse}. Following the classification framework \cite{duan2021metaverse}, we present a taxonomy for the above gaming Metaverse according to their features in Table~\ref{table:example}.
	\begin{figure*}[t]
		\centering
		%\subfigure[Campuses are rebuilt in Minecraft to host graduation ceremonies for students from hundreds of other universities~\cite{gradu}]{
			%\includegraphics[width=0.3\linewidth, height=0.2\linewidth]{grad.png}
			%   \label{fig:grad}
			%}
		%\quad
		\subfigure[Screenshot of AltspaceVR platform~\cite{altspace}.]{
			\includegraphics[width=0.3\linewidth, height=0.2\linewidth]{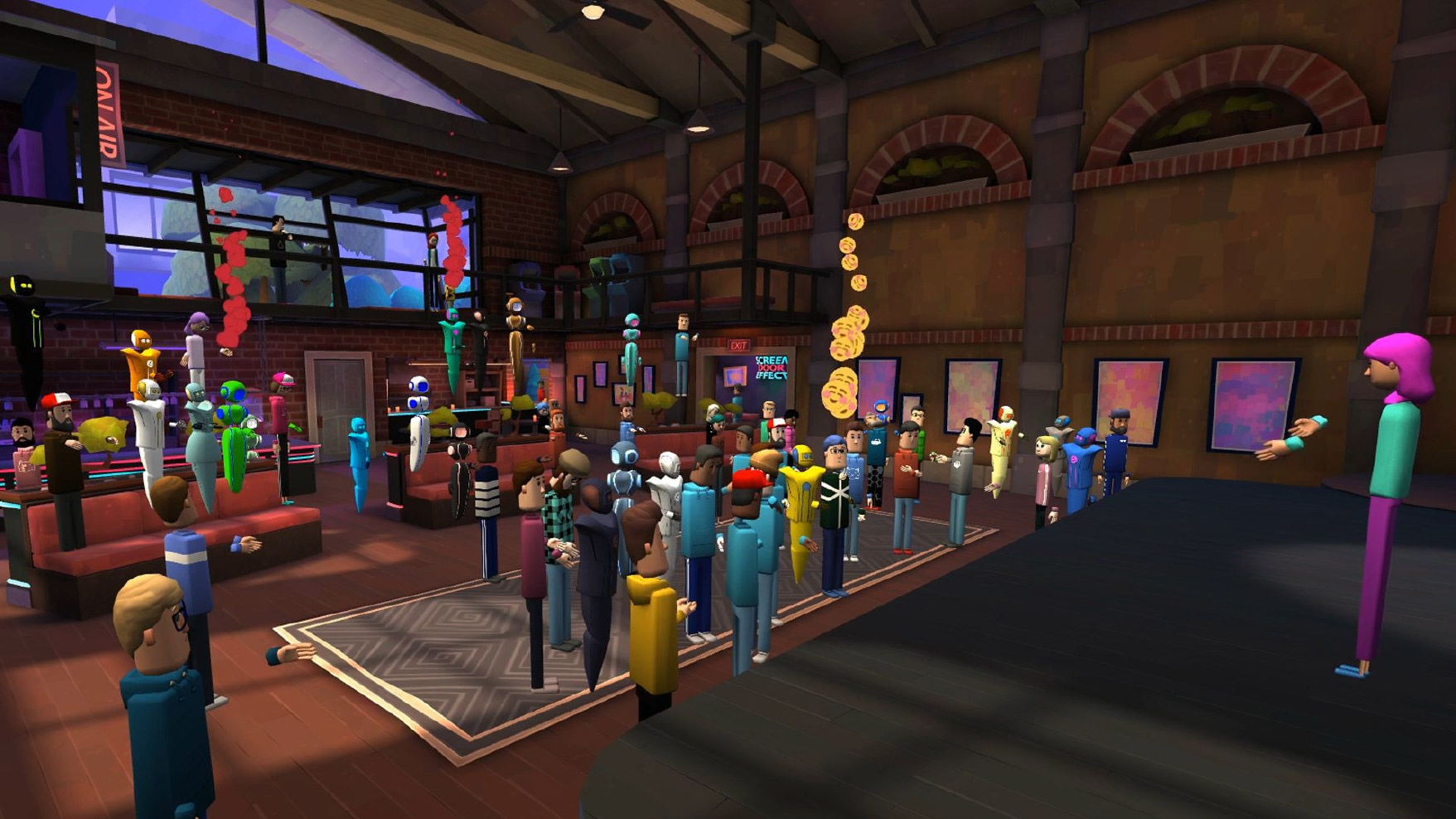}
			\label{fig:altspace}
		}
		%\quad
		%\subfigure[Xirang's UI when creating users' avatars~\cite{def}.]{
			%\includegraphics[width=0.3\linewidth, height=0.2\linewidth]{xirang.jpeg}
			%   \label{fig:xirang}
			%}
		%\quad
		%\subfigure[The Chinese University of Hong Kong, Shenzhen Metaverse is built using Unity %in~\cite{duan2021metaverse}.]{
			%\includegraphics[width=0.3\linewidth, height=0.2\linewidth]{unity.PNG} 
			%   \label{fig:unity}
			%} 
		\quad
		\subfigure[Screenshot of the non-fungible LAND tokens constitute the Decentraland Metaverse~\cite{decen}.]{
			\includegraphics[width=0.3\linewidth, height=0.2\linewidth]{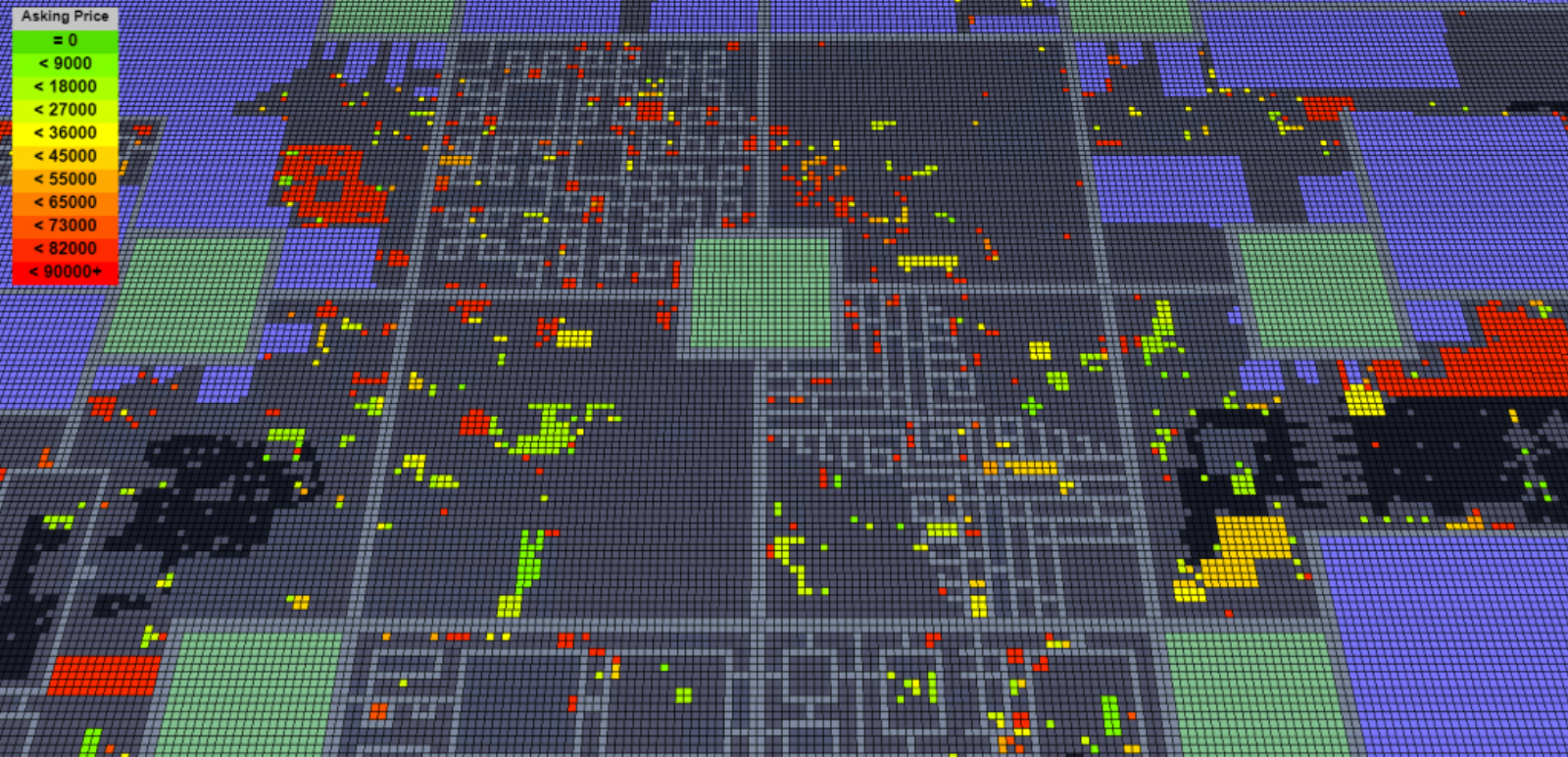}
			\label{fig:decen}
		} 
		\quad
		\subfigure[Nvidia Omniverse for modeling physics, materials, and real-time path tracing~\cite{nvidia}.]{
			\includegraphics[width=0.3\linewidth, height=0.2\linewidth]{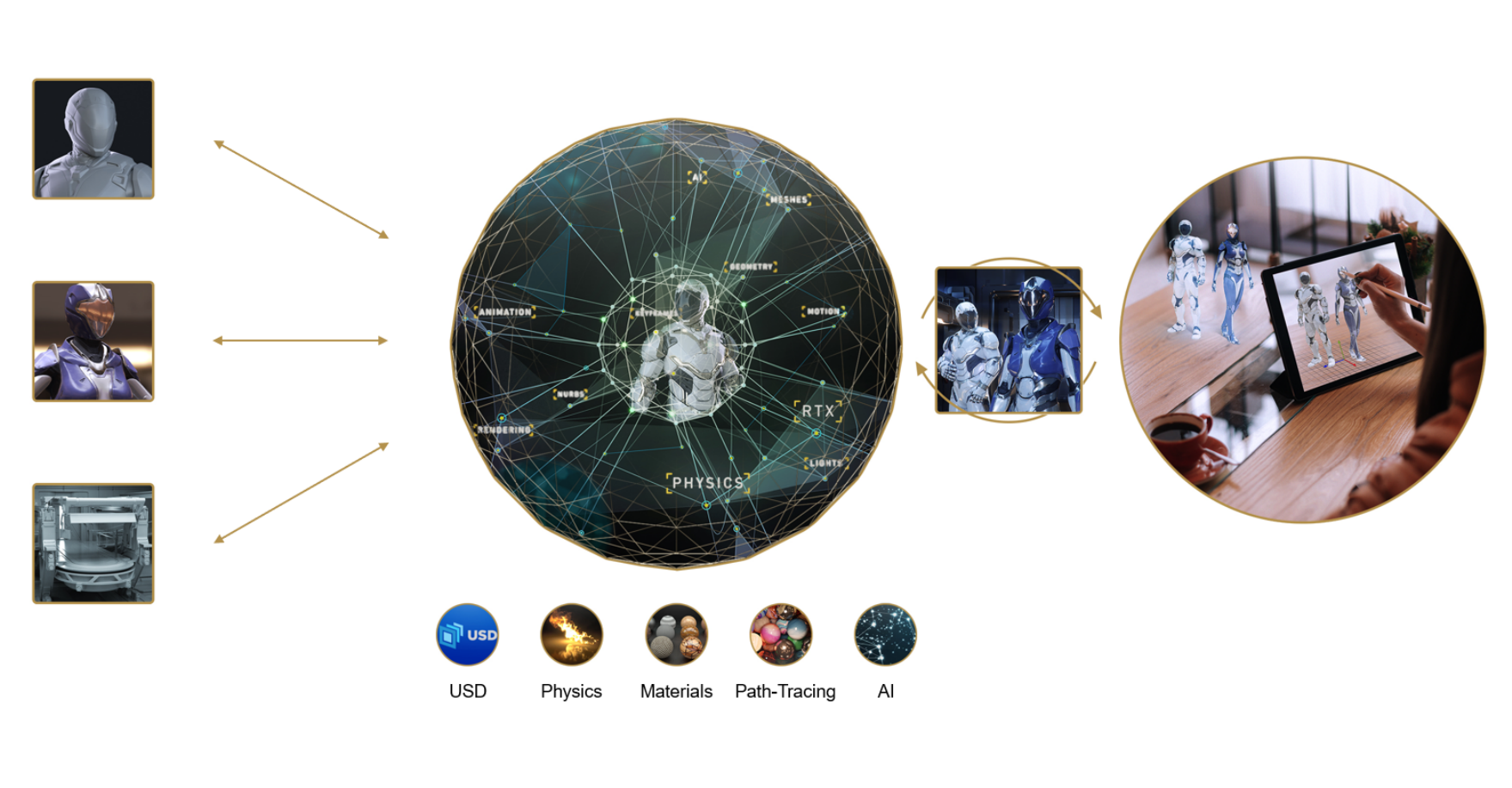}  
			\label{fig:omniverse}
		} 
		\caption{Screenshots of (a) AltspaceVR, (b) Decentraland, and (c) Nvidia Omniverse.}
		\label{fig:fig1}
	\end{figure*}
	\subsubsection{Applications of the Metaverse} Beyond the gaming industry, multiple entities, from governments to tech companies, have announced their interest in establishing a presence in the Metaverse. This includes the smart cities~\cite{zhang2017security}, entertainment industry, education institutions, working environments, and healthcare (Fig.~\ref{fig:fig1}).
	\begin{itemize}
		\item \textbf{Metaverse Seoul:} The Seoul Metropolitan Government plans to release Metaverse Seoul in Seoul Vision 2030~\cite{lim2022realizing}. This platform contains a virtual city hall, tourism destinations, social service centers, and many other features. The local users can meet with avatar officials for consultations and provisions of civil services. The foreign users can tour vividly replicated locations and join in celebrating festival events while wearing their VR headsets. This aims to boost the tourism industry.
		\item\textbf{Barbados Metaverse Embassy:} Barbados, an island nation, is prepared to declare digital real estate sovereign property by establishing a Metaverse embassy~\cite{lim2022realizing}. The government works with multiple Metaverse companies to aid the country in land acquisition, the design of virtual embassies and consulates, the development of infrastructure to provide services such as ``e-visas", and the construction of a ``teleporter" that will allow users to transport their avatars across worlds.
		\item \textbf{AltspaceVR:} AltspaceVR is a virtual meeting space designed for live and virtual events, and it is compatible with a variety of operating systems~\cite{altar}. Basically, the users of AltspaceVR are allowed to meet and engage online through the use of avatars. It recently implemented a safety bubble feature to reduce the risk of inappropriate behavior and harassment.
		\item\textbf{Decentraland:} Decentraland is built on the Ethereum blockchain~\cite{decen}. It is a decentralized virtual world that allows users to buy and sell virtual land. Users can make their own micro-worlds complete with virtual trees, planes, and other features. Everything that is available in the physical world can be found in Decentraland, from transportation systems to hotels.
		\item\textbf{Xirang:} The Chinese tech giant Baidu has released a Metaverse application named Xirang (which means land of hope), allowing the users to explore the virtual worlds using portable gadgets such as smartphones and VR devices~\cite{def}. In addition, personalized avatars can be automatically generated from the user's uploaded images. Xirang hosted a three-day virtual AI developers’ conference in 2021 that supports 100,000 users at once. This is an important step towards realizing the shared Metaverse.
		\item\textbf{The Chinese University of Hong Kong, Shenzhen (CUHKSZ):} CUHKSZ Metaverse is a virtual copy of the physical campus. It is powered by the blockchain to provide the students with an interactive Metaverse, a hybrid environment in which students' real-world behaviors can have an impact on the virtual world, and vice versa~\cite{duan2021metaverse}.
		\item\textbf{Telemedicine:} Telemedicine is immersion in healthcare industry, allowing patients to connect virtually with healthcare professionals when they are physically apart~\cite{Health}. Doctors and their patients can meet in the virtual world through VR to provide telemedicine consultations. While physical follow-ups, medical scans, and tests can then be booked and performed in a facility near the patient, the results can be sent to any preferred doctor.
	\end{itemize}
	
	\begin{table*}
		\small\centering
		\caption{The summary of the development tools, platforms, and frameworks that support the Metaverse ecosystem.}
		\begin{tabular}{|m{.13\textwidth}<{\centering}|m{.8\textwidth}<{\raggedright}|} 
			\hline
			\textbf{Tools} & \textbf{Features} \\ [0.5ex] 
			\hline
			Unity & A platform for creating real-time 2D/3D content, including 3D rendering, physics, collision detection, and etc.\\
			\hline
			Unreal Engine & Real-time 3D creation platform with features like photorealistic rendering, dynamic physics and effects, lifelike animation, and robust data translation\\
			\hline
			Roblox & Similar to Unity and Unreal Engine, provides a platform for content creators to create content in the Metaverse
			\\
			\hline
			Nvidia Omniverse & A platform for 3D simulation and design based on Nvidia RTX technologies\\
			\hline
			Meta Avatars & A avatar system that can be implemented in various platforms to ease the creation of authentic avatars\\
			\hline
			Meta’s Haptics Prototype & A glove that allows the users to feel when the virtual object is touched\\
			\hline
			Hololens2 & A head-mounted mixed reality device\\
			\hline
			Oculus Quest2 & A head-mounted VR device\\
			\hline
			Xverse & Blockchain-based virtual ecosystem, where users may use their digital assets to play, create, share, and work\\
			\hline
			AnamXR & Cloud-based virtual e-commerce platform based on Unreal Engine to transform physical objects into pixels\\
			\hline
			Microsoft Mesh & A platform that enables presence and shared experience from any device through mixed reality applications\\
			\hline
			Structure Sensor & A 3D sensor that can be mounted onto mobile devices for 3D scanning\\ 
			\hline
			Canvas & Software for 3D room scanning, creating accurate 3D models using Structure Sensor\\
			\hline
			Gaimin & Allow game/content developers to build gaming tokens into the Metaverse\\
			\hline
		\end{tabular}
		\label{table:software}
	\end{table*}

	\subsection{Tools, Platforms, and Frameworks}
	% Currently, the development of the Metaverse is still at the early stage. The full version of the Metaverse is still decades away. It is still required tremendous technological developments, content creation, and protocol standardization. Therefore, 
	In this section, we will discuss a variety of key tools, platforms, and frameworks used currently in the development of the Metaverse. Some examples are AnamXR, Microsoft Mesh, Structure Sensor, Canvas, and Gaimin. Table~\ref{table:software} provides a summary of the features of these development tools that support the Metaverse ecosystem. Some of the screen captures of these tools are also shown in Fig.~\ref{fig:fig1}.
	\begin{itemize}
		\item \textbf{Unity: } Unity is a 3D content creation environment that includes a fully integrated 3D engine as well as studio design~\cite{metaversetools}. Virtual reality, augmented reality, and Metaverse experience can be created using the aforementioned features. Decentralization options in the Metaverse are provided by the Unity components store, which includes features such as edge computing, AI agents, microservices, and blockchain.
		\item \textbf{Unreal Engine: } Unreal Engine is a Metaverse development tool. It includes designing studios, i.e., MetaHuman Creator, and an asset marketplace~\cite{metaversetools}. Using MetaHuman Creator, the required time of user avatar creation process can be reduced from months to hours with unprecedented quality, fidelity, and realism.
		\item \textbf{Roblox: } While Roblox is known by many as a game, it has slowly evolved into becoming a platform that provides multitudes of development tools such as avatar generation\footnote{\url{https://developer.roblox.com/en-us/api-reference/class/Tool}} and teleportation mechanisms~\cite{metaversetools}. Roblox offers a proprietary 3D engine connected with the design platform and a marketplace where creators may trade assets and code.
		\item \textbf{Nvidia Omniverse: }The graphics and AI chips company Nvidia is going to build a Metaverse platform called Omniverse~\cite{nvidia}. Omniverse is a scalable, multi-GPU real-time reference development platform for 3D simulation and design collaboration~\cite{nvidia}. Content producers can use the Omniverse platform to integrate and expedite their 3D workflows, while developers may quickly construct new tools and services by plugging into the platform layer of the Omniverse stack. Nvidia will simulate all factories in the Omniverse to reduce the amount of waste. Nvidia's CEO Jensen Huang created his own avatar with AI and computer graphics modeling and presented it at a press conference in April 2021.
		\item\textbf{Meta Avatars:} Meta Avatars is an avatar system to offer support to all Unity developers on Quest, Rift, and Windows-based VR platforms~\cite{metaavatar}. Users may customize their Meta Avatars by using a platform-integrated editor, ensuring that all apps that use the Meta Avatars SDK reflect them consistently and authentically. Hand tracking, controller tracking, and audio input are all used by Meta Avatars to produce expressive Avatar representations.
		\item\textbf{Meta Haptics Prototype:} Meta’s haptic glove allows the users to feel an object and it also acts as a VR controller~\cite{airpockets}. The glove contains inflatable plastic pads to fit along the wearer's palm, underside of fingers, and fingertips. When the VR application runs, the system will regulate the level of inflation and exert pressure on different portions of the hand. If the user contacts an object, the user will experience the sensation of a virtual object pressing into the skin. When the user grips a virtual item, the lengthy finger actuators tighten, giving the user a resistive feeling.
		\item\textbf{Hololens2: } HoloLens2 is a head-mounted mixed reality device and comes with an onboard computer and built-in Wi-Fi~\cite{holo}. It has more computing power, better sensors, and longer battery life when compared with its predecessor. Hololens2 can be used together with Microsoft Mesh to connect users worldwide. The user can communicate, maintain eye contact, and observe other users' facial expressions who are located elsewhere.
		\item\textbf{Oculus Quest2: } Oculus Quest2 is a head-mounted VR device from Meta~\cite{metaoculus}, which can be connected to a desktop computer with USB or Wi-Fi to run compatible VR applications.
	\end{itemize}
	
	\subsection{Lessons Learned}
	\subsubsection{The Metaverse cannot be realized by a standalone technology}
		From the aforementioned definition, we can learn that the Metaverse is built from the convergence of multiple engines for its physical synchronization, i.e., AR/VR, the tactile Internet, DT, AI, and blockchain-based economy. However, the Metaverse cannot be considered as only one or some of them. In detail, the lessons learned are listed as follows:
		\begin{itemize}
			\item AR/VR: AR/VR is only one of the most popular methods to immerse users in 3D virtual worlds during the V2P synchronization of the Metaverse. In detail, AR/VR will offer Metaverse users with 3D virtual/auditory content.
			\item Tactile Internet: As another popular method to immerse users in 3D virtual worlds during the V2P synchronization of the Metaverse, tactile Internet can offer tactile/kinesthetic content to users in the Metaverse as the supplementary to visual/auditory content of AR/VR.
			\item Digital twin: DT is the foundation during the P2V synchronization of the Metaverse that provides ubiquitous connectivity between physical and virtual entities. Based on DT of the physical world, the Metaverse can support a better V2P synchronization but cannot directly provide the immersive experience to users.
			\item AI: The Metaverse is considered to be AI-native, where ubiquitous AI exists in the physical and virtual worlds. AI can make the physical-virtual synchronization become more efficient. However, it cannot facilitate the synchronization directly.
			\item Blockchain-based Economy: Blockchain-based Economy, a.k.a, Web 3.0, is envisioned as the decentralized Internet. Evolving from the read Web 1.0, the read/write Web 2.0, the Web 3.0 is read/write/own. However, the Metaverse is the next important stage and long-term vision of the continuous digital transformation, which is more informative that the discrete definition of Web 1.0/2.0/3.0.
		\end{itemize}
	
	\subsubsection{Implementation challenges}
	The representative Metaverse examples discussed in Table \ref{table:example} may be ``prototypes" or early development examples of the Metaverse. Still, many of them, e.g., Grand Theft Auto Online, already require high-specification computers to run. Moreover, the application examples such as Metaverse Seoul are ambitious projects that require computation-intensive rendering, real-time updating of DTs, and low-latency reliable communications to be realized. Clearly, the implementation challenges of the Metaverse have to be addressed. We do so from the communication and networking, computation, as well as blockchain perspectives in the subsequent sections. 
	\subsubsection{Standardized Protocols, Frameworks, and Tools} 
	Each company is building its own Metaverse based on different protocols, frameworks, and tools. Collaborative efforts have to be made to standardize the Metaverse protocols, framework, and tools. Once standardized, it will be easier for the content creators to create content as the content can be easily transferable among various virtual worlds, even if they are developed by different companies. Furthermore, the users can experience a smooth transition when they are ``teleported” across different virtual worlds in the Metaverse. The interconnected virtual worlds will be the Metaverse's main difference/advantage as compared to the MMORPG games discussed previously.
	
	\subsubsection{Metaverse Ecosystem and Economics} 
	Many of the examples given in this section rely on and enable UGC and open-source contributions. For example, Roblox allows game creation by any user, and Meta Avatars supports user-friendly game development. The functioning of the Metaverse ecosystem will therefore leverage the contribution of users worldwide. The UGC and services provided by users, therefore, can be supported and valued with cryptocurrency-promoted transactions without the reliance on third-party entities. Due to safety and security reasons, the transactions in the Metaverse are mainly based on cryptocurrencies. However, the current payment gateway interface is complex/not user-friendly to the general public. A simple interface can have better support in digital currency exchanges and payment across fiat currencies and cryptocurrencies. We revisit challenges and solutions for blockchain technology subsequently.
	%The Metaverse involves the resources from the virtual worlds and the physical world. Whenever a user performs a certain action in the physical world, a corresponding real-time update has to be reflected in the virtual worlds or vice versa. For example, when the lecture theatre has reached its maximum capacity, this information should be reflected in the virtual worlds immediately to avoid any inconvenience. The rest of the users can attend the lecture online, virtually.

	\section{Communication and Networking}
	\label{sec:communication}
	The early representative developments of the Metaverse discussed in Table \ref{table:example} are mostly ambitious projects that will require high-specification edge devices to run. The accessibility of the Metaverse is, therefore, a significant challenge. The mobile edge networks are expected to provide users with efficient communication and networking coverage with high-speed, low-delay, and wide-area wireless access so that users can immerse themselves in the Metaverse and have seamless and real-time experience~\cite{duan2021metaverse}. To achieve this, stringent communication requirements have to be met (Table~\ref{table:requirements}). Unlike the traditional transmission of 2D images, the need to transmit 3D virtual items and scenarios in the Metaverse to provide users with an immersive experience places a significant burden on the current 5G network infrastructure. 
	For example, the Metaverse for mobile edge networks proposed in~\cite{khan2022metaverse} demonstrates that massive connectivity of Metaverse enablers, e.g., AR/VR, tactile Internet, avatars, and DT, require large bandwidth and URLLC to support immersive content delivery and real-time interaction. In addition, a Metaverse-native communication system based on the blockchain is proposed in \cite{xu2022metaverse} to provide decentralized and anonymous connectivity for all physical and virtual entities with respect to encrypted addresses.
	% For example, massive user interactions require large bandwidth to support the delivery of high-resolution content to players and URLLC to facilitate real-time and extensive player interactions with few interruptions\cite{khan2022metaverse}. 
	Virtual concerts and executions of manufacturing decisions have to be in real-time. Overall, the following characteristics of the Metaverse present significant new challenges to mobile edge networks in providing Metaverse services:
	
		\begin{enumerate}
			\item \textit{Immersive 3D streaming:} To blur the boundary of the physical and virtual worlds, 3D streaming enables the immersive experience of users in the Metaverse to be delivered to millions of users at mobile edge networks.
			\item \textit{Multi-sensory communications:} The Metaverse consists of multiple 3D virtual worlds~\cite{what2021nivida}, where users can immerse as avatars with multi-sensory perception, including visual, auditory, haptic, and kinesthetic. Therefore, edge users need to have ubiquitous connectivity to the Metaverse to access the 3D multimedia services, such as AR/VR and the tactile Internet, anytime and anywhere.
			\item \textit{Real-time interaction:} Massive types of interactions, including human-to-human (H2H), human-to-machine (H2M), and machine-to-machine (M2M) communications, form the basis for user-user and user-avatar real-time interactions in the Metaverse. Therefore, social multimedia services in the Metaverse have to be delivered under stringent requirements in real-time interactions, such as motion-to-photon latency~\cite{zhao2017estimating}, interaction latency, and haptic perception latency. % This way, user embodied experience is satisfied in the virtual worlds compared with the life-like experience in the physical world. 
			\item \textit{Seamless physical-virtual synchronization:} To provide seamless synchronization services among the physical and virtual worlds, physical and virtual entities need to communicate and update their real-time statuses with each other. The importance and valuation of synchronizing data may be affected by the age of information (AoI) or the value of information (VoI)~\cite{popovski2021internet}.
			\item \textit{Multi-dimensional collaboration:} Maintenance of the Metaverse entails a multi-dimensional collaboration of VSPs and PSPs in both the physical and virtual worlds~\cite{pan2021network}. For example, as discussed in Section \ref{sec:archi}, the physical entities in edge networks, e.g., sensor networks operated by PSPs, need to collect information in the physical world regularly to ensure the freshness of information for P2V synchronization. In return, decisions in the virtual worlds can be transformed into actuation in the physical world. This loop requires the collaboration of a multitude of physical and virtual entities over multiple dimensions, e.g., time and space.
		\end{enumerate}

\begin{figure}
		\centering
		\includegraphics[width=1\linewidth]{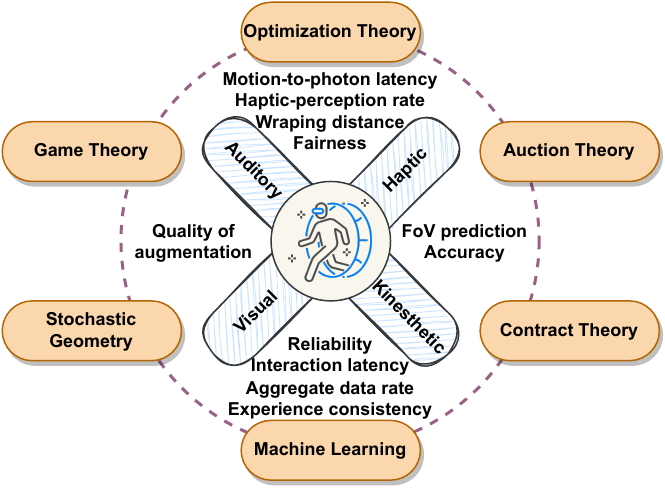}
		\caption{Mathematical tools and human-oriented metrics for designing multi-sensory multimedia networks.}
		\label{fig:multisensory}
	\end{figure}

	To address these issues, we provide a review of cutting-edge communication and networking solutions for the edge-enabled Metaverse. First, users must immerse in the Metaverse via the delivery of 3D multimedia services seamlessly in Section~\ref{sec:multimedia}. This includes the support of VR streaming and AR adaptation, allowing users and physical entities to synchronize with the virtual worlds seamlessly. Beyond conventional content delivery networks, communication and networking that supports the Metaverse must prioritize user-centric considerations in the content delivery process as shown in Section~\ref{sec:human}. The reason is that the Metaverse is full of contextual and personalized content-based services like AR/VR and the tactile Internet. Moreover, the explosive growth of data and limited bandwidth necessitates a paradigm shift away from the conventional focus of classical information theory. In response, we review semantic/goal-oriented communication solutions~\cite{yang2022semantic} that can serve to alleviate the spectrum scarcity for next-generation multimedia services. Finally, real-time bidirectional physical-virtual synchronization, i.e., DT, is essential for the construction of the Metaverse, which is discussed in Section~\ref{sec:synchronization}. The sensing and actuation interaction between the physical and virtual worlds will have to leverage intelligent communication infrastructures such as reconfigurable intelligent surface (RIS) and unmanned aerial vehicle (UAV). We present the instrumental mathematical tools and human-oriented metrics in Fig.~\ref{fig:multisensory}, as well as a brief summary of the reviewed papers, in terms of scenarios, problems, performance metrics, and mathematical tools in Table~\ref{table:com_sum}. %Note that in the Metaverse, users can travel freely through their digital avatars across different cyberspaces at a speed that depends on the speed of the digital signal transmission, i.e. the speed of light ($3\times 10^8$m/s). This is the first time that human beings have the ability to move at sub-light speeds.

	\begin{table*}[t]
		\caption{Communication requirements of services in the Metaverse\cite{alves2021beyond}.}
		\small\centering
		\begin{tabular}{|l|l|l|l|l|}
			\hline
			\textbf{Services}         & \textbf{Reliability} (\%) & \textbf{Latency} (ms) & \textbf{Data Rate} (Mbit/s) & \makecell[l]{\textbf{Connection Density} \\(devices/km$^2$)}\\ \hline
			VR entertainment             &        1 - 10$^{-5}$      &   7-15           &      250        &   1,000-50,000   \\ \hline
			\makecell[l]{The tactile Internet}              &         1 - 10$^{-6}$         &        1      &       1      &    1,000-50,000   \\ \hline
			Digital twin of smart city &        1 - 10$^{-5}$          &        5-10     &     10     &   100,000       \\ \hline
			AR smart healthcare &        1 - 10$^{-6}$          &        5     &     10,000     &   50       \\ \hline
			Hologram education (point could) &        1 - 10$^{-5}$          &        20     &     500-2,000     &   1,000       \\ \hline
			Hologram education (light field) &        1 - 10$^{-5}$          &        20     &     1 $\times$10$^5$-1 $\times$10$^6$    &   1,000       \\ \hline
		\end{tabular}\label{table:requirements}
	\end{table*}
	
	\subsection{Rate-Reliability-Latency 3D Multimedia Networks}
	\label{sec:multimedia}
	The seamless and immersive experience of embodied telepresent via AR/VR brought by the Metaverse places high demands on the communication and network infrastructure of mobile edge networks in terms of transmission rate, reliability, and latency~\cite{saad2019vision}. In particular, the high volume of data exchanged to support AR/VR services that traverse between the virtual worlds and the physical world requires the holistic performance of the edge networking and communication infrastructure that optimizes the trade-off among rate, reliability, and latency. 
	
	The first dimension of consideration is the data rate that supports round-trip interactions between the Metaverse and the physical world (e.g., for users and sensor networks aiding in the physical-virtual world synchronization)~\cite{letaief2019roadmap,quang2022virtual, mao2022rate}. Second, interaction latency is another critical challenge for users to experience realism in the AR/VR services~\cite{you2021towards}. Mobile edge networks, for example, allow players to ubiquitously participate in massively  multiplayer online games that require ultra-low latency for smooth interaction. The reason is that latency determines how quickly players receive information about their situation in the virtual worlds and how quickly their responses are transmitted to other players. Finally, the third dimension is the reliability of physical network services, which refers to the frequency of BIP for users connecting to the Metaverse~\cite{tataria20216g}. Moreover, the reliability requirements of AR/VR applications and users may vary dynamically over time, thereby complicating the edge resource allocation for AR/VR services provision of VSPs and PSPs.
	
	% However, 3D wireless communication design is challenging as it should satisfy two conflicting requirements from a physical layer transport perspective~\cite{bennis2018ultrareliable}: (i) Minimizing latency requires the use of short packets, which in turn leads to a severe degradation of channel coding gain; (ii) Ensuring reliability requires more resources (e.g. parity, redundancy and retransmission) while increasing latency (especially for time domain redundancy).
	
	\subsubsection{\textbf{Resource Allocation for VR Streaming}}
	
	 In the Metaverse, the most direct way for wireless users to access the virtual worlds at mobile edge networks is through VR streaming. Edge servers are expected to stream VR content to users immersed in the Metaverse with high-performance guarantees in terms of rate, reliability, and latency~\cite{bastug2017toward}.
	% Users accessing the 3D worlds via VR require data-oriented edge networks to provide high-performance guarantees of rate, reliability, and latency~\cite{bastug2017toward}.
	However, spectrum resources of edge networks and wireless devices' energy resources are limited, leading to challenging resource allocation problems for VR streaming services. Performance analysis and optimization of VR streaming services in the Metaverse require a trade-off between multiple objectives and a shift from simple averaging to fine-grained analysis to address spurs, distribution, and quality of service (QoS) issues. For example, to address the spurs of heterogeneous users, mobile edge networks provide a variety of alternative VR transmission modes, allowing users to choose the service mode that best suits their needs. Meanwhile, users can dynamically decide the quality and utility of multimedia services based on their own demand and cost of distribution. Finally, the multimedia service providers in the mobile edge network dynamically adjust their 3D services and user field-of-view (FoV), i.e., the user attention window, prediction to maximize the utility of all Metaverse users in edge networks.% Some recent research in~\cite{bennis2018ultrareliable, durisi2016toward} provides the first step toward building high-performance wireless networks to support next-generation Internet services.
	
	\paragraph{VR streaming model selection} The VR content of the Metaverse is expected to be delivered in 6G heterogeneous networks (HetNets)~\cite{tsiropoulos2017cooperation, feng2020smart}. First, VR content delivered in marcocell broadcasting can provide the largest coverage but identical QoS to all VR users within the coverage of marcocells~\cite{gimenez20195g}. Second, to improve the performance of marcocell broadcasting, small cell unicasting VR streaming~\cite{elbamby2018edge} can help marcocells to extend their coverage and thus improve QoS of the VR streaming system. To efficiently utilize the available communication resources of mobile devices, VR users can also adaptively participate in device-to-device (D2D) multicast clusters and send shared reused tracking signals to service providers. In this way, the performance of the VR steaming system can be improved by increasing the total bit rate within the cell while ensuring short-term equality among users~\cite{militano2015single}. To provide accurate indoor positioning services to VR users, the authors in~\cite{wang2022meta} develop a VR service provisioning model over terahertz (THz) /visible light communication (VLC) wireless networks. The THz/VLC wireless networks~\cite{chaccour2022can} are capable of providing larger available bandwidth for the VR content transmission requiring unprecedentedly high data rate~\cite{abbas2010constructing}. In the proposed framework, VR users can 
turn on a suitable VLC access point and select the base station with which they are associated. Aiming at the quick adaptation of new users' movement patterns, they propose a meta reinforcement learning (RL) algorithm to maximize the average number of successfully served users.
	Finally, to integrate these three types of streaming schemes, the authors in~\cite{feng2020smart} propose an intelligent streaming mode selection framework to improve  the throughput of VR streaming systems, using unsupervised learning for D2D clustering and RL for online transmission mode selection. From their performance evaluation results, the proposed hybrid transmission mode selection framework can achieve more than twice the average data rate as compared to the baseline schemes that only rely on small cell unicasting or macrocell broadcasting. 
	
	\paragraph{Field-of-view prediction in VR streaming} In addition to the dynamic selection of transmission modes for VR services, tiling and prefetching (e.g., from edge server caches) of VR content are promising solutions to enhance the user experience in 360$^\circ$ VR video streaming~\cite{hu2021virtual}. For example, tiling of VR content can save bandwidth by only transmitting the pixels that are watched. Moreover, prefetching VR content before they are rendered can reduce the latency of 3D multimedia services. However, the performance of tiling and prefetching techniques depends on the allocated resources and the FoV prediction accuracy. Therefore, it is important to consider how to balance the allocation of communication resources and the fetched VR scenes when streaming to multiple wireless users. To address this complex problem, the authors in ~\cite{zhang2021buffer} propose a federated learning (FL)-based approach to train VR viewpoint prediction models in a distributed and privacy-preserving manner. In this way, VR viewpoint prediction models with good performance can reduce the download of VR tiles that are not rendered, thereby saving a large portion of the data transfer rate and improving the resource efficiency of mobile edge networks~\cite{chung2018hand}. In detail, the authors consider the impact of viewpoint prediction and prefetching in multi-user scenarios and develop a VR stochastic buffering game for the resource allocation problem in wireless edge networks. In this game formulation, VR users independently decide the requested data rate depending on the dynamics of the local network and their current viewing states. Since each user has limited information, the analytical solution of this game cannot be obtained. The users' observations, actions, and utility are modeled as a partially observable Markov decision process, with long-term utility maximized using the Dueling Double Deep Recurrent Q-Network~\cite{ou2021autonomous} algorithm. The experimental results show that the proposed learning-based algorithm can improve the viewing utility by 60\% over the baseline algorithm with the exact prediction accuracy and double the utility with the same number of resource blocks. 
	%For example, visual traffic requires high data rate and relatively low reliability with packet error rate (PER) on the order of $10^{-1}\sim 10^{-3}$~\cite{shi2009effects, park2017revisiting}. This requirements can be supported mostly through enhanced mobile broadband (eMBB) links~\cite{series2015imt}. On the other hand, Haptic traffic, by contrast, should guarantee a fixed target rate and high reliability with PER on the order of $10^{-4} \sim 10^{-5}$~\cite{steinbach2012haptic, zhang2018towards}, which can be satisfied via ultra-reliable and low latency communication (URLLC) links~\cite{popovski2018wireless, bennis2018ultrareliable}. 

	\paragraph{Scalable VR streaming systems} The scalability and self-adaptability of VR streaming systems are essential considerations for the Metaverse at mobile edge networks. In~\cite{guo2020adaptive}, the authors propose a joint optimization algorithm to maximize the quality of experience (QoE) of VR users. Specifically, they classify the VR service problem into VR user association, VR service offloading, and VR content caching. Due to the intractability of this joint optimization problem, they propose a two-stage offline training and online exploitation algorithm to support the delivery of high-quality VR services over mmWave-enabled mobile edge networks. Different from the conventional transmission of 2D objects, VR content is 360$^\circ$ and panoramic. As users move through the 3D world in the Metaverse, their eyes focus on specific FoV. To accurately optimize for high-quality delivery of content, the movements of users' eyes and heads, or hands for interaction with the virtual worlds, have to be closely tracked. The authors in~\cite{chen2018virtual} propose a joint metric that accounts for tracking accuracy and processing and transmission latency, based on the multi-attribute utility theory. The multi-attribute utility theory model assigns a unique value to each tracking and delay component of VR QoS while allowing the proposed sum utility function to capture the VR QoS effectively. They aim to rationally allocate resources in the small-cell networks to meet users' multi-attribute utility requirements. Therefore, they formulate this resource allocation problem of VR as a non-cooperative game among base stations. To obtain a hybrid policy equilibrium for VR transmission, they propose a low-complexity ML algorithm to capture the multi-attribute utility of VR users for the service.

	\paragraph{Multimodal VR streaming} Beyond visual content, multimodal content delivery is essential to provide users with immersive experience in the Metaverse. The authors in~\cite{park2018urllc} investigate the problem of simultaneously supporting visual and haptic content delivery from a network-slicing perspective. While high-resolution visual content delivery demands medium reliability and maximum data rate, haptic perception requires a fixed data rate and very high reliability. Therefore, visual content needs to be provided to VR users over eMMB networks, and haptic content has to be delivered over URLLC networks. Moreover, to mitigate the problem of self-interference in multimodal VR services, they employ successive interference cancellation (SIC)~\cite{meredith2015study} for visual-haptic VR services in different transmission power domains. To derive the expression for the average rate in this eMMB-URLLC multiplexed VR service model, the authors use stochastic geometry to analyze the orthogonal multiple access (OMA) and non-orthogonal multiple access (NOMA) downlinks in large-scale cellular networks.
	
	\paragraph{Edge wireless VR service marketplace} As the aforementioned studies leverage edge resources to support the high-quality delivery of content, economic consideration of service pricing of scarce resources at the edge is also essential. The economic mechanisms for pricing services for entities involved in Metaverse, such as VSP, PSP, and data owners, are necessary to facilitate collaborative efforts to achieve Metaverse. In~\cite{xu2022wireless}, the authors design a learning-based double Dutch auction scheme to price and match users with VR service providers (e.g., base stations) in the Metaverse. In this scheme, VR users and service providers can make fast transactions within the call market to meet the short-term demand for VR services. Experimental results show that this learning-based scheme can achieve near-optimal social welfare while reducing the auction information exchange cost.
	
	\subsubsection{\textbf{Resource Allocation for AR Adaptation}}
	
	In addition to VR streaming, another way to access the Metaverse is to adapt to the real-world environment through AR or mobile AR (MAR)~\cite{siriwardhana2021survey}. At mobile edge networks, users use edge-supported AR devices to upload and analyze their environment. After AR customization, they download the appropriate AR content from edge servers and access the Metaverse. However, resources on the edge networks are limited, while the customization and communication resources used for AR adaptation are enormous~\cite{qiu2018avr, liu2019edge,huang2022joint}. Therefore, the resource allocation problem has to be solved for efficient AR adaptation. Specifically, users need to balance a trade-off between accuracy of adaptation and latency of interaction, i.e., the longer the adaptation time, the higher accuracy of AR adaptation.
	
	\paragraph{Adaptive configuration of MAR services} Mobile edge networks can provide two typical types of MAR services to Metaverse users, namely the precise AR and the approximate AR~\cite{liu2018dare}. The precise AR detects all possible objects in uploaded images using computer vision algorithms to provide a rich AR interaction experience. In contrast, the approximate AR detects only the targets where visual attention of AR users are focusing, thereby reducing the computation and communication overhead on edge servers. Therefore, the precise AR usually requires higher computing capacity and service delay than the approximate AR. The reason is that the high-resolution image data needs to be uploaded for the precise AR requires more communication resources to transmit and more computation resources to process.
	The QoA for MAR services reflects the accuracy of the analysis and the latency of AR adaptation services. To maximize the QoA for users under different network conditions and edge servers with various computing capacities, the authors in~\cite{liu2018dare} propose a framework for dynamically configuring adaptive MAR services. In their proposed framework, MAR users can upload the environmental content that needs to be processed to the edge server, depending on their service requirements and wireless channel conditions. Based on the computation resources, AR service providers select the object detection algorithms and allocate computation resources to process the environmental information uploaded by MAR users after analyzing MAR requests (e.g., the size of video frames). % AR service providers send the AR configuration information back to MAR users according to computation capacities. Based on the AR service provider's description of its processing model and computation resources, the MAR user adjusts the video frame size and frame rate according to its requirements and tolerable latency and uploads the modified AR configuration information back to the AR service provider. Finally, the AR service provider processes the content uploaded by the AR user according to its preferred configuration and sends the resulting AR content back to the MAR user. In this way, by dynamically adjusting the video frame size and video frame rate of the MAR user, the AR service provider can provide the desired processing model and computing resources to AR users.
	
	\paragraph{Tradeoff between latency and AR accuracy} In general, the higher the latency of AR services (i.e., the longer the processing time of the AR service provider and the longer the upload time of the MAR user), the higher the analysis accuracy of the service. To improve performance over edge MAR systems, the authors in~\cite{liu2018edge} develop a network coordinator to balance the trade-off between analysis accuracy and service latency. They rely on the block coordinate descent method~\cite{grippo2000convergence} to develop a multi-objective analysis algorithm for MAR services that enables quick and precise target analysis of AR content. This way, the AR server's resource allocation efficiency is improved, and it can better serve MAR users with different requirements in frame size and frame rate. 
	
	\paragraph{Shared multi-user AR} Multiple users in popular areas simultaneously request AR services from AR service providers. To extend MAR to multi-user scenarios, the authors in~\cite{ren2020edge} propose a multi-user collaborative MAR framework in edge networks called Edge AR X5. They present a multi-user communication and prediction protocol that considers cross-platform interaction, efficient communication, and economic computation requirements of heterogeneous AR users at mobile edge networks. They propose a multi-user communication protocol and a prediction-based key-frame selection algorithm. In detail, AR users and service providers can dynamically create appropriate communication schemes in the multi-user communication protocol according to their respective network environments. Moreover, the key frames of AR applications can be predicted based on the movement information of AR users. Moreover, a peer-to-peer co-processing scheme solves the initialization problem when AR users join. To enhance the interactivity and sharing of MAR, the authors in~\cite{ran2019sharear} propose a shared AR framework, where AR user providers can experience the same AR services for their range of content MAR users and interact with each other. Through this type of shared MAR service, MAR service providers can significantly enhance their users' immersion and social acceptance in the Metaverse~\cite{duan2021metaverse,9772313}.
	
	\subsubsection{\textbf{Edge Caching for AR/VR Content}}
	
	Real-time AR/VR delivery is required to immerse users in the shared 3D virtual worlds of the Metaverse. At mobile edge networks, AR/VR content caches can be deployed at edge servers and edge devices to reduce transmission overhead~\cite{hennessy2011computer, paschos2018role, li2018survey, gao2020design}. However, these shared 3D worlds present a new challenge for distributed edge caching. The question of efficiently using the edge cache while ensuring the uniformity of Metaverse content across geographically distributed edge servers has yet to be resolved. The base stations can leverage their caching resources to improve the QoE of wireless users~\cite{wang2017survey}. In the Metaverse vision, most of the services provided are closely related to the context in which the content is provided, such as AR/VR and DT. Context-aware services for these applications can benefit from caching popular content on edge servers to ensure low latency and efficient allocation of network resources. In the Metaverse, where users are immersed in a ubiquitous, real-time shared and consistent digital world, caching devices at the edge receive large amounts of heterogeneous edge stream data. Building an integrated Metaverse with distributed caching is a formidable challenge for mobile edge networks.
	
	At mobile edge networks, edge servers can cache panoramic VR videos to reduce bandwidth consumption through an adaptive 360$^\circ$ VR streaming solution based on the users' FoV~\cite{mahzari2018fov}. The FoV refers to the scene that the user's attention is focused on currently. The authors in~\cite{mahzari2018fov} allow the edge servers to use probabilistic learning to develop appropriate FoV-based caching policies for VR video delivery using historical VR video viewing data and current user preferences. Their experimental results show that their proposed FoV-aware caching scheme can improve the cache hit ratio and bandwidth saving by a factor of two, thus reducing network latency and relieving pressure on the network bandwidth of VR services. To fully utilize the uplink bandwidth of users in VR services, the authors in~\cite{huang2018d2d} design a point-to-point VR content distribution system serving the edge network with high-density devices. They propose a pre-caching algorithm for VR videos to maximize the QoE of VR services. In detail, the algorithm perceives the probability of caching content based on the user's interest and popularity to improve the accuracy and efficiency of edge caching.
	
	The QoS of VR may also be affected by the behavior of the VR user (e.g., position or rapid head movements). In particular, if the prediction of these movements is not taken into account, the streaming VR content will experience undesirable jitter and degradation of the playback quality. The authors in~\cite{zhang2019exploiting} propose a VR content caching scheme that uses wireless HMD caching to address this problem. In this scheme, a linear regression-based motion prediction algorithm reduces the motion-to-photon latency of VR services. The HMD of the VR user renders the image based on the predicted location and then stores and displays it on the HMD. The authors use a new VR service metric called Warping Distance to measure the quality of the VR video provided. Since mobile devices used to display AR display are usually resource-constrained, the QoS of the MAR cache requires a trade-off between communication resources and energy latency. To optimize mobile cache size and transmission power, the authors in~\cite{seo2021novel} provide a framework for AR content cache management for mobile devices, introducing an optimization problem in terms of service latency and energy consumption to provide satisfactory MAR services at mobile edge networks.
	
	\subsubsection{\textbf{Hologram Streaming}}
	
	In addition to accessing the Metaverse via AR/VR services, users can also project every detailed information about themselves into the physical world via wireless holographic communication. Holography~\cite{huo2017wireless} is a complex technique that relies on the interference and diffraction of objects with visible light to record and reproduce the amplitude and phase of optical wavefronts, including optical holography and computer graphics holography (CGH), and digital holography (DH). The authors in~\cite{huo2017wireless} propose a forward error correction coding scheme based on unequal error protection to optimize the QoS of wireless holographic communication at different bit levels of coding rates. Furthermore, the authors in~\cite{huo2017wireless} introduce an immersive wireless holographic-type communication mode for remote holographic data over wireless networks and rendering (e.g., point clouds~\cite{mekuria2016design} and light fields~\cite{karafin2017support}) interaction.
	
   Point clouds are the representation of immersive 3D  graphics consisting of a collection of points with various attributes, e.g., color and coordinate~\cite{clemm2020toward}. To stream high-quality point clouds to mobile users, the authors in~\cite{li2020demo} propose a point cloud streaming system over wireless networks. Based on the information of user devices and available communication and computation resources, their system can reduce unnecessary streaming while satisfying the experience of users. However, real-time point cloud streaming requires adaptive resource scheduling according to dynamic user requirements and network conditions. Therefore, the authors in~\cite{huang2021aitransfer} leverage the power of ML and propose an AI-empowered point cloud streaming system at mobile edge networks. In the system, intelligent edge servers can extract key features of 3D point clouds based on FoV and network conditions of edge devices. As such, immersive 3D content is streamed to users in a cost-efficient manner. Since the privacy concerns on Metaverse users' FoV information and network conditions, the authors in~\cite{gao2022fras} propose a federated RL-based encoding rate selection scheme to optimize the personalized QoE of users.
		
Different from the point clouds with only color and intensity values, light fields consist of multiple viewing angles of 3D graphics and allow users to interact with the same 3D objects from different angles in parallel~\cite{karafin2017support}. The light fields can provide users with a glasses-free 3D experience when they are immersing in the Metaverse. In \cite{kara2018evaluation}, the authors study the QoE evaluation of the dynamic adaptive streaming for light field videos with both spatial and angular resolution. In addition to viewpoint selection in point cloud streaming, view angle selection is also important to improve the quality of light field streaming. Therefore, the authors in \cite{wang2019region} study the user‑dependent view selection and video encoding algorithm for personalized light field streaming to provide ground truth light field videos while reducing around 50\% required bitrate. To adapt to  real-time demands of edge users, the authors in~\cite{hu2021multiple} propose a graph neural network-based solution for light field video streaming systems under best-effort network conditions. This intelligent streaming solution can improve the reliability and stability of light field streaming systems in heterogeneous networks.

	\begin{table*}[htbp]
		\small\centering
		\caption{Summary of scenarios, problems, performance metrics, and mathematical tools for AR/VR, the tactile Internet, hologram streaming, and semantic-communication.}
		\label{tab:commsumm}
		\begin{tabular}{|m{.05\textwidth}<{\centering}|m{.2\textwidth}<{\centering}|m{.2\textwidth}<{\centering}|m{.2\textwidth}<{\centering}|m{.2\textwidth}<{\centering}|}
			\hline
			\rowcolor[HTML]{C0C0C0} 
			\textbf{Ref.} &
			\textbf{Scenarios} &
			\textbf{Problems} &
			\textbf{Performance Metrics} &
			\textbf{Mathematical Tools} \\ \hline
			\cite{gimenez20195g} &
			MBMS over 5G &
			Physical layer design and analysis &
			VR service flexibility, capacity, and coverage &
			Flexible numerology and configurations \\ \hline
			\cite{elbamby2018edge} &  Interactive mmWave-enabled VR gaming &
			mmWave resource allocation &
			Interaction latency &
			Matching theory \\ \hline
			\cite{militano2015single} &
			D2D-enhanced MBMS with single frequency &
			D2D radio resource management &
			Aggregate data rate and short-term fairness &
			Greedy algorithm and iterative   search algorithm \\ \hline
			\cite{wang2022meta} &
			Indoor VR over THz/VLC wireless network &
			VAP selection and user association &
			VR  service reliability and bitrate &
			Meta RL\\ \hline
			\cite{feng2020smart} &
			VR streaming in 5G HetNet &
			Mode selection for streaming  &
			Algorithm convergence and VR streaming throughput &
			Unsupervised learning and reinforcement learning \\ \hline
			\cite{zhang2021buffer} &
			Buffer-aware VR video streaming &
			Personalized and private viewport prediction &
			User-wise accuracy and streaming utility &
			FL and DRL \\ \hline
			\cite{park2018urllc} &
			Downlink VR multimodal perceptions &
			Multiplexing design under OMA   and NOMA &
			Integrated resolution and haptic rate &
			Stochastic geometry \\ \hline
			\cite{guo2020adaptive} &
			mmWave-enabled wireless VR systems &
			Task offloading for real-time VR rendering &
			QoE and convergence time &
			DRL and game theory \\ \hline
			\cite{chen2018virtual} &
			Wireless VR over small cell networks &
			Joint uplink and downlink resource allocation &
			Tracking accuracy, processing delay, and transmission delay &
			Game theory and deep learning \\ \hline
			\cite{xu2022wireless} &
			Non-panoramic wireless VR &
			Market and mechanism design &
			Social welfare and auction information exchange cost &
			Auction theory and DRL\\ \hline
			\cite{liu2018dare} &
			MAR with dynamic edge resource &
			MAR service configuration &
			Quality   of augmentation &
			Optimization  mechanisms \\ \hline
			\cite{liu2018edge} &
			Edge-based MAR system &
			System design and analysis &
			Analytics  accuracy and service latency &
			Optimization   mechanisms \\ \hline
			\cite{ren2020edge} &
			Mobile Web AR over 5G networks &
			Adaptive   motion-aware key frame selection &
			Communication   efficiency and feature extraction-tracking accuracy &
			Planning mechanisms \\ \hline
			\cite{ran2019sharear} &
			Shared multi-user AR &
			System implementation and evaluation &
			Virtual Object pose accuracy, jitter and drift, and end-to-end   latency &
			Fine-grained control mechanism \\ \hline
			\cite{mahzari2018fov} &
			Adaptive 360$^\circ$ VR streaming &
			FoV-aware   caching replacement &
			Number   and duration of rebuffering events, accuracy of FoV. &
			Probabilistic learning \\ \hline
			\cite{huang2018d2d} &
			D2D-assisted VR video distribution systems &
			Video pre-caching deployment &
			Stability and QoE &
			Optimization mechanism \\ \hline
			\cite{zhang2019exploiting} &
			Real-time cinematic VR service &
			VR caching replacement on HMD &
			Warping distance and  consistency of VR experience &
			Linear   regression \\ \hline
			\cite{seo2021novel} &
			MEC-assisted MAR services &
			Joint mobile cache and power management &
			Service   latency and energy consumption &
			Optimization mechanism \\ \hline
			\cite{huang2021aitransfer} &
			Point cloud video streaming over wireless networks &
			Key feature extraction &
			Point cloud compression ratio &
			AI-powered transmission mechanism \\ \hline
			\cite{gao2022fras} &
			Adaptive point cloud video streaming &
			Encoding rate selection &
			Privacy preservation and QoE &
			Federated RL \\ \hline
			\cite{kara2018evaluation} &
			Adaptive streaming for light field videos &
			Spatial and angular mode selection&
			QoE &
			Subjective tests \\ \hline
			\cite{wang2019region} &
			Personalized light field streaming &
			View angle selection &
			QoE and required bitrate &
			Personalized compression method \\ \hline
			\cite{hu2021multiple} &
			Light field video streaming systems in heterogeneous networks &
			Streaming mode selection &
			Reliability and stability &
			Graph neural network \\ \hline
			
			\cite{boabang2021machine} &
			The 5G-enabled tactile Internet remote surgery &
			Delayed and lost content prediction &
			Relative error and prediction time &
			Gaussian process regression \\ \hline
			\cite{hou2018burstiness} &
			URLLC in the tactile Internet &
			Burstiness-aware bandwidth reservation &
			Bandwidth efficiency &
			Unsupervised learning \\ \hline
			\cite{aijaz2018toward} &
			Haptic remote operation over mobile networks &
			Radio resource allocation &
			Algorithm complexity and QoE &
			Canonical duality theory and Hungarian method \\ \hline
			%\begin{comment}
			\cite{yan2022resource} &
			Semantic-aware wireless networks &
			Semantic-aware resource allocation &
			Semantic spectral efficiency &
			Optimization mechanism \\ \hline
			\cite{liew2022economics} &
			Semantic communication system in wireless IoT &
			Energy allocation problem &
			Sentence similarity score and BLEU score&
			DL-based auction \\ \hline
			%\end{comment}
		\end{tabular}%
		\label{table:com_sum}
	\end{table*}
	\subsection{Human-in-the-loop Communication}\label{sec:human}
	The Metaverse is envisioned as a human-centric 3D virtual space supported by massive machine-type communication. The future of wireless networks will be driven by these human-centric applications, such as holography AR social media, immersive VR gaming, and healthcare with the tactile Internet~\cite{6755599,8482445}. Creating highly simulated XR systems requires the integration of not only engineering in service provision (communication, computation, storage) but also the interdisciplinary considerations of human body perception, cognition, and physiology for haptic communication~\cite{simsek20165g,hirche2012human,antonakoglou2018toward}. The haptics in haptic communication refers to both kinesthetic perception (information of forces, torques, position, and velocity that is sensed by muscles, joints, and tendons of the human body) and tactile perception (knowledge of surface texture and friction that is felt by different types of mechanoreceptors in the human skin). To further evaluate the performance of the service providers in the Metaverse, some new metrics, such as Quality of Physical Experience (QoPE)~\cite{saad2019vision} are proposed to evaluate the quality of human-in-the-loop communication. The trend of these metrics is to blend the physical factors, such as physiological and psychological perception, of human users with the classical inputs of QoS (latency, rate, and reliability) and QoE (average opinion score). The factors affecting QoPE are influenced by various human-related activities such as cognition, physiology, and gestures. 
	
 \subsubsection{\textbf{URLLC for the tactile Internet}}
		
		Unlike traditional Internet multimedia services (audio and video), multimedia services in the Metaverse allow users to immerse themselves in the 3D world in a multi-sensory embodied manner. The Metaverse requires not only 360$^{\circ}$ visual and auditory content for the immersive experience but also haptic and kinematic interaction. Fortunately, the tactile Internet is paving the way for the emergence of the Metaverse that provides users with haptic and kinematic interaction. However, this interaction of the tactile Internet requires extremely reliable and responsive networks for immersive experience during real-time interaction~\cite{sharma2020toward}. The Gaming experience in the Metaverse, such as, Assassination, for example, is an important task in Role-playing games (RPGs) of the fantasy or action genre. The assassination usually takes place from the blind spot of the victim. As such, the assassinated does not receive feedback from visual and auditory information but is aware of being assaulted from behind by the tactile sensation delivered to his game controller. Therefore, humans' tactile and kinematic information in the bilateral remote sensing operating system is exchanged via wireless networks. In the Metaverse's teleoperation, the transmission of tactile information is more sensitive to the stability and latency of the system, which requires a very strict end-to-end latency, at around 1ms~\cite{lawrence1993stability}, as a supplement to visual and auditory information. Beyond the gaming industry, such a breakthrough in haptic-visual feedback control will also change the way humans communicate worldwide, i.e., human-in-the-loop communication~\cite{duan2017human}.
	
	% \textbf{Edge Testbeds for the Tactile Internet:} Although the tactile Internet has the potential to revolutionize the future of wireless communication, it is still far from being deployed on a large scale. Research on the tactile Internet is still in the testing phase, and there are two major obstacles: (i) the lack of consensus on the performance of the tactile Internet, and (ii) the different levels of development among different disciplines of the tactile Internet. To address these issues, the authors in~\cite{gokhale2020tixt} design a universal, modular, and scalable testbed. To provide a proof of concept for their proposed testbed, they give two real-world examples of the tactile Internet in real and virtual worldss. Under the guidelines of the IEEE 1918.1 haptic communication protocol, the authors design the OVGU-HC, a testbed for haptic communication at the Otto-von-Guericke University of Magdeburg, and describe its components and structure in~\cite{engelhardt2020ovgu}. The main features of the OVGU-HC are its experimental description language (DES-Cript), its experimental automation services, and the various hardware available for conducting experiments. Finally, an open Tactile Cyber-Physical Systems (TCPS) testbed is developed in~\cite{polachan2019towards}, whose main purpose is rapid prototyping and evaluation of TCPS applications. Different from previous testbeds without assessment means, the engineers equip their testbed with tools for evaluating latency and management performance, which are key features of TCPS.
	The extremely demanding of reliability and latency in the tactile Internet presents new challenges for resource allocation at mobile edge networks. As a real-time transmission medium for touch and motion, the tactile Internet, together with AR/VR, will provide multi-sensory multimedia services for Metaverse users~\cite{lim2022realizing}. However, unlike eMBB required by AR/VR, the tactile Internet requires URLLC for the human-human and human-avatar interaction. Noticing that the tactile Internet applications are sensitive to network jitters~\cite{sarathchandra2021enabling}, the authors propose a forecasting algorithm for delayed or lost data in~\cite{boabang2021machine} that uses scalable Gaussian process regression (GPR) for content prediction. To support telerobotic surgery, they build an online learning model using GPR coefficients and a low-rank matrix decomposition for service context. 
	
	In addition, the very bursty arrival process of packets on the tactile Internet complicates the allocation of resources in the wireless edge network. Therefore, the authors in~\cite{hou2018burstiness} formulate an optimization problem for minimizing the reserved bandwidth under service and reliability constraints. Based on unsupervised learning, they propose a model- and technical data-based approach to cluster the fractional arrival process of users to reduce the classification error and improve the longitudinal spectrum efficiency in wireless networks. Considering the perceptual coding and symmetric design in the tactile Internet, the authors in~\cite{aijaz2018toward} decompose the resource allocation problem in the tactile Internet and propose a two-step solution. They offer a low-complexity heuristic algorithm and a near-optimal approach to the wireless resource allocation algorithm to overcome the analytical and computational difficulties. 
	
	% \subsubsection{User-centric communication}
	
	% With the increasing popularity of mobile devices such as smartphones and tablets, the past few years have witnessed a tremendous growth of multimedia applications in wireless systems. To meet the explosive demand for delivering high-definition video streams over cellular networks, the authors in~\cite{cao2015social} design a social-aware video multicase system leveraging D2D communication. In this system, clients can obtain missing packets from other clients and restore incomplete video frames via social ties, i.e., social trust and social reciprocity. Based on these social ties among devices, a coalitional game is formulated to devise a distributed algorithm to obtain the grouping solution and the resource allocation scheme is proposed for the BS to handle D2D radio resource requests from client groups.
	
	\subsubsection{\textbf{Semantic/Goal-aware Communication}}
	\begin{figure}
		\centering
		\includegraphics[width=1\linewidth]{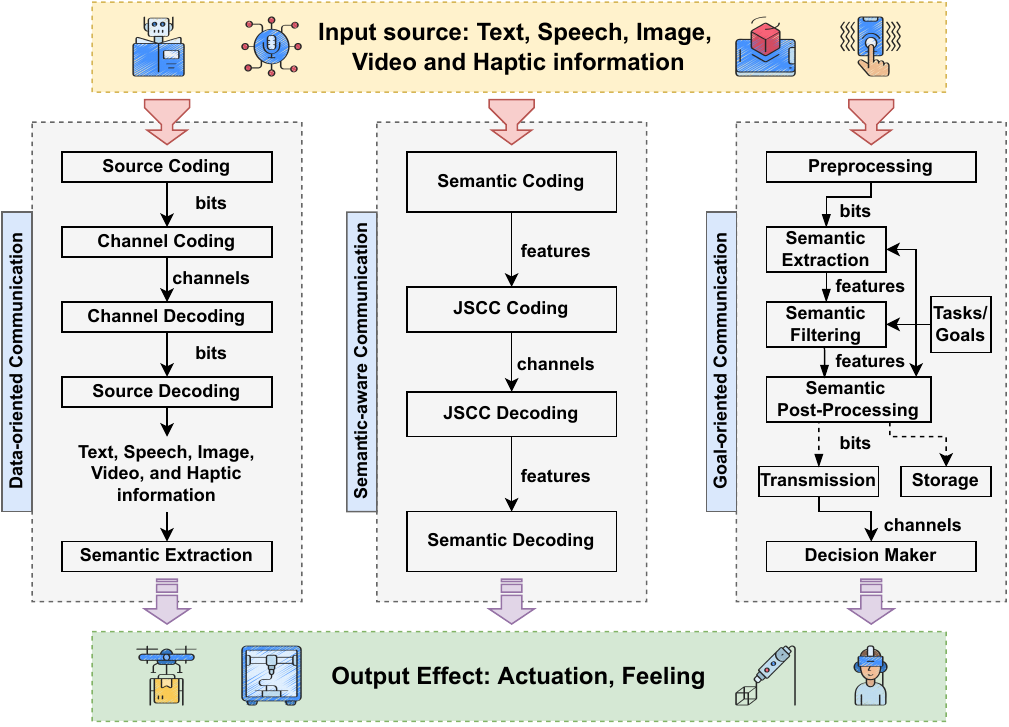}
		\caption{Comparison of data-oriented, semantic-oriented, and goal-oriented communication systems.}
		\label{fig:semantic}
	\end{figure}
	
	The Metaverse is a comprehensive multimedia system consisting of many semantically oriented applications. Metaverse content dissemination services can place a massive burden on today's data-oriented communication~\cite{strinati20216g}, where real-time traffic requires a channel with infinite capacity. Therefore, applications in the Metaverse require service diversity and service-level optimization to reduce the load on the wireless channels at mobile edge networks. In human-in-the-loop communication, communication is not simply the presentation of original data in bits to the user but the presentation of information in semantic structures. Semantic communication combines knowledge representation and reasoning tools with ML algorithms, paving the way for semantic recognition, knowledge modeling, and coordination.
	
	As shown in Fig.~\ref{fig:semantic}, there are three \textit{Semantic Communication Components} of a semantic-aware communication system~\cite{qin2021semantic,yang2022semantic}:
	\begin{itemize}
		\item The semantic encoder (source) detects and identifies the semantic information, i.e., knowledge, of the input source signal and compresses or removes the irrelevant information from the current conversation.
		\item The semantic decoder (target) decodes the received semantic information and restores it in a form understandable to the target user.
		\item The semantic noise interferes with semantic information in communication and causes the receiver to misunderstand or misperceive it. It is common in semantic encoding, transmission, and decoding.
	\end{itemize}
	
	As one of the promising approaches to overcome the Shannon capacity limit~\cite{qin2021semantic}, semantic communication, one of the intelligent solutions on the physical layer, can transmit only the information required for the respective task/goal~\cite{qin2019deep}.
	For example, in signal compression and signal detection in wireless networks, deep neural networks (DNN) can extract knowledge about the environment in different scenarios based on historical training data. Therefore, the authors propose a robust end-to-end communication system in~\cite{qin2019deep} aiming to facilitate intelligent communication at the physical layer. For goal-oriented communication~\cite{kalfa2021towards} shown in Fig.~\ref{fig:semantic}, the authors in~\cite{bourtsoulatze2019deep} propose a joint source-channel coding (JSSC) scheme for extracting and transmitting semantic features in messages to perform specific tasks. In this scheme, the semantic sender removes and transmits the semantic elements in the sent message. In contrast, the semantic receiver only needs to complete the corresponding actions in the semantic features without restoring them in the source message.
	
	In contrast to the conventional bit error rate in data-oriented communication, the performance metrics of semantic communication must consider the factor of human perception. In~\cite{xie2021deep}, the authors propose a metric called sentence similarity to measure the semantic error of transmitted sentences in semantic communication systems. Based on this metric, they apply natural language processing to physical layer communication's encoding and decoding process. In particular, deep learning (DL)-based machine translation techniques are used to extract and interpret the transmitted messages. Experimental results show that the proposed system can achieve higher information transmission efficiency than the reference system, especially with a low signal-to-noise ratio. 
	
	Using visual question answering (VQA) as an example, the authors in~\cite{xie2021task} extend the previous model to a semantic communication system with multimodal data. However, IoT devices are often resource-constrained, making it difficult to support computation and energy-intensive neural networks in semantic communication. Therefore, in~\cite{xie2020lite}, the authors propose a lightweight semantic communication system for low-complexity text transmission in IoT. In this system, the authors use neural network pruning techniques to remove redundant nodes and weights from the semantic communication model to reduce the computational resources required for the semantic communication models. This way, IoT devices can afford semantic communication and significantly reduce the required communication bandwidth between them and the cloud/edge. On the other hand, the authors in~\cite{tong2021federated} propose a scheme for information semantics of audio signals. They use a wave-to-vector structure in which convolutional neural networks form the autoencoder to convert auditory information to semantic features. To reduce the communication overhead associated with co-training the semantic communication model, they also introduce an FL-based approach that allows edge devices to jointly train a federated semantic communication model without sharing private information.
	
	In parallel to the design of the semantic communication system, its deployment in the mobile edge network requires efficient resource allocation to reduce the consumption of computation resources in semantic communication~\cite{yang2022semantic}. The authors in~\cite{yan2022resource} note that resource allocation problems for semantic networks remain unexplored and propose an optimization mechanism to maximize semantic spectral efficiency (S-SE). Their simulation results show that the proposed semantic resource allocation model is valid and feasible to achieve higher S-SE than the baseline algorithms. From the perspective of economics-driven semantic communication system design, the authors in~\cite{liew2022economics} propose a DL-based auction for energy allocation for IoT devices. They propose an energy evaluation scheme based on the sentence similarity score and the bilingual evaluation score (BLEU) for semantic communication systems. The results show that the semantic service provider's revenue is maximized when individual rationality (IR) and incentive compatibility (IC) are satisfied.
	
	\subsection{Real-time Physical-Virtual Synchronization}
	\label{sec:synchronization}
	
	\begin{figure}
		\centering
		\includegraphics[width=1\linewidth]{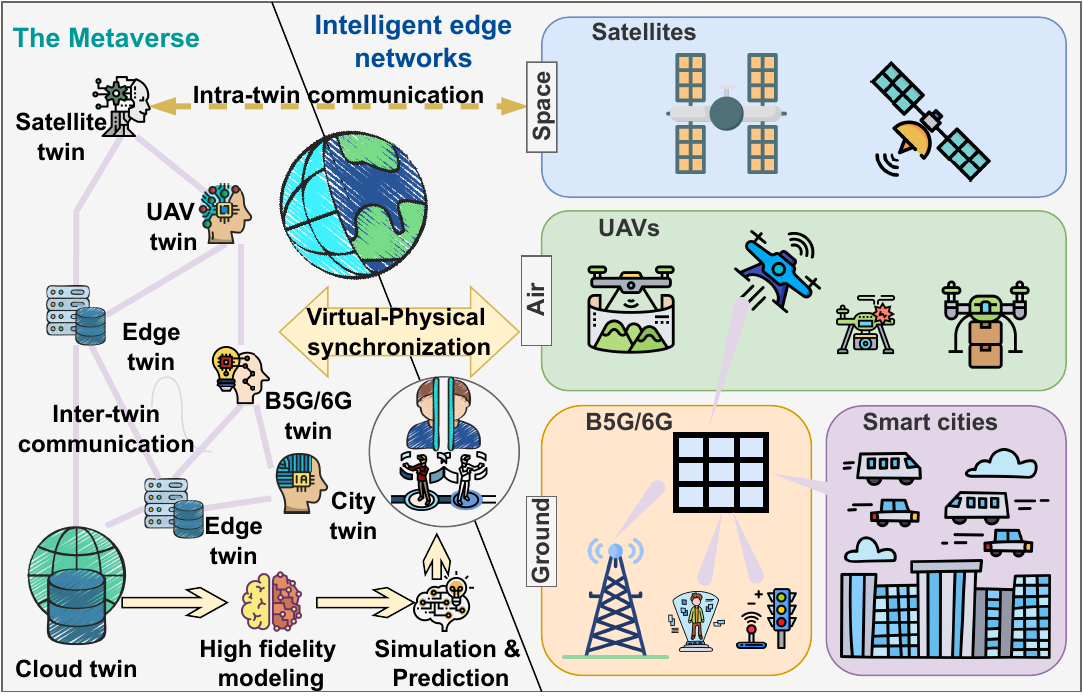}
		\caption{An illustration of real-time physical-virtual synchronization between the Metaverse and Intelligent edge networks.}
		\label{fig:dt}
	\end{figure}
	The Metaverse is an unprecedented medium that blurs the boundary of the physical and virtual worlds~\cite{duan2021metaverse}. 
		%The Metaverse is a set of virtual worlds based in part on the physical world~\cite{duan2021metaverse}. 
		For example, by maintaining a digital representation of the physical wireless environment in the virtual worlds~\cite{khan2022metaverse}, the Metaverse can improve the performance of mobile edge networks via offline simulation and online decision-making.
	For example, employees can build digital copies of their physical offices using DTs to facilitate work-from-home in the Metaverse. 
	To maintain bidirectional real-time synchronization between the Metaverse and the physical world, widely distributed edge devices and edge servers are required for sensing and actuation~\cite{zhang2022holographic}. As shown in Fig.~\ref{fig:dt}, one of the possible solutions for such virtual services is the DT~\cite{han2022dynamic}, i.e., digital replications of real-world entities in the Metaverse, e.g., city twins and copies of smart factories.  Based on the historical data of digital replications, the DT can provide high fidelity modeling, simulation, and prediction for physical entities. Though intra-twin communication and inter-twin communication, the Metaverse can be used to improve the efficiency of the edge networks. Specifically, by monitoring the DTs of edge devices and edge infrastructure such as RIS, UAV, and space-air-ground integrated network (SAGIN)~\cite{cheng2019space,cheng2018air,xu2022quantum}, users can instantly control and calibrate their physical entities in the physical world through the Metaverse.
	
	\subsubsection{\textbf{Resource Allocation for Physical-Virtual Synchronization}}
	
	Efficient resource allocation in internal communication between physical and virtual entities in the mobile edge network during synchronization of physical and virtual worlds is key to seamless real-time immersion in the Metaverse~\cite{shen2021holistic}. To this end, the authors in~\cite{han2022dynamic} design a data collection incentive mechanism for self-interested edge devices. In this mechanism, IoT devices act as PSPs to collect data in the physical world. This mechanism uses a dynamic evolutionary approach that allows different device owners to enter into various contracts to update their policies. Eventually, service providers converge to a policy that copes with time-varying resource supply and demand to maximize the dynamic rewards. To extend ~\cite{han2022dynamic} into dynamic IoV scenarios, the authors in~\cite{sun2021dynamic} propose a distributed incentive mechanism based on the alternating direction method of multipliers to incentivize vehicles contributing their resources to the physical-virtual synchronization in the Metaverse. This mechanism can maximize the energy efficiency of vehicular twin networks in dynamic environments.
	
	Moreover, the complex network topology in DT networks poses challenges to the design of resource allocation schemes~\cite{khan2022digitals}. The authors in~\cite{dai2020deep} consider the probabilistic arrival of synchronization tasks in DT networks to propose efficient methods for offloading computation and allocating resources. In particular, they combine the problems of offloading decisions, transmission power, network bandwidth, and computational resource allocation to balance resource consumption and service delay in DT networks. However, since this joint problem is non-convex, they transform the original problem into a long-term revenue maximization problem using the advantage actor-critic algorithm and Lyapunov optimization techniques. At mobile edge networks, the spatiotemporal dynamics of mobile physical entities require that the service delivery framework of existing edge networks provide resilient synchronization resources. To respond to the random synchronization requests of edge devices on time, the authors propose a congestion control scheme based on contract theory in~\cite{lin2021stochastic}. The scheme uses Lyapunov optimization to maximize the long-term profit maximization of DT service providers while satisfying individual rationality and incentive compatibility by taking into account the delay sensitivity of the services offered. Moreover, in~\cite{lu2020communication}, the privacy-preserving approaches such as FL are also employed to develop an effective collaborative strategy for providing physical-virtual synchronization using DNN models.
	
	\subsubsection{\textbf{Intelligent Edge Network-aided Virtual-Physical Synchronization}}
	
	The DT of the physical edge networks that supports the Metaverse provides a possible intelligence solution to the traditional edge networks for improving their performance through physical-virtual synchronization~\cite{khan2022digital, li2021slicing, li2021joint,9846954}. For example, the energy efficiency of real-world edge networks can be enhanced by communication and coordination of DT within the Metaverse. The authors in~\cite{dong2019deep} design a DT that mirrors the mobile edge computing systems by modeling the network topology and the channel and queueing model. The optimization goal of this system is to support URLLC services and delay-tolerant services in this system with maximum energy efficiency. To reduce energy consumption in this system, the authors use the DT data to train a DL architecture to jointly determine user grouping, resource allocation, and offloading probability. 
	Moreover, as the complexity of edge networks and the diversity of mobile applications increase, slicing-edge networks can provide fine-grained services to mobile users. In~\cite{wang2020graph}, the authors propose an intelligent network slicing architecture based on DT. In this architecture, graph neural networks are used to capture the interconnection relationships between slices in different physical network environments, and DT are used to monitoring the end-to-end scalability metrics of the slices.
	
	 In traditional edge networks, wireless resources are often allocated by sensing network conditions and then passively adjusting the information and power transmitted. Fortunately, RIS technologies in 6G can transform the wireless environment by adaptively phasing RIS elements~\cite{di2019smart,9771079}. This way, RIS can actively change the environment of the wireless network so that the transmission is not limited by physical conditions~\cite{wu2019intelligent}. In addition, RIS can also be leveraged to stream immersive content~\cite{liu2021learning}.  However, the optimal phase configuration of RIS and power allocation of beamforming signals is still challenging. In~\cite{sheen2020digital}, the authors propose an Environmental Twin (Env-Twin) scheme, where the infrastructure in wireless networks is virtualized into a DT network. By monitoring and controlling different cell layouts (macrocells, small cells, terrestrial and non-terrestrial networks), spectrum ranges (including millimeter waves, terahertz, and even visible light communication), and RISs in complex edge networks, the wireless edge environment can be optimized into an intelligent edge environment. In this Env-Twin, the techniques from DL predict the most appropriate RIS configuration for each new receiver site in the same wireless network to maximize spectrum efficiency in edge networks. To enable faster convergence of intelligent wireless environments to a satisfactory performance, the authors in~\cite{di2019smart} propose an ambient AI approach to control RIS through RL. In this approach, the RIS acts as a learning agent to learn the optimal strategies during interaction with intelligent edge networks without prior knowledge of the wireless environment.
	
	To provide ubiquitous coverage to Metaverse users, IoV, UAV, and RIS in the wireless edge network need to form a SAGIN with satellites~\cite{zhou2020deep, centenaro2021survey,9854055}. In~\cite{yin2021physical}, the authors set up a DT network in the virtual space for SAGIN to enable secure vehicle communication in SAGIN. By communicating with virtual entities in the virtual space, physical entities can protect the efficient and secure information transmission between vehicles. Besides, while spectrum sharing can improve spectrum utilization in SAGIN, it is still vulnerable to eavesdropping and malicious threats. To address this issue, they propose a mathematical model for the secrecy rate of the satellite-vehicle link and the secrecy rate of the base station-vehicle link. They solve the secure communication rate maximization problem in SAGIN using relaxation and planning methods.

	\subsection{Lessons Learned}
	
	% - highlight the key research issues and where more materials can be found
	
	% - which issues have been studied more/less, key observation, remaining challenges, and connection to the next section
	
		\subsubsection{Efficient Immersive Streaming and Interaction}
		The Metaverse is driven by an amalgamation of VR/AR/tactile Internet and involves multimodal content delivery involving images, audio, and touch. Moreover, different from traditional VR/AR, which tends to focus on single-user scenarios, the Metaverse will involve a massive number of users co-existing and interacting with each other in the virtual worlds within. As such, there exists a crucial need for efficient resource allocation to optimize the efficiency of service delivery to a massive number of users at the edge. Besides high-performance communication networks, the real-time user interactions and rendering of 3D FoVs will require computation-intensive calculations. We revisit the computation challenges and edge-driven solutions in Section \ref{sec:computation}.
		\subsubsection{AI for the Intelligent Edge Communication}
		The complex and dynamic network environment necessitates the use of AI methods to efficiently manage communication resources. The ``AI for edge" approach in the works that we review utilizes AI to improve the allocation efficiency of resources at the edge, e.g., for efficient task offloading and bandwidth allocation. AI can be used to supplement conventional optimization tools, e.g., in reducing the communication cost and speeding up the convergence of auctions as means to price the services of PSPs. However, the concern is that such AI models are computationally costly to train and store on resource-constrained devices. In Section \ref{sec:computation}, we discuss the edge-driven solutions and techniques to manage the computation cost of AI deployment at the edge.
		\subsubsection{Context-aware Immersive Content Delivery} In the Metaverse immersive content distribution network, VSPs deploy a cache of immersive content to PSPs based on the current user's contextual preferences, such as current FoVs, view angles, and interacting objects. Consequently, the user-perceived latency becomes lower while the total bandwidth required for immersive streaming and interaction is reduced. In Section~\ref{sec:computation}, we will describe how to accurately predict recommendations for the immersive content delivery.
		\subsubsection{Self-sustainable Physical-Virtual Synchronization} From Section~\ref{sec:synchronization}, we can learn that the physical-virtual synchronization is self-sustainable. On one hand, in the P2V synchronization, the physical world maintains DTs in the virtual worlds, which requires a large overhead in communication resources. On the other hand, in the V2P synchronization, the DYs in the virtual worlds can transform the physical environment into smart wireless networks to reduce the communication resources in return.
		\subsubsection{Decentralized Incentive Mechanism} The Metaverse is based on the vast amounts of data provided by wireless sensors in the physical world. VSPs and PSPs are typically self-interested and unwilling to provide their communication resources and energy for data collection. The right incentives for wireless sensors at scale will help maintain the long-term sustainability of the Metaverse.

	\section{Computation}
	\label{sec:computation}
	\begin{figure*}[t]
		\centering
		\subfigure[Meta collaborates with leading museums to provide the users with a brand new perspective when connecting with the culture~\cite{fb}.]{
			\includegraphics[width=0.30\linewidth, height=0.22\linewidth]{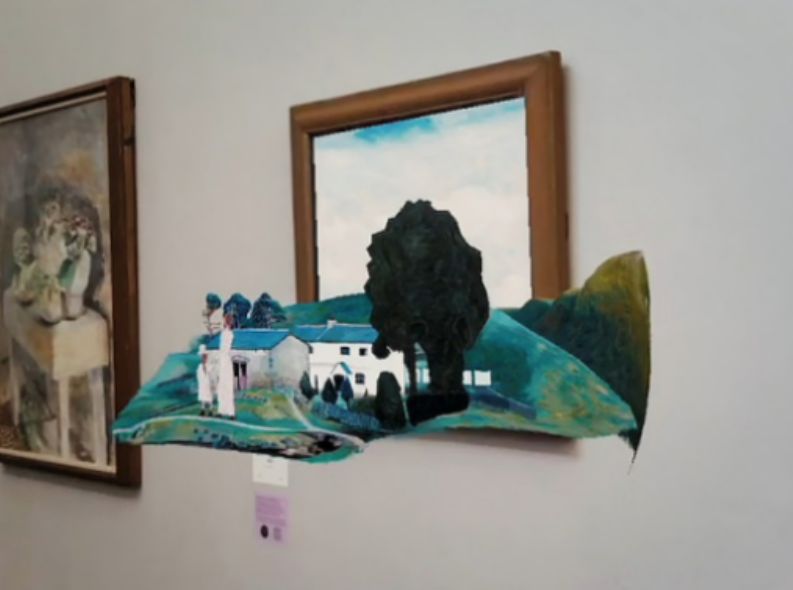}
			\label{fig:metaar}
		}
		\quad
		\subfigure[Meta's demo of AI generated content though audio inputs and 3D object ouputs~\cite{aigc}.]{
			\includegraphics[width=0.30\linewidth, height=0.22\linewidth]{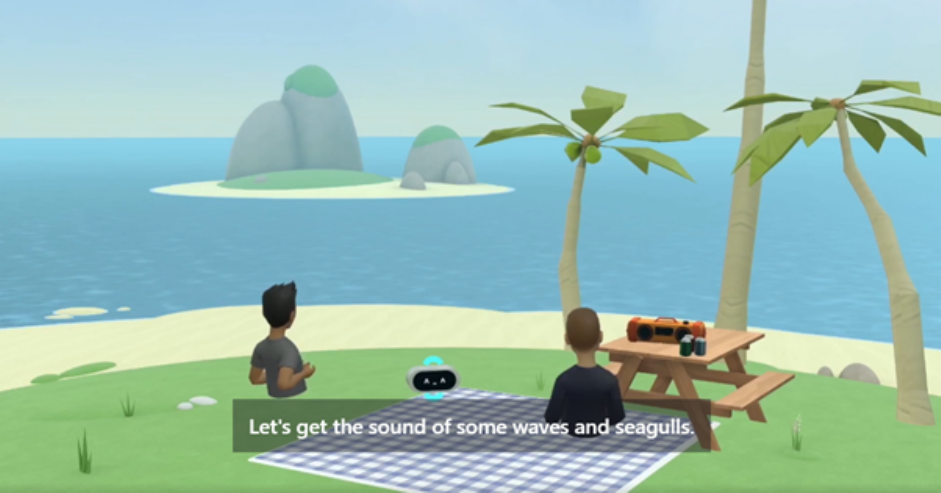}
			\label{fig:aigc}
		}
		\quad
		\subfigure[AI technology combines hand gestures and AR to copy and paste files across computers~\cite{shi2015offloading}.]{
			\includegraphics[width=0.30\linewidth, height=0.22\linewidth]{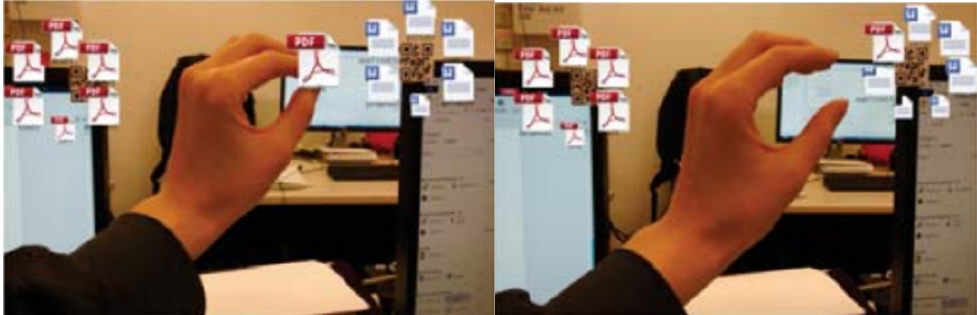}
			\label{fig:copy}
		}
		% \quad
		%\subfigure[Screenshot from Microsoft Flight Simulator trailer~\cite{mfs}.]{
			%  \includegraphics[width=0.44\linewidth, height=0.27\linewidth]{flight.pdf}  
			%   \label{fig:flight}
			%} 
		\caption{Screenshots of Meta AR, AIGC demo, and gesture AR.}
		\label{fig:fig2}
	\end{figure*}
	%\textcolor{blue}{The people and objects in the physical world are constructed as holographic image to display on the various terminal devices.}
	Beyond reliable and low-latency communication systems, the resource-intensive services provided by the Metaverse will require costly computation resources to support~\cite{duan2021metaverse}. Some screenshots of exemplary Metaverse applications, such as Meta AR, hand gesture AR, and AI-generated content (AIGC) applications, are illustrated in Fig.~\ref{fig:fig2}. In specific, the Metaverse requires the execution of the following computations:
 \begin{itemize}
			\item \textbf{High-dimensional Data Processing:} The physical and virtual worlds will generate vast quantities of high-dimensional data, such as spatiotemporal data~\cite{wang2022survey}, that allows users to experience realism in the Metaverse. For example, when a user drops a virtual object, its behavior should be governed by the laws of physical for realism. The virtual worlds with physical properties can be built upon physics game engines, such as the recently developed real-time Ray Tracing technology to provide the ultimate visual quality by calculating more bounces in each light ray in a scene~\cite{unityray}. During these activities, this generated high-dimensional data will be processed and stored in the databases of the Metaverse.
			\item \textbf{3D Virtual World Rendering:} Rendering is the process of converting raw data of virtual worlds into displayable 3D objects. When users are immersing in the Metaverse, all the images/videos/3D objects require a rendering process to display on AR/VR devices for users to watch. For example, Meta operates research facilities that set up collaborations with museums to transform 2D painting into 3D~\cite{fb} as illustrated in Fig.~\ref{fig:metaar}. In addition, Metaverse are expected to produce and render 3D objects in an intelligent way. For example, as shown in Fig.~\ref{fig:aigc}, 3D objects, e.g., clouds, islands, trees, picnic blankets, tables, stereos, drinks, and even sounds, can be generated by AI (i.e., AIGC) through audio inputs~\cite{aigc}.
			\item \textbf{Avatar Computing: } Avatars are digital representations of user in the virtual worlds that require ubiquitous computation and intelligence in avatar generation and interaction. On one hand, the avatar generation is based on the ML technologies, such as computer vision (CV) and natural language processing (NLP). On the other hand, the real-time interaction between avatars and users is computationally intensive to determine and predict the interaction results. For example,  AI can synergize with AR applications to provide avatar services such as retrieving text information from the users' view, text amplification on paper text, and using a hand gesture to copy and paste files between computers~\cite{shi2015offloading} in Fig.~\ref{fig:copy}. 
	\end{itemize}

	The Metaverse should be built with ubiquitous accessibility for users at mobile edge networks. Therefore, computation resources are required not only by service providers of the Metaverse, e.g., for user analytics but also from users' edge devices to access the Metaverse at mobile edge networks. For example, using data on mobile devices, a AR/VR recommender system can be trained to recommend immersive content that satisfies users~\cite{lam2021a2w}. Second, AR/VR rendering requested users to have an immersive experience. In the following, we present the key computation considerations:

		\begin{enumerate}
			\item \textit{Ubiquitous computing and intelligence:} Users equipped with various types of resource-constrained devices can access the Metaverse anytime and anywhere, requiring ubiquitous computing and intelligence services to surround and support them~\cite{huynh2022artificial}. On one hand, users can offload computation-intensive tasks to edge computing services for mitigating straggler effects to improve performance. On the other hand, stochastic demand and user mobility require adaptive and continuous computing and intelligence services at mobile edge networks.
			\item \textit{Embodied telepresent:} Users in different geographic locations can be simultaneously telepresent in the virtual worlds of the Metaverse, regardless of the limits of physical distance~\cite{lee2021all}. To provide users with embodied telepresent, the Metaverse is expected to render 3D objects in AR/VR efficiently with ubiquitous computation resources at mobile edge networks.
			\item \textit{Personalized AR/VR recommendation:} Users have personal preferences for different AR/VR content in the Metaverse~\cite{duan2021metaverse}. Therefore, content providers need to calculate personalized recommendations for AR/VR content based on users' historical and contextual information. In this way, the QoE of users is improved.
			\item \textit{Cognitive Avatars:} Users control their avatars, which are endowed with cognitive abilities, to access the Metaverse. The cognitive avatars~\cite{kocur2020effects}, which are equipped with AI, can automatically move and act in the Metaverse by mirroring the behavior and actions of their owners. In this way, users are able to control multiple avatars simultaneously in different virtual worlds. In addition, users can be relieved from some simple but heavy tasks that can be done by the cognitive avatars.
			\item \textit{Privacy and security:} The Metaverse will be supported by ubiquitous computing servers and devices for various types of computation tasks from AR/VR rendering to avatar generating. However, the reliance on distributed computing paradigm, e.g., mobile cloud computing, implies that users may have to share their private data with external parties. The Metaverse will collect extensive users' physiological responses and body movements~\cite{di2021metaverse}, which includes personal and sensitive information such as user habits and their physiological characteristics. Therefore, privacy and security in computation is an indispensable consideration that prevent data from leaked to third parties during the user's immersion in the Metaverse. 
	\end{enumerate}
	In this section, we survey are based on two categories of solutions. For the first category, we begin by providing the readers with a concise background of the cloud-edge-end computation paradigm in Section \ref{paradigm}. Then, we describe the computation schemes for resource-efficient AR/VR cloud-edge-end rendering in Section~\ref{vr_render}. For the second category, we discuss AI-driven techniques. In Section~\ref{ai_model_training}, we discuss studies that propose to compress and accelerate AI model training. In Section~\ref{security}, we discuss privacy and security measures for AI in the Metaverse.

	\subsection{The Cloud-Edge-End Computing Paradigm}
	\label{paradigm}
	
\begin{figure}[t]
		\centering
		\includegraphics[width=1\linewidth]{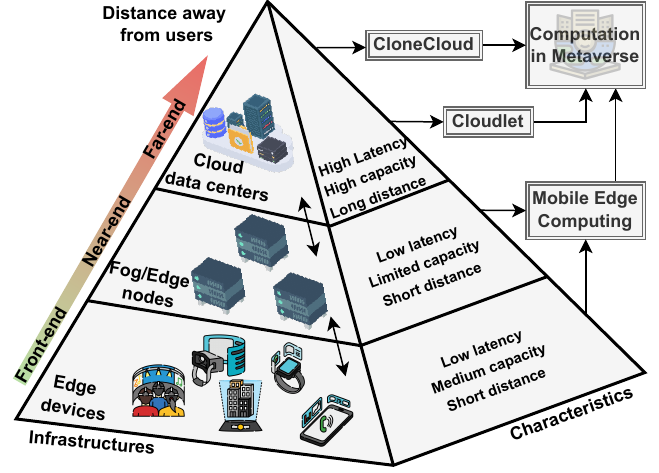}
		\caption{Various types of computing infrastructure to support the computation in the Metaverse and their characteristics.}
		\label{fig:computation}
	\end{figure}

	In this subsection, we first introduce the mobile cloud-edge-end collaborative computing paradigm, as shown in Fig.~\ref{fig:computation}, that provides ubiquitous computing and intelligence for users and service providers in the Metaverse.

	\textbf{Mobile Cloud Computing: }\label{cloud} Several of the world's tech behemoths have made the full dive Metaverse as their newest macro-goal. In the next decade, billions of dollars will be invested in cloud gaming~\cite{han2020virtual}, with the notion that such technologies will underlie our online-offline virtual future~\cite{abc}. Cloud gaming integrates cloud computing to enable high-quality video games. Mobile cloud computing provides computing services to either mobile devices or mobile embedded systems. The cloud provider installs computing servers in their private infrastructure, and services are provided to clients via virtual machines~\cite{li2020virtual}. However, the most cost-effective approach for an organization is to combine cloud and in-house resources rather than choosing one over the other~\cite{huang2012mobile}. Fortunately, a few system architectures such as Cloudlet~\cite{fernandez2018fog} and CloneCloud~\cite{chun2009augmented} are proposed to manage mobile cloud computing and provide support for the device with weak computation so that they can access the Metaverse. % Microsoft Flight Simulator is one of the applications that combines both cloud and in-house resources. In Microsoft Flight Simulator, there are 2 trillion individually rendered trees, 1.5 billion buildings, and almost all the roads, infrastructure, and airports of international cities. Microsoft Flight Simulator provides the users with a realistic flying experience, as shown in Fig.~\ref{fig:flight}. However, Microsoft Flight Simulator requires over 2.5 petabytes of data to run. It is even more challenging to run on mobile devices such as portal gaming devices or smartphones. Therefore, most simulator data is stored and processed in the cloud. The gameplay content (e.g., bird's eye view of cities) is directly streamed to the user device (i.e., personal computers). The local device only stores a core amount of data, with computations offloaded where necessary.

	\textbf{Edge Computing:} The Metaverse is a black hole of computational resource. It will always require as much computation capacity as possible and the mobile users accessing the Metaverse require the latency to be as low as possible to deliver real-time, advanced, computation-intense capabilities like speech recognition and AR/VR~\cite{zhang2019dynamic}. Therefore, edge computing provides low latency, high efficiency, and security to sustain the Metaverse. The concept of edge computing is similar to fog computing. Both the edge and fog computing observe the computation capability in the local network and compute the task that could have been carried out in the cloud, thereby reducing latency and burden on backbone networks. However, in edge computing, the additional computation capacity and storage of devices on edge are used. We can treat the fog layer as a layer between the edge and the mobile cloud, which extends the mobile cloud closer to the nodes that produce and act on IoT data~\cite{bierzynski2017cloud}.
	
	\textbf{Fog Computing:} The Metaverse is envisioned to provide ubiquitous connectivity for millions of users and IoT devices. This will cause an unprecedented generation of huge amounts and heterogeneous modalities of data~\cite{hu2017survey}. Therefore, mobile cloud computing can no longer support the high demands for real-time interaction of immersive and social applications~\cite{you2019fog} given the unavoidable transmission latency to distant cloud servers. As a consequence, fog computing is first proposed in~\cite{bonomi2012fog} to extend cloud services closer to users in order to reduce the latency when the computation tasks are offloaded. The authors in~\cite{sun2022enabling} show theoretically that fog computing results in a lower latency as compared to traditional cloud computing. Different from mobile cloud computing, fog computing deploys many fog nodes between the cloud and the edge devices~\cite{tan2020uav}. For example, mobile users commuting on trains are unable to access the Metaverse as they cannot attain a stable connection with the cloud. Fog computing nodes can therefore be installed in the locomotive to improve the connectivity. Fog nodes could be the industrial gateways, routers, and other devices with the necessary processing power, storage capabilities, and network connection~\cite{santos2021srfog}. To reduce the latency, fog nodes can connect edge devices/users through wireless connection modes, such as B5G/6G, and provide computation and storage services. % Privacy loss is mitigated by keeping the detailed personal information within the fog node, and it only transfers the anonymized data to the cloud during further Metaverse applications~\cite{bermbach2017research}. 
	\subsection{Efficient AR/VR Cloud-Edge-End Rendering}\label{vr_render}
	\label{sec:render}
	The edge-enabled Metaverse can be accessed using various types of mobile and wired devices, including HMD and AR goggles, VR allows the users to experience a simulated virtual world while in the physical world. The VR devices generate sensory images that form the virtual worlds~\cite{duan2021metaverse}. Real-time rendering is required in the Metaverse so that the sensory images can be produced quickly enough to form a continuous flow rather than discrete events. For example, education platforms can incorporate VR to improve lesson delivery in the virtual world. Unlike VR, AR allows the user to interact with the physical world that has been digitally modified or augmented in some way. AR applications are best suited for scenarios in which users are physically present in the physical world, such as on-the-job training and computer-assisted tasks~\cite{mill}. However, users' devices have limited computation capability, memory storage, and battery. Intensive AR/VR applications that are required for the immersive Metaverse cannot run smoothly on such devices.   Fortunately, based on the cloud-edge-end collaborative computing paradigm, users' devices can leverage ubiquitous computing resources and intelligence for remote rendering and task offloading~\cite{chen2019energy} to overcome these challenges.
	
	% Edge computing can achieve the lowest latency among all three~\cite{sunyaev2020fog} as it is the closest to the mobile users.
	
	%\begin{figure}[t]
	%\centering
	%    \includegraphics[width=8.5cm, height=9cm,trim={0cm 17.8cm 10.5cm 0},clip]{edge_arch.pdf}
	%  \caption{A typical architecture of edge computing networks.}
	%  \label{fig:edge_arch}
	%\end{figure}
	%which requires at least 5ms end-to-end latency
	
	%Mobile edge computing (MEC) is the key solution to support real-time rendering~\cite{alshahrani2020efficient}. It is based on the concept of edge computing~\cite{mec} and is regarded as one of the viable options to deliver real-time VR videos via wireless network~\cite{yang2018communication}. 
	  At mobile edge networks, the cloud-edge-end computing architecture allows the computations, such as high-dimensional data processing, 3D virtual world rendering, and avatar computing, to be executed where data is generated~\cite{sunyaev2020fog}. For example, when the user is interacting with avatars in the Metaverse from a vehicle to perform task computation, the vehicle can send the data to nearby vehicles or roadside units to perform the computation instead of cloud offloading, thereby dramatically reducing end-to-end latency. The proposed cloud-edge-end collaborative computing paradigm is a promising solution to reduce latency for mobile users~\cite{beck2014mobile} by leveraging computing capacities at the mobile edge networks~\cite{abbas2017mobile}. This is important as mobile users should be able to access Metaverse anytime and anywhere. Furthermore, mobile edge computing lightens the network traffic by shifting the computation from the Internet to the edge. At mobile edge networks, edge servers are equipped with edge servers that can provide computing resources to mobile users~\cite{beck2014mobile}. For example, the authors in~\cite{7762913} partition a task into smaller sub-tasks and decide if the mobile users should perform binary offloading to edge servers or perform minimum offloading. Hence, VR content can be transmitted to nearby edge servers for real-time rendering to improve the QoS of the users. We discuss some of the AR/VR rendering challengers in the following.  Table~\ref{table:summary1} summarizes the approaches to efficient AR/VR cloud-edge-end rendering. According to the surveyed approaches, it can be observed that the most common advantage of them is that they can adapt to various environments and requirements of AR/VR rendering. However, they have different disadvantages, such as they might be myopic, unfair, unscalable, or inaccurate solutions.
	% {|m{.05\textwidth}<{\centering}|m{.2\textwidth}<{\centering}|m{.2\textwidth}<{\centering}|m{.2\textwidth}<{\centering}|m{.2\textwidth}<{\centering}|}
	\begin{table*}
		\small\centering
		\caption{Summary of the approaches in efficient AR/VR cloud-edge-end rendering.}
		\begin{tabular}{|m{.05\textwidth}<{\centering}|m{.15\textwidth}<{\centering}|m{.4\textwidth}<{\raggedright}|m{.1\textwidth}<{\centering}|m{.1\textwidth}<{\centering}|} 
			\hline
			\textbf{Ref.} & \textbf{Issue} & \makecell[c]{\textbf{Main Idea}}& \makecell[c]{\textbf{Pros}}& \makecell[c]{\textbf{Cons}} \\ 
			\hline
			\cite{ng2022unified} & Stochastic demand and network condition &  Selecting the optimal decision based on demand uncertainty& Adaptive allocation & General framework\\
			\hline
			%\cite{braud2020multipath} & Stochastic demand and network condition &  \makecell[l]{Scheduling algorithm to allocate multi-path and multi-server task offloading}\\
			\cite{shi2015offloading} & Stochastic demand and network condition &  Computational offloading guidelines are proposed for AR applications& Offloading guideline & Myopic solution \\
			\hline
			\cite{yu2017polynomial} & Stragglers in edge networks &  Proposed coding technique Polynomial codes to use recovery threshold to mitigate stragglers& Large-scale offloading & Unfair\\
			\hline
			\cite{yuna} & Stragglers in edge networks &  Selecting workers based on reputation and game-theoretic scheme& Reliable & Unscalable\\
			\hline
			\cite{dutta2019optimal} & Stragglers in edge networks  &  Following~\cite{yu2017polynomial}, MatDot code is proposed to reduce the recovery threshold at the expense of high communication cost and PolyDot code to characterize the trade-off between the recovery threshold and communication costs& Latency guaranteed & Inaccurate\\
			\hline
			\cite{ng2021optimal} & Stragglers in edge networks &  Adopt PolyDot code in~\cite{yu2017polynomial} to mitigate the stragglers in edge computing& Energy Efficient & Collaboration of UAVs\\
			\hline
			\cite{baek2020heterogeneous} & Heterogeneous tasks &  Multiple fog nodes are used to replace remote servers to minimize the average latency& Long-term optimization & Unfair\\
			\hline
			\cite{wang2020fast} & Heterogeneous tasks &  Meta RL to improve task offloading adaptation& Fast adaptation & Complex\\
			\hline
			%\cite{8975727} & Heterogeneous tasks &  Three computation methods to determine the best offloading strategy\\
			%\hline
		\end{tabular}
		\label{table:summary1}
	\end{table*}
	
	\subsubsection{\textbf{Stochastic Demand and Network Condition}} The Metaverse will host millions of users concurrently, and try to satisfy each user has a different demand schedule and usage requirements. When an AR/VR device is used, the AR/VR device may associate with an edge server to provide computation support for real-time services. However, the optimal number of edge computation services cannot be precisely subscribed in advance if the user's demand is uncertain. Moreover, given the stochastic fluctuations in demand schedule, it is not practical for the  users' demand to be assumed as a constant. If the Metaverse VSP subscribes to insufficient edge servers for the AR/VR devices to offload, the AR/VR applications would be unable to provide real-time services. The authors in~\cite{ng2022unified} propose two-stage stochastic integer programming (SIP) to model the uncertainty in user demand. The simulation results show that despite the uncertainty regarding user demand, the optimal number of edge servers required to meet the latency requirement can be obtained while minimizing the overall network cost.
	%Besides uncertainty in users' demand, the network condition should also be taken into consideration. When the number of users in the same area increases, it will cause traffic congestion and reduce the access data rate for all users~\cite{band}. As a result, the round trip latency reduces the feasibility of offloading. Furthermore, reliable computation offloading should not only rely on edge computing. Due to the mobility of users or the congestion of an area, it is possible that there are no available edge servers nearby for offloading. In response, the authors in~\cite{braud2020multipath} present an offloading scheme for multi-server, multi-path device-to-device, edge, and cloud offloading. This scheme allows the AR applications to offload their computations in parallel over the available edge devices, i.e., through D2D communication. If there are no edge devices available, the task is offloaded to the next edge-cloud units. The transmission delay is further minimized by considering multiple available communication links. 
	A more in-depth offloading guideline is proposed by the authors in~\cite{shi2015offloading}, and the simulations are conducted based on running AR applications on AR devices such as Google Glass. It has been suggested in the offloading scheme that the AR device should compute the task locally if the network condition is poor or offload AR rendering tasks to edge servers across edge networks. % Basically, the AR devices should offload the task to nearby edge devices to reduce latency using WIFI and Bluetooth. Then, the edge devices should only accept the offload if and only if their battery level is high.

	\subsubsection{\textbf{Stragglers at Edge Networks}} Next, we look at some of the AR device offloading schemes that further reduce the overall latency and the required device resources at mobile edge networks.   In the Metaverse, all users are immersed simultaneously in the shared virtual worlds. The stragglers on edge networks with poor connections can affect the immersive experience of other users by causing delayed responses to social interaction.
	The reliance on the AR device can be reduced when parts of the task are offloaded to other edge devices. % Other than offloading to one edge device, a single task can be partitioned into segments and offloaded to multiple edge devices to further reduce the latency. 
	The authors in~\cite{zhou2018joint} propose a parallel computing approach that leverages the computation capabilities of edge servers and nearby vehicles. This approach partitions the AR task into two sub-tasks at a calibrated proportion so that the edge server and the nearby vehicle can compute the two sub-tasks simultaneously. A joint offloading proportion and resource allocation optimization algorithm is proposed to identify the offloading proportion, communication, and computation resource allocation. Whenever the computational task complexity is high, most of the computations are performed within the edge server rather than in the vehicle so that the computing resources in edge servers can be fully utilized.
	
	% However, some of edge servers may fail to return the result to the application due to delays. As a result, the whole computation process is slowed down. This results in increased computation latency, i.e., the user has to wait a longer time for the computation result, or reduced quality of rendering, e.g., the next VR scene only appears after a long loading time. These edge servers that are delayed are known as straggling nodes or stragglers. Parallel computing can be further extended to mitigate stragglers by adopting a coded distributed computing (CDC) technique. Coded distributed computing consists of a master/fusion node (a task allocator or orchestrator) and worker nodes (nearby edge servers). The master node encodes the computation sub-tasks before offloading them to the worker nodes. The worker nodes perform the sub-task computation and return the completed sub-task back to the fusion node. Unlike parallel computing, the fusion node does not require receiving all the computed sub-tasks to retrieve the final output. The final result can be recovered as long as the number of returned sub-tasks is greater than or equal to the recovery threshold. 
	As several AI-driven AR/VR rendering tasks in the Metaverse, e.g., training DL models, are mainly performing matrix multiplications, the computation process can be sped up if we can accelerate the matrix multiplication process. The authors in~\cite{yu2017polynomial} propose a coding scheme known as polynomial codes to accelerate the computation process of large-scale distributed matrix-matrix multiplications. It can achieve the optimal recovery threshold, and it is proven that the recovery threshold does not increase with the number of worker nodes.   In the vehicular Metaverse, the authors in~\cite{yuna} propose a collaborative computing paradigm based on CDC for executing Metaverse rendering tasks collaboratively among vehicles. To improve reliability and sustainability of the proposed paradigm, they leverage the reputation-based coalition formulation and the Stackelberg game to selection workers in the vehicular Metaverse. In addition, they propose a blockchain-based secure reputation calculation and management scheme for physical entities, i.e., vehicles, to store historical records during CDC. Their experimental results demonstrate the effectiveness and efficiency in alleviating stragglers with malicious workers. 
	
	  However, the large number of physical entities in the Metaverse increases the encoding and decoding cost in CDC. In response, various types of coding schemes, such as MatDot codes and PolyDot codes are proposed in~\cite{dutta2019optimal}, can be adopted in CDC of the Metaverse. These coding techniques can be implemented in edge computing to achieve smooth performance degradation with low-complexity decoding with a fixed deadline~\cite{kim2020coded}. MatDot codes have a lower recovery threshold than the polynomial codes by only computing the relevant cross-products of the sub-tasks, but it has a higher communication cost. PolyDot codes use polynomial codes and MatDot codes as the two extreme ends and construct a trade-off between the recovery threshold and communication cost. By adopting PolyDot codes, the authors in~\cite{ng2021optimal} propose a coded stochastic offloading scheme in the UAV edge network to mitigate stragglers. The offloading scheme makes use of SIP to model the stragglers' shortfall uncertainty to achieve the optimal offloading decisions by minimizing the UAV energy consumption. When comparing the fixed (i.e., when the occurrence of straggler nodes is a constant) and SIP schemes, it can be identified that the SIP scheme has a lower cost. The reason is that the fixed schemes do not account for the shortfall uncertainty of stragglers.
	
	\subsubsection{\textbf{Heterogeneous Tasks}}   The diversity of the Metaverse AR/VR applications, including working, infotainment, and socializing, generates different computational tasks with various delay requirements, e.g., rendering foreground VR scenes has much more stringent latency requirements than rendering background VR scenes. The design of AR/VR remote rendering schemes should therefore consider the coexistence of various tasks in the Metaverse in order to improve the rendering efficiency. %Many DRL-based methods have been proposed to learn offloading policies.  %Therefore, edge computing should work together with fog computing to lower the network latency. 
	% As a single edge node has limited computing capability, the limitation can be covered by using multiple edge nodes. 
	The authors in~\cite{baek2020heterogeneous} adopt a deep recurrent Q-network (DRQN) to classify the offloaded tasks according to their performance requirements to maximize the number of completed tasks within the time limit of completion. Basically, DRQN combines deep Q-learning (DQN) with a recurrent layer to provide a decision when there is only partial observation. The proposed DRQN further adopts a scheduling method to reduce the number of incorrect actions in the training process. In this computation offloading scheme, multiple edge nodes cooperate with each other to guarantee the specific QoS of each task. Each edge node executes an individual action with a joint goal of maximizing the average rewards of all nodes. % Comparing DRQN with other baselines, it can achieve the lowest average task delay by selecting a neighboring edge node that minimizes transmission delay as well as waiting time in the buffers. 
	  However, AR/VR rendering algorithms that rely on DRL often have low adaptability, as the DRL model has to be retrained whenever the environment changes. This is especially applicable for Metaverse applications in which the services provided to users are consistently evolving. In response, the authors in~\cite{wang2020fast} propose an offloading framework in edge computing that is based on a meta RL algorithm. Different from DRL, meta RL utilizes a limited amount of gradient updates and samples that can quickly adapt to different environments. The AR/VR applications can be modeled as Directed Acyclic Graphs (DAGs) and the offloading policy by a sequence-to-sequence neural network. %Using different datasets, meta RL-based offloading framework can always achieve the lowest latency as it can adapt to new tasks much more quickly than the fine-tuned DRL method.
	%As AR/VR applications generate heterogeneous tasks randomly in real-time, the edge servers may be overloaded during one-off random events that result in a high usage demand of service, e.g., due to ad-hoc demand spikes driven by social media influences. The authors in~\cite{8975727} use the cloud-edge-end collaborative computing framework to minimize the average latency in the network when the task generated is heterogeneity. 
	%Each task can have three computing modes to support the required latency. 
	% In this framework, AR/VR devices can determine the best offloading strategy by comparing the computing ability and transmission time of tasks from local devices to the edge and remote cloud servers. The framework reformulates the problem and decomposes it into a series of subproblems that can each be handled by a mobile smart device using Lyapunov optimization and duality theory. Compared to baseline algorithms, the latency of these baselines increases during high usage demand scenarios since the computing resources are misallocated and the capacity of servers is exceeded.
	%However, the proposed framework can always achieve the lowest latency regardless of how many tasks are generated due to the adaptive offloading strategy.

	\begin{table*}
		\small\centering
		\caption{Summary of the approaches in scalable AI model training.}
		\begin{tabular}{|c|c|c|c|c|c|c|c|c|c|} 
			\hline
			\multirow{2}{*}{\textbf{Approach}} & \multirow{2}{*}{\makecell[c]{\textbf{Compression/}\\\textbf{Acceleration}}} & \multirow{2}{*}{\makecell[c]{\textbf{Loss in} \\\textbf{Accuracy}}}& \multicolumn{7}{c}{\textbf{Models}}\vline \\ \cline{4-10}
			& &  & AlexNet & VGG-S &VGG-16 & GoogLeNet & ResNet18 & ResNet34& physical world \\
			\hline
			\cite{han2015learning} &Compression& No& $9\times$ &- &$13\times$ & -&-&-&-\\
			\hline
			%\cite{li2018optimization}&Compression &No& $82\times$ &-&- & -&-&-&-\\
			%\hline
			\cite{han2015deep}&Compression &No& $35\times$ &- &$49\times$ & -&-&-&-\\
			\hline
			\cite{jaderberg2014speeding}&Acceleration &$1\%$ drop& - &- &- & -&- &-&$4.5\times$\\
			\hline
			\cite{kim2015compression}&Compression &Yes& $5.46\times$&$7.4\times$  &$1.09\times$  & $1.28\times$&- &-&-\\
			\hline
			\cite{kozyrskiy2020cnn}&Compression &No& -&-  &-  & -&$2.5\times$ &$7.5\times$&-\\
			\hline
			\cite{hinton2015distilling}&Compression &Yes& -&-  &-  & -&- &-&$1.5\times$\\
			\hline
			\cite{romero2014fitnets}&Compression &No& -&-  &-  & -&- &-&$10.4\times$\\
			\hline
			%\cite{wu2019multi}&Compression &$1.79\&$& -&-  &-  & -&- &-&$2.4\times$\\
			%\hline
		\end{tabular}
		\label{table:summary2}
	\end{table*}
	
	\subsection{Scalable AI Model Training}\label{ai_model_training}
	
	  The virtual services, e.g., AR/VR recommender systems and cognitive avatars, offered by the Metaverse should be intelligent and personalized. To achieve this vision at scale, AI, such as in speech recognition and content generation, will be a key engine.
	The virtual services in the Metaverse can even mimic the cognitive abilities of avatars by using AI techniques so that non-player characters (NPCs) in the Metaverse are smart enough to make intelligent decisions. Even tools can be provided to the users to build their avatars in the Metaverse. However, today's complex DL models are often computation- and storage-intensive when performing training or computing for inference results. It is especially so for consumer-grade edge devices with resource constraints. Therefore, besides the offloading techniques discussed above, the computation efficiencies of the edge devices can also be increased through DL model compression and acceleration. Model compression improves the processing speed and also reduces the storage and communication cost~\cite{cheng2017survey} such that it is more feasible for model training and inference to be performed locally on edge devices with limited resources, e.g., HMDs.
	
	A complex DL model requires high demand for memory and CPUs/GPUs. It is difficult to run real-time Metaverse applications on edge devices when the inference speed of AI model is slow and the computing consumption is high. Therefore, it is essential to convert the complex DL model into a lightweight one.   Lightweight DL models have been successfully implemented in HMDs~\cite{lee2019predicting}, e.g., to predict the FoVs of users. Together with AR, a walk-in-place navigation interface can appear in front of the user to assist the navigation in the Metaverse. In the food and beverage industry, AI-enabled HMDs can provide personalized AR/VR recommendations to users with regard to the users' preferences~\cite{ahn2015supporting, tan2022towards,zhou2022usst}. Furthermore, through AR, non-physical information about the selected 3D objects can be displayed via virtual AR layers that can help users make informed purchasing decisions~\cite{gutierrez2018phara}. As edge intelligence is ubiquitous in the Metaverse at mobile edge networks, it is crucial to reduce the DL models' sizes when the number of AI-enabled services increases and AI models become more complex. In the following, we discuss selected model compression and acceleration methods. The reviewed approaches are summarized in Table~\ref{table:summary2}.
	
	\begin{figure*}[t]
		\centering
		\begin{multicols}{2}
			\includegraphics[width=8.5cm, height=5.5cm,trim={0cm 0cm 0cm 0},clip]{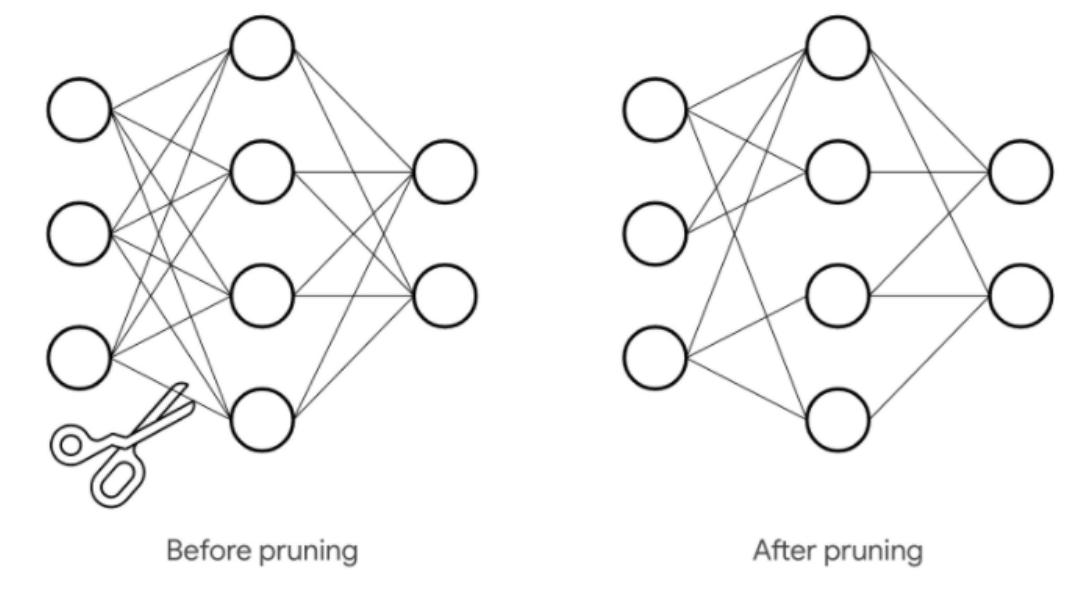}
			\caption{A visualization of connection pruning in a neural network~\cite{prune}.}
			\label{fig:prune}
			\includegraphics[width=7.5cm, height=5.5cm,trim={0cm 22.8cm 13.5cm 0},clip]{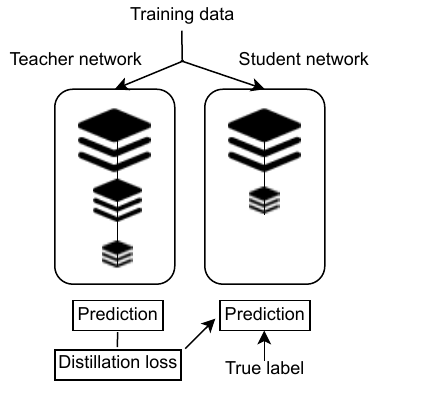}\par
			\caption{A visualization of the student-teacher network paradigm~\cite{wang2021knowledge}}
			\label{fig:kd}
		\end{multicols}
	\end{figure*}
	
	%\begin{figure}[t]
	%\centering
	%    \includegraphics[width=8cm, height=4.5cm,clip]{prune.PNG}
	%  \caption{A visualization of connection pruning in a neural network~\cite{prune}.}
	%  \label{fig:prune}
	%\end{figure}
	
	\subsubsection{\textbf{Parameter Pruning and Quantization}} 
	
	  In the Metaverse, the DL models are expected to provide QoE analysis, evaluation and prediction for multitude Metaverse applications~\cite{du2022rethinking,du2022exploring,du2022attention,liu2022slicing}. However, the Metaverse applications should not contain only simple DL models for QoE evaluation. Compared with complex DL models, simple DL models can only provide applications with limited features. However, complex DL models may comprise millions of weights, most edge devices may be unable to train or store such models. A solution is to prune the unimportant or redundant weights/connections/neurons with zero, such that the model size can be reduced without compromising on the performance. Fig.~\ref{fig:prune} shows a visualization of connection pruning in a neural network. The magnitude-based pruning (MP) is proposed by the authors in~\cite{han2015learning}. MP mainly consists of three steps. The important connections in the network are first learned. Then, the magnitudes of the weights are compared with a threshold value. The weight is pruned when the magnitude is below the threshold. The third step will retain the network to fine-tune the remaining weights. MP can reduce the number of parameters in the AlexNet by a factor of 9 times without affecting the accuracy. MP can also be extended to layer-wise magnitude-based pruning (LMP) to target the multi-layer nature of DNNs. For each layer, LMP uses MP at the different threshold values.
	
	% However, finding the right threshold for parameter pruning is difficult. Most of the time, the threshold is handcrafted based on the practitioner's experience. As such, the authors in~\cite{li2018optimization} transform the threshold tuning problem into a constrained optimization problem to minimize the size of the model but retain the model accuracy. For every iteration, optimization layer-wise magnitude-based pruning is used to prune the model with the use of a subset of the data. Then, the pruned model is adjusted using the rest of the data to preserve the model's accuracy. Finally, the iteration stops when the accuracy is converged. The simulation results show that this approach can reduce the AlexNet-style network up to 82 times while maintaining the same accuracy. Moreover, the threshold does not have to be estimated by the practitioner. However, the convergence of this algorithm is challenging for more complex tasks.
	
	Besides pruning, another method known as quantization is to reduce the storage size of the weights in the DL model by replacing floating points with integers. Once the weights and inputs are converted to integer types, the models are less costly to store in terms of memory and can be  processed faster. However, there is a trade-off with quantization. The model accuracy may drop when the quantization is more aggressive. Typically, the DL models are able to deal with such information loss since they are trained to ignore high levels of noise in the input. The authors in~\cite{han2015deep} proposed a three-stage pipeline that involves pruning, trained quantization, and Huffman coding. Pruning reduces the number of weights by ten times, quantization further improves the compression rate by reducing the number of bits, and Huffman coding provides even more compression. Throughout the compression, the accuracy of the model is not lost. As a result, the DL model can be stored on edge devices without affecting the model accuracy. This method manages to reduce the storage requirement of AlexNet by 35 times and VGG-16 by 49 times. The lightweight models that do not result in a significant compromise in inference accuracies will be instrumental in running intelligent applications on mobile devices that can access the Metaverse ubiquitously.

	\subsubsection{\textbf{Low-rank Approximation}}
	  It is crucial to design smaller models that can be conveniently stored on edge devices to make inferences locally to support the real-time physical-virtual synchronization between the physical and virtual worlds. Besides pruning redundant parameters to reduce the model's size, we can perform decomposition to factorize a large matrix into smaller matrices, which is known as low-rank approximation. It compresses the layers in this way so that the memory footprint of AI models required in Metaverse applications can be reduced. The authors in~\cite{jaderberg2014speeding} take advantage of cross-filter redundancy to speed up the computation within the layers. Basically, it reduces the computational complexity in the convolution process by approximating the filters with a high rank with a linear combination of lower rank filters. This scheme can help to speed up text character recognition applications in AR/VR applications. From the simulation result, the scheme can speed up 2.5 times and ensure zero loss in accuracy.
	
	As explained in Section~\ref{vr_render}, DL should be able to be deployed in AR/VR devices to enhance user experience in the Metaverse. The authors in~\cite{kim2015compression} proposed a scheme called one-shot whole network compression so that the DL model can be deployed in edge devices. This scheme involves three steps: (1) rank selection, (2) Tucker decomposition, and (3) fine-tuning to restore accuracy. Variational Bayesian matrix factorization is applied on each kernel tensor to determine the rank. Then, Tucker-2 decomposition is used from the second convolutional layer to the first fully connected layers, while Tucker-1 decomposition is used for the rest of the layers to compress the network. Finally, back-propagation is used to fine-tune the network to recover the loss of accuracy. Simulations are conducted on edge devices based on AlexNet, VGGS, GoogLeNet, and VGG-16 models, and the scheme is shown to reduce the run time and model size significantly. Apart from using aforementioned DL models, the authors in~\cite{kozyrskiy2020cnn} used other DL models, such as ResNet18 and ResNet34, to combine low-rank approximation and quantization to enable DL in edge devices. Simulations are performed based on multiple datasets, and the compressed models can produce a slightly higher test accuracy than the non-compressed model. This is because low-rank approximation and quantization are orthogonal as the former utilizes redundancy in over-parameterization, and the latter utilizes redundancy in parameter representation. Therefore, combining these two approaches results in a significant compression without reducing the accuracy.
	%Related work:~\cite{denton2014exploiting}
	%\begin{figure}[t]
	%\centering
	%    \includegraphics[width=8cm, height=6cm,trim={0cm 22.8cm 13.5cm 0},clip]{kd.pdf}
	%  \caption{A visualization of the student-teacher network paradigm~\cite{wang2021knowledge}}
	%  \label{fig:kd}
	%\end{figure}
	
	\subsubsection{\textbf{Knowledge Distillation}} 
	% The Metaverse applications face restrictions on computational resources when the DL models are trained in the edge devices. The DL model extracts as many features as possible from the input data in the training phase. As a result, trained models are often large, and the inference time is longer than expected. As a result, the Metaverse platform must compress the model when they are placed in the edge devices.
	
	The concept of knowledge transfer is proposed by~\cite{bucilua2006model} to transfer the knowledge of a large model (teacher) to a smaller model (student). The authors in~\cite{hinton2015distilling} name this concept knowledge distillation. Figure~\ref{fig:kd} shows an illustration of the student-teacher network paradigm. The edge devices do not have to store or run the actual DL model locally with this knowledge transfer. Instead, the teacher can be trained, whereas only the student is deployed to the edge devices. The teacher transfers the knowledge of predicted class probabilities to the student as a soft target. The distillation process occurs when a hyperparameter temperature $T$ is included in the softmax function. When $T=1$, it will be a standard softmax function. When $T$ increases, the probability distribution of the output generated by the softmax becomes softer, this means that the model can provide more information to which classes the teacher found more similar to the predicted class~\cite{hinton2015distilling}. The student learns the class probabilities generated by the teacher, known as soft targets. It is pointed out in~\cite{hinton2015distilling} that hyperparameter $T$ should not be very high, i.e., a lower $T$ usually works better. This is reasonable as when $T$ is high, the teacher model will incorporate more knowledge, and the student model cannot capture so much knowledge because it is smaller in size of parameters.
	
	The authors in~\cite{romero2014fitnets} extend the work in~\cite{hinton2015distilling} and find that using the teacher's intermediate hidden layers of information can improve the final performance of the student. With this information, the authors propose FitNets. Unlike~\cite{hinton2015distilling}, the student in FitNets is deeper and thinner than the teacher so that it can generalize better or run faster. Simulation results confirm that the deeper models generalize better, and the thinner model significantly reduces the computational burden on resource-constrained devices. In the smallest network, the student is 36 times smaller than the teacher, with only a $1.3\%$ drop in performance. Although there is a slight trade-off between efficiency and performance, the student is always significantly faster than the teacher, and the student is small in size.   This shows the effectiveness of knowledge distillation in deploying lightweight models on resource-constrained mobile devices. As such,  the cognitive avatars in the Metaverse can interact with human players in real-time and release their power to help users handle some simple but tedious work.
	
	% Instead of using only one prominent model, the authors in~\cite{wu2019multi} propose to use multiple teachers to transfer the knowledge to a single student. The student is trained using the average output of the teachers. The authors explain that the student learns better than using a single teacher as multiple teachers with different input types can pass more comprehensive knowledge to the student. Using information from various applications in the Metaverse, the model in the edge device can then be trained to learn more in-depth knowledge to increase the model accuracy.
	
	\subsection{Privacy and Security}\label{security}
	
	While the emergence of AR/VR will enhance the QoS delivery in the Metaverse, this also implies that the data can be collected in new modalities. For example, since HMDs instead of smartphones are used by the user, the eye tracking data of users may be captured. While this data may be vital in improving the efficiency of VR rendering, the data can also be used by companies to derive the attention span of users in order to market their products better~\cite{zhang2022no}. As a result, a risk assessment framework can be designed to investigate the negative outcome of any privacy or security intrusion and predict the probability of these events~\cite{8651847}. With the help of the framework, the threats can be prioritized accordingly, and administrators can have better visibility about VR system component vulnerability based on the score values.
	
	AR has the potential to work together with Metaverse applications to alter our lifestyles and revolutionize several industries. However, it is key for policies to be implemented to protect users as this technology becomes more widely adopted. The value of security and data protection and the new ways in which users may be impacted must not be neglected. One way to enforce the security of edge devices is to restrict the sensor data accessed by the AR applications~\cite{jana2013enabling}. As AR content delivery requires the users to upload images or video streams, sensitive information can be leaked when this information is shared with a third-party server. As such, only the most necessary data, such as gestures and voice commands, should be accessed by third parties in order to protect the user while maintaining an immersive experience. Even so, proper authentication is required to validate the correct inputs and block the undesired content. Without authentication, the attacker can copy voice commands and physical gestures to access the user's data. For voice access protection, a voice-spoofing defense system can be used to identify whether the voice command is coming from the environment or from the person who is using the device~\cite{8824980}. For gesture access protection, a GaitLock system can be used to authenticate the users based on their gait signatures~\cite{8276563}. GaitLock does not require additional hardware such as a fingerprint scanner. It only uses the onboard inertial measurement units.
	
	The privacy and security in AR/VR technology should be enhanced from two aspects, computational offloading (using trusted execution environment and FL) and ML (using adversarial ML). In the following, we discuss trusted execution environment (TEE), FL, and adversarial ML techniques.   Table~\ref{table:summary3} summarizes the approaches discussed in this section. During edge training and inference, the approaches can protect sensitive data and models of edge devices, while they will add extra overhead or degraded performance.
	
	\begin{table*}
		\small\centering
		\caption{Summary of the approaches in computing privacy and security.}
		\begin{tabular}{|m{.05\textwidth}<{\centering}|m{.13\textwidth}<{\centering}|m{.4\textwidth}<{\raggedright}|m{.13\textwidth}<{\centering}|m{.13\textwidth}<{\centering}|} 
			\hline
			\textbf{Ref.} & \textbf{Technique} & \textbf{Main Idea} &\textbf{Pros} & \textbf{Cons}\\ [0.5ex]
			\hline
			\cite{costan2016intel} & Trusted Execution Environment & Encrypt a portion of CPU memory to isolate specific application in memory& Separate data portions & Switching overhead \\
			\hline
			\cite{ning2018preliminary} & Trusted Execution Environment & Evaluate the feasibility of deploying TEE on edge device with different CPU& Feasible for edge devices& Switching overhead\\
			\hline
			% \cite{li2020task} & Trusted Execution Environment & Following~\cite{costan2016intel}, the data are encrypted in the TEE before they are offloaded to other edge servers\\
			% \hline
			\cite{wei2020federated} & Federated Learning & Integrate differential privacy into FL to improve the protection level & Prevent privacy leakage & Degraded performance\\
			\hline
			\cite{geyer2017differentially} & Federated Learning & Following~\cite{wei2020federated}, randomized mechanisms are added together with DP to hide users’ contributions during training &Personalized&Lower accuracy\\
			\hline
			\cite{paudice2018detection} & Adversarial ML & Distance between data points and distribution-estimation based outlier detection algorithms are used to defense against poisoning attack&Attacker-proof& Attacker discovery overhead\\
			\hline
			\cite{srisakaokul2018muldef} & Adversarial ML & Propose to use multiple models to form a model family so that it is more robust in white-box attack scenario&Multi-model defense in low-cost& Risky for spare networks\\
			\hline
			%\cite{wang2021you} & Adversarial ML & Propose to use label change rate to protect the pruned neural network against adversarial sample\\
			%\hline
		\end{tabular}
		\label{table:summary3}
	\end{table*}
	
	\subsubsection{\textbf{Trusted Execution Environment}}
	As mentioned in Section \ref{sec:render}, AR/VR devices can be treated as edge devices to perform the computation locally or offload to edge servers or fog/mobile cloud servers. There can be two types of processing environments within an AR/VR device, a trusted execution environment (TEE) and a rich execution environment~\cite{ekberg2014untapped}. A TEE is a trusted and secure environment that consists of processing, memory, and storage capabilities. In contrast, the device operating system and Metaverse applications run in a rich execution environment~\cite{tee}. The operating system will partition the Metaverse applications to restrict sensitive operations and data to the TEE whenever the security is compromised. Implementing TEE can remove hardware tokens for authentication. For example, tokens can be used to wirelessly open doors in buildings or automobiles. The service provider's cost can be reduced without lowering the security level.
	
	\begin{figure}[t]
		\centering
		\includegraphics[width=8cm, height=4cm,trim={0cm 24cm 6.5cm 0},clip]{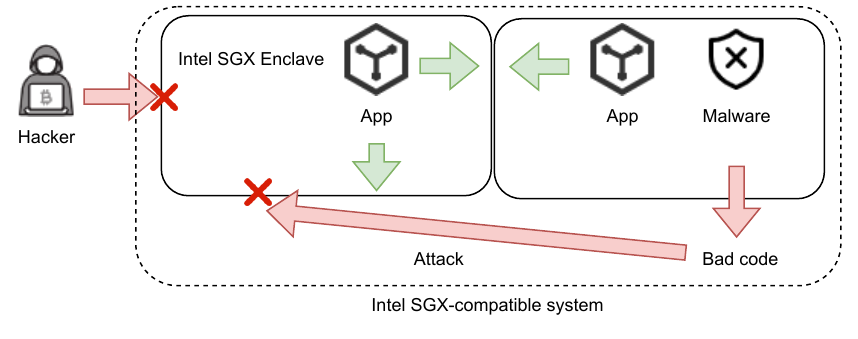}
		\caption{Applications can be protected by placing them into enclaves.}
		\label{fig:enclaves}
	\end{figure}
	
	Intel software guard extensions (SGX) are implemented in Intel architecture that adopts TEE to isolate specific application code and data in memory~\cite{costan2016intel}. The secured memory is known as enclaves. Basically, the AR/VR applications can be split into two parts, the secure and non-secure portions. Sensitive data is placed inside the secure portion, and the application code can only access them within the enclaves. An encrypted communication channel can be created between an enclave and an edge device if the enclave successfully attests itself to the edge device. Sensitive data can then be provided to the enclave. Fig.~\ref{fig:enclaves} shows the visual illustration of an Intel SGX-compatible system. The Metaverse platform owner can set up a list of requirements to prevent the users from using edge devices that are not using SGX-enabled Intel CPU. With the help of SGX, edge devices can trust the data running on a platform with the latest security updates. The authors in~\cite{ning2018preliminary} analyze the feasibility of deploying TEE on the edge platform. It first analyzed the switching time, i.e., the time required to enter and quit the enclave on the edge device. Then, it analyzes the amount of overhead to move the sensitive computation into the enclave. The result shows that the CPU performance in normal mode and enclave are similar, and the overhead of moving the sensitive computation depends on the switching overhead. Finally, it shows that the performance of the non-sensitive computation on the edge device is not affected when the sensitive computation is running inside the TEE.

	% Although TEE can enforce security in task execution, task information can be leaked before execution in TEE on the edge device. Therefore, the offloading policy should include encryption and decryption operations. However, these operations also consume certain computation and energy resources, and they have to be considered when making secure offloading decisions. The authors in~\cite{li2020task} introduce an offloading strategy that includes encryption and decryption operations to further enhance the security in task offloading with TEE-enabled. All the edge devices with TEE are taken into consideration to minimize the total completion time under the energy budget constraint. The proposed framework is more effective than the baseline when the energy budget varies. When the energy budget is low, the edge device has no choice but to offload the task to edge servers. As the energy budget increases, the edge device tends to perform the local computation to avoid data transmission and encryption/decryption, thus achieving a shorter total completion time.

	\subsubsection{\textbf{Federated Learning}}
\begin{figure}[t]
		\centering
		\includegraphics[width=8cm, height=6cm,trim={0cm 20.5cm 9cm 0},clip]{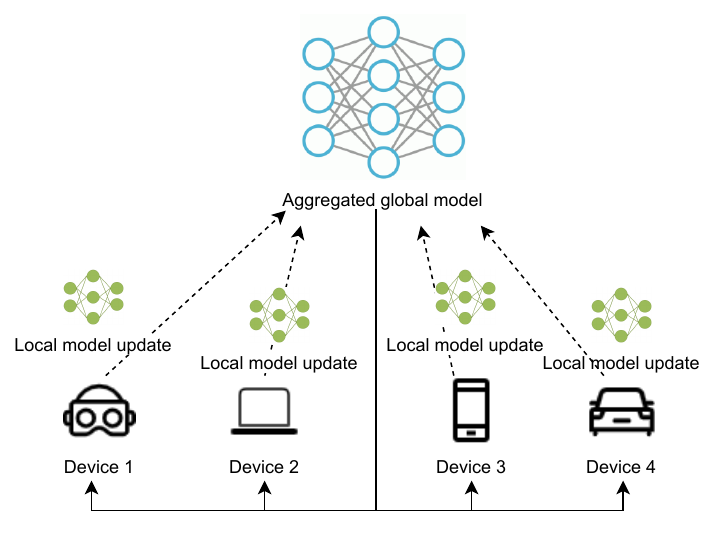}
		\caption{Federated learning in edge networks.}
		\label{fig:fl}
	\end{figure}

	Federated learning~\cite{yang2019federated, 8832210 ,zhang2022challenges} is similar to distributed learning. Both processes speed up the ML process for applications in the Metaverse by splitting computation across many different edge devices/servers. However, in distributed learning, the data from Metaverse applications are first collected in a centralized location, then shuffled and distributed over computing nodes~\cite{konevcny2016federated}. In FL, ML models are trained without moving the data to a centralized location. Fig.~\ref{fig:fl} illustrates an example of FL at mobile edge networks. FL will send the initial ML model parameters to each of the edge devices, which are trained based on their own local data~\cite{konevcny2016federated}. Edge devices send the trained model parameters to update the global model in the server. The global model can be updated in many ways, such as federated averaging, to combine the local models from the edge devices. Once the global model is updated, a new global model is then sent back to the edge devices for further local training. This process will repeat itself until the global model reaches a certain accuracy.
	
	FL provides several benefits for ML model training to enable Metaverse applications. First, FL reduces communication costs since only the model parameters are transmitted to the parameter server rather than the raw data. This is important, especially when the training data is 3D objects or high-resolution video streams \cite{lim2021decentralized}. Second, FL models can be constantly trained locally to allow continual learning. ML models will not be outdated since they are trained using the freshest data that is detected by the edge devices' sensors. Third, FL enables privacy protection as the user data does not have to be sent to a centralized server. For data-sensitive applications, e.g., in healthcare~\cite{sheller2020federated}, data privacy laws may not allow data to be shared with other users. These applications can therefore leverage FL to improve the model quality. Fourth, since personal data is used to train the ML model, personalization can be built into FL to deliver tailored services to users \cite{arivazhagan2019federated}.
	
	Even though personal data protection is enabled when the users' Metaverse data is only trained on the edge devices, FL still has limitations in data privacy.
	\begin{itemize}
		\item \textbf{Data Poisoning:} Users of the Metaverse may train ML models using their devices such as smartphones or HMDs. The highly distributed nature of FL  means that the attack surfaces inevitably increase. Malicious users may train the model on mislabeled or manipulated data to mislead the global model.
		\item \textbf{Inference Attacks :} FL training usually involves updating the global model using multiple local models located on the user's devices. For every update, the adversarial participants can save the model parameters from the local models, and much information can be leaked through the model update, e.g., unintended features about participants' training data. Therefore, the attackers can infer a significant amount of private information, e.g., recover the original training samples of users~\cite{lyu2020threats}.
	\end{itemize}
	% There should be a strong privacy and security protection framework to prevent users from having any negative experience when they are in the Metaverse. For example, when the user is creating an avatar, the attacker can modify the leaked data. As a result, the avatar that is created is different from what the user designed.
	
	A natural way to prevent privacy leakage from shared parameters is to add artificial noise into the updated local model parameters before the global aggregation phase. This is known as differentially private (DP) techniques~\cite{wei2020federated}, where noise is added to the parameter such that an attacker is unable to distinguish the individual contributor's ID. The authors in~\cite{wei2020federated} add noise before sending the model parameters to the server for aggregation. A trade-off is derived between the privacy protection and the model performance, i.e., a better model performance leads to a lower level of privacy protection. Using the $K$-random scheduling strategy, the proposed framework obtains the optimal number of edge devices to participate in the model aggregation in order to achieve the best performance at a fixed privacy level. From the simulations, if the number of edge devices participating in the model aggregation is sufficiently large, the framework can achieve a better convergence performance as it has a lower variance of the artificial noise.
	
	\subsubsection{\textbf{Adversarial Machine Learning}}
	The authors in~\cite{geyer2017differentially} also introduce a DP approach to FL. At the same time, it is also shown that the edge device's participation can be hidden in FL at the expense of a minor loss in model performance with a randomized mechanism. The randomized mechanism consists of random sub-sampling and distorting. The edge device's data is protected if the attacker does not know if the edge device is participating in the training. A drawback of DP techniques in FL is that when the number of edge devices participating in the model aggregation is small, the model accuracy is significantly below the non-DP model performance. However, when there are more devices participating in the training, the model accuracies are similar.
	%Since adding noises will cause a trade-off between the accuracy and protection. Therefore, to prevent adding unnecessary noise, the authors in~\cite{andrew2019differentially} designed a scheme based on the adaptive gradient to limit the amount of noise penetrating the gradient. During training, the clipping threshold is tuned to track a given quantile of the update norm distribution. The scheme is compatible with other technologies such as compression and secure aggregation. When the threshold is clipping to the median, the proposed scheme outperforms fixed clipping to the best fixed clip chosen in hindsight.

	DL models in the Metaverse are often large and complicated, and they are known for being black boxes, which means that we lack an understanding of how they derive predictions. As a result, attackers may be able to find and exploit hidden issues when these black-box models are deployed in a distributed manner across edge devices. For example, the attackers could deceive the model into making inaccurate predictions or revealing confidential user information. In addition, without our knowledge, fake data could be exploited to distort models. Specifically, the attacks may target the following parties:
	\begin{itemize}
		\item \textbf{Metaverse platforms:} The Metaverse platforms mostly rely on data to provide better services to the users. The attackers can alter the data to manipulate the AR/VR applications, e.g., by targeting recommender systems to push undesirable content to users~\cite{zhao2017unified}.
		\item \textbf{Virtual service providers:} The VSPs may utilize ML as a service for users who have insufficient computation capacity~\cite{MLAAS}. Attackers can eavesdrop on the parameters exchanged to infer the membership of users or contaminate the AI model. 
		\item \textbf{Users :} The AR/VR/haptic devices are the devices used by the users to connect with the Metaverse. Once the attacker controls the devices, they can inject adversarial training examples to contaminate the AI model. 
	\end{itemize}
	
	Adversarial ML is a field that seeks to address these issues. The adversarial attack can be prevented by using adversarial training. The DL models should be trained to identify the injected adversarial examples. The authors in~\cite{paudice2018detection} propose a defense mechanism that allows the edge devices to pre-filter the training dataset to remove the adversarial examples. This defense mechanism is built based on outlier detection techniques. It assumes that a small portion of data points are trusted. If all the data points in the Metaverse are untrusted, then defending against adversarial examples is even harder to discover these adversarial examples. The removal of outliers is crucial for VR devices in user authentication mechanisms. The VR device should only be unlocked by the owner or authorized users, while outliers are prohibited from access~\cite{zhu2020blinkey}. Different outlier detection techniques are used to compare the model accuracy. The distribution-based approach outlier detection algorithms~\cite{breunig2000lof} have the best detection accuracy when the outlier detection threshold is at the 99 percentile. 
	
	The security can be further enhanced by deploying multiple DL models within AR/VR applications. Considering the unique features of DL models, the adversarial examples generated for a specific Metaverse application cannot transfer to other applications. The edge devices can store multiple DL models to act as an illusion to the attacker. The authors in~\cite{srisakaokul2018muldef} propose a multi-model defense technique known as MULDEF. The defense framework consists of two components, i.e., the model generator and runtime model selection. The model generator will train multiple models to form a model family to achieve robustness diversity so that the adversarial examples can only successfully attack one DL model. The runtime model selection has a low-cost random strategy to randomly select a model from a model family. From the simulation result, it has been clearly indicated that the edge devices are more robust against adversarial examples by having more models in MULDEF. More models mean that the chance of selecting a robust model is higher.

	%However, deploying multiple DL models in edge devices consumes too much memory and computing resources. In addition, the edge device user does not want a single application to consume too much power. The authors in~\cite{wang2021you} detect the adversarial samples via pruning models. The framework is mainly tested with four pruning methods, i.e., random channel pruning, $L_1$ norm pruning, lottery ticket hypothesis pruning, and subnetwork extraction. A simple defense mechanism is used to identify the true label of the adversarial sample with no additional computational cost. The pruned models are smaller in size, more sensitive to adversarial examples, and more feasible in practice to deploy multiple models in edge devices. The result shows that the pruned models can obtain high accuracy while reducing the model to $48\%$ of the original size on the CIFAR10 dataset.
	
	\subsection{Lessons Learned}
	
	%\subsubsection{human-centric computing}
	%The Metaverse involves of people all around the world. They could be requesting different services depending on their locations. It is not important 
	 
		\subsubsection{Adaptive AR/VR Cloud-Edge-End Rendering} Leveraging the ubiquitous computing and intelligence in mobile edge networks, users can offload AR/VR tasks to available computing nodes in the network for rendering. With the cloud-edge-end collaborative computing paradigm, the performance bottlenecks in AR/VR rendering, such as stochastic demand and network condition, stragglers at edge networks, and heterogeneous tasks, can be adaptively addressed through efficient coordination of ubiquitous computing and intelligence.
		\subsubsection{On-demand and Generalized Model Compression} Model compression can reduce the required hardware requirements of the local devices to access the Metaverse. However, there will be devices still facing resource constraints such as storage, computational power, and energy. Therefore, implementing an additional on-demand model compression is vital to compress the model further. The trade-off between the accuracy and users' QoE should be studied to strike the balance of the on-demand model compression~\cite{wang2021deepnetqoe}. On the other hand, 
		%the Metaverse services will consist of numerous DL models. Therefore, it is impractical to apply a different model compression or acceleration framework for each model. Instead, 
		a generalized model compression technique can be introduced to derive the most suitable compression technique or level of compression for heterogeneous tasks for the computational interoperability in the Metaverse.
		\subsubsection{User-centric Computing} The Metaverse should be enforced in a trustworthy and privacy-preserving manner if it is to be accepted by users. The protection can be enforced from two aspects, i.e., the virtual and physical worlds. In virtual worlds, the applications should protect the sensitive information located in the edge devices when the applications perform the computation. In the physical world, information such as GPS locations, the voice of the users, and eye movements captured by external gadgets like AR/VR devices should also be protected as they can reveal additional information about the users \cite{hongyang}. While data-hungry AI algorithms need a lot of such data input, it is important to ensure that users' privacy is protected through data minimum principles.
		\subsubsection{Secure Interoperable Computing} 
		In this section, we have discussed computation paradigms from a general perspective. In practice, however, the data of users may be stored in separate edge/cloud servers. The development frameworks utilized for the virtual worlds may also differ from another, thereby complicating the process of edge or cloud offloading. As such, it is important to consider distributed storage solutions to ensure users can traverse among the physical and virtual worlds. We will revisit this in Section \ref{sec:blockchain}.
		\subsubsection{From Distribution to Decentralization}
		The distributed computation paradigms discussed in this section involve cloud/edge servers or centralized parameter servers. While distributed computing technologies, such as FL, discussed can mitigate privacy loss, the involvement of centralized servers is at risk of a single point of failure. In Section \ref{sec:blockchain}, we will discuss the decentralized data storage/sharing and edge offloading techniques that can address this issue.

	\section{Blockchain}
	\label{sec:blockchain}
	
	\begin{figure}[b]
		\centering
		\includegraphics[width=1\linewidth]{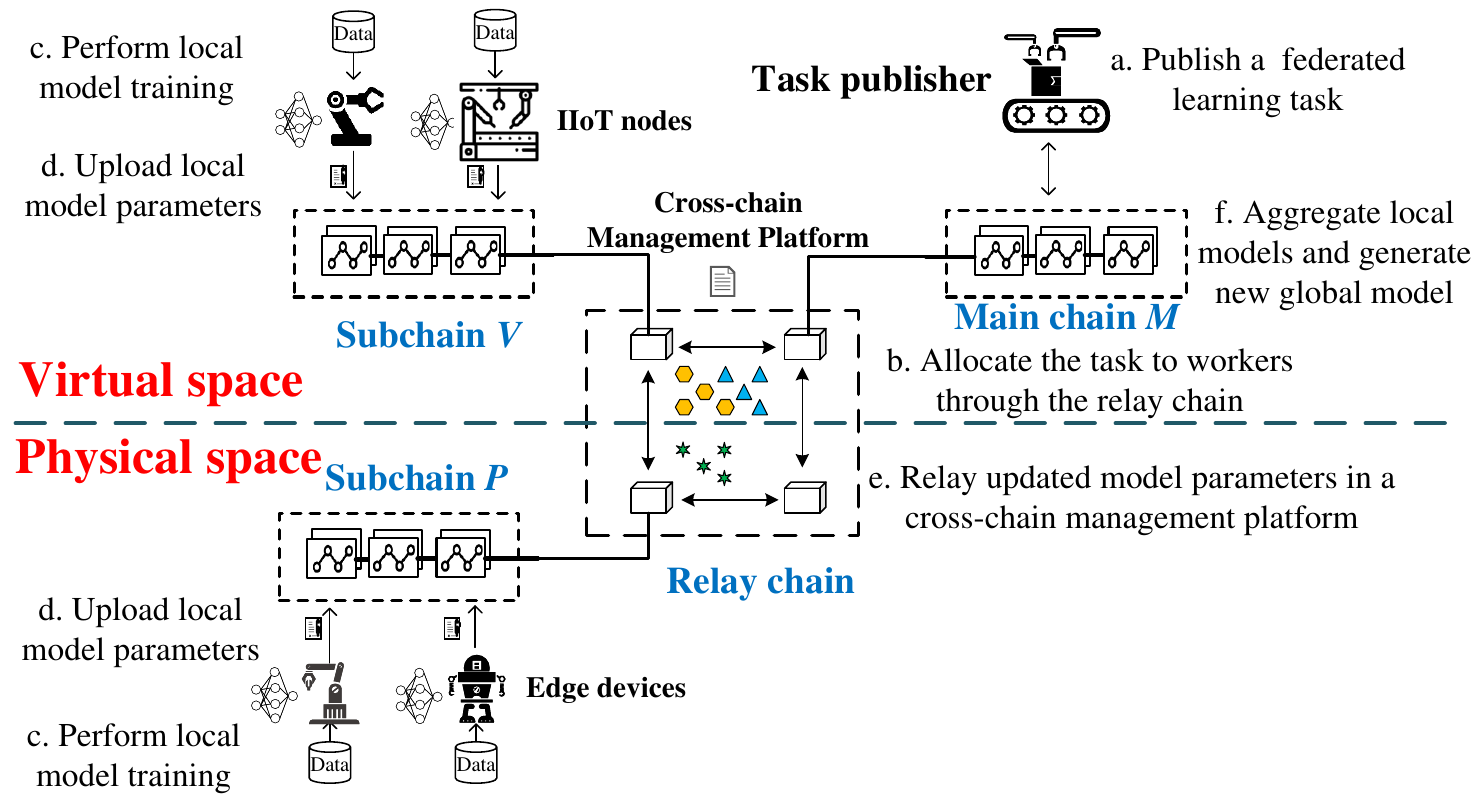}
		\caption{Decentralized FL empowered by a cross-chain system for the Metaverse~\cite{kang2022blockchain}.}
		\label{fig:kangfl}
	\end{figure}
	
	% As the quantity and value of data and knowledge in the Metaverse increase exponentially~\cite{lee2021all}, the necessity of data management reliability and security at mobile edge networks is growing rapidly. Therefore, wireless blockchain plays an indispensable role as one of the fundamental technologies for metaverse  that are required to ensure the reliability and security of Metaverse data and services at mobile edge networks~\cite{jeon2021blockchain}. There are three main kinds of wireless blockchains in the implementation of the Metaverse with different privacy and security levels~\cite{wang2021synergy}, i.e., public blockchain, private blockchain, and consortium blockchain. In details, public blockchain is suitable for safeguarding and protecting the privacy and security of open applications, such as social media. Since public blockchain is a permissionless system, it allows numerous edge devices and edge nodes to participate equally in the Metaverse. On the other hand, consortium blockchain is appropriate to secure commercial applications in the Metaverse as a permissioned system, operated by a group of pre-selected organizations. Finally, private blockchains are used as a permissioned system operated by a single organization. Thus, private blockchain provides a viable way for some organizations and applications to handle sensitive data and records. 
	
	Security and privacy are critical concerns for embodied users and service providers in the Metaverse at mobile edge networks, as discussed in Section~\ref{security}. The blockchain, as a decentralized and tamper-proof database, is another promising feasible solution that can protect the assets and activities of both Metaverse users and service providers without third-party authorities~\cite{zheng2018blockchain,yuna}. With outstanding properties such as immutability, auditability, and transparency, the blockchain can support proof-of-ownership for goods and services in the Metaverse~\cite{dai2019blockchain,yang2022fusing}. First, users immerse in virtual worlds as avatars, which contain a large amount of private user data. The users can generate virtual content during their activities and interactions in the Metaverse, such as gaming items and digital collectibles. After the blockchain stores the metadata and media data in its distributed ledger, this virtual content becomes digital assets and can be proved for ownership and uniqueness. Second, during the construction and maintenance of the Metaverse, physical-virtual synchronization supported by PSPs, such as IoT devices and sensors, is also recorded as transactions in the blockchain. As a result, the avatar-related data and edge resources can be governed by the blockchain in a secure and interoperable manner. Furthermore, blockchain-based interoperability allows economic systems in virtual worlds to synchronize the economic system in the physical world with more robust and fluent circulation.   For example, as shown in Fig.~\ref{fig:kangfl}, the authors in \cite{kang2022blockchain}  user-defined decentralized FL framework for industrial IoT devices in the Metaverses. To seamlessly utilize data generated in the physical and virtual worlds, a hierarchical blockchain architecture with a main chain and multiple subchains is proposed to aggregate the global model securely. In detail, the components of blockchain as an infrastructure for the Metaverse are shown in Fig.~\ref{fig:lego}. To achieve a secure and reliable Metaverse at mobile edge networks, the following criteria should be satisfied:
	\begin{figure}[t]
		\centering
		\includegraphics[width=1\linewidth]{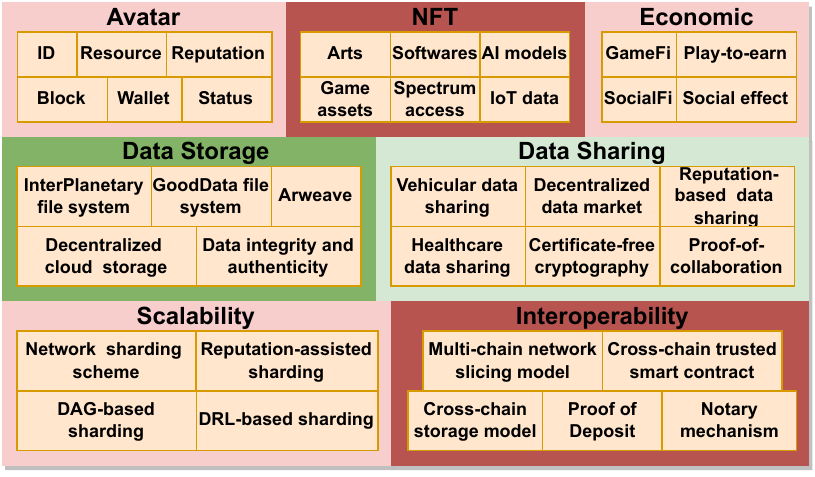}
		\caption{Components of Blockchain-as-an-Infrastructure for the Metaverse at mobile edge networks.}
		\label{fig:lego}
	\end{figure}
	 
		\begin{enumerate}
			\item \textit{Decentralized consensus and trust:} With the smart contracts and consensus algorithms in the blockchain~\cite{yang2022fusing}, the blockchain can ensure security and trust for data and activities in the Metaverse in a decentralized and tamper-proof manner without third-party authorities.
			\item \textit{Proof-of-ownership:} The digital content created by users in the Metaverse can be permanently recorded and stored on the blockchain as NFTs~\cite{wang2021non}, and thus the ownership of UGC is provable and unique. This motivates users to actively generate contents that will populate the Metaverse. The digital content can be traded for digital or fiat currencies in the Metaverse Economic system.
			\item \textit{Scalability and Interoperability:} The Metaverse is expected to blur the boundary of the physical world and virtual worlds via seamless synchronization among massive physical and virtual entities. Therefore, the blockchain needs to provide scalability and interoperability for various virtual services~\cite{lee2021all}. The scalability allows massive devices and users to access the Metaverse at the same time, while the interoperability enables avatars in different virtual worlds to interact seamlessly.
			\item \textit{Automatic edge resource management:} Through various types of smart contracts, the blockchain allows IoT devices, sensors, and PSPs, to automatically manage resources at mobile edge networks. In addition, the blockchain can motivate PSPs to support the physical-virtual synchronization by distributing incentives~\cite{xu2022wireless}.
			\item \textit{Transparency and Privacy:} In the blockchain, information of goods and services is transparent and anonymous for all participating entities~\cite{duan2021metaverse}. In this way, all information in the Metaverse is open-access in the blockchain, overcoming the asymmetric information of traditional networks. Moreover, anonymous users and providers in the blockchain allow the blockchain to protect their privacy to the greatest extent possible.
	\end{enumerate}

	\subsection{The Blockchain-based Metaverse Economic System}
	
	% The Metaverse consists of a collection of shared virtual worlds where users can engage in life-like activities in the physical world, such as socializing, working, and doing business~\cite{lee2021all}.
	  The Metaverse economic system is a merge of the physical economic system with fiat currencies and the virtual economic systems with cryptocurrency, stablecoins, central bank digital currencies (CBDCs)~\cite{fung2016central}.
		Based on the economic systems in the physical world, the Metaverse economic systems are essential to facilitate the circulation of digital and physical currencies, goods, and services in the Metaverse~\cite{duan2021metaverse}. However, there is no centralized government in the Metaverse. Therefore, the virtual Metaverse economic systems built upon the blockchain are expected to provide transparent and verifiable economic operation without third-party authorities. 
		To support \textit{the creator economy} in the Metaverse at mobile edge networks, users can be creators of UGC in the Metaverse based on the communication and computation resources provided by PSPs~\cite{lee2021creators}. Next, the uniqueness and ownership of UGC can be protected via the wireless blockchain by minting the UGC as NFTs for collecting and trading. Finally, these tradable NFTs in games and social activities can be sold for revenue in the form of digital currencies or physical currencies.
	% As a well-known component of economic characteristics of the Metaverse, \textit{the creator economy} allows participants in the Metaverse to become creators of digital content in 3D worlds, i.e., UGC~\cite{lee2021creators}. 

		\subsubsection{\textbf{User-generated Content}} 
		
		User-generated content is one of the most important features of the Metaverse, in which users at mobile edge networks can be creators of virtual assets, e.g., videos, music, and profile pictures, rather than having the platform's developers/operators offer them for sale in a marketplace \cite{duan2021metaverse}. 
		% For example, in the popular Metaverse game named Roblox, more than 115 million monthly active players enjoy more than two billion hours of UGC playtime per month. 
		The development of Metaverse applications inevitably generates huge amounts of data, which poses data processing and storage challenges for Metaverse PSPs~\cite{ning2021survey}. Fortunately, UGC can be stored in distributed cloud and edge servers. For example, the authors in \cite{8726068} propose a blockchain-empowered resource management framework for UGC data close to the data source.
		A shared edge structure is proposed that enables the adaptable use of heterogeneous data and available edge resources, reducing service delay and network load, but without considering the single point of failure of edge servers in UGC storage.

	\subsubsection{\textbf{Non-Fungible Tokens}} 
	
	Non-fungible tokens are digital assets that link ownership to unique physical or digital items, which can help to tokenize UGC like arts, music, collectibles, video, and in-game items~\cite{kiong2021defi}. 
	The NFTs ensure uniqueness by permanently storing encrypted transaction history on the blockchain. Each NFT has a unique ID value, which is used to verify the ownership of digital assets and assign a value to future transactions~\cite{wang2021non}. NFTs are mainly used to commemorate special moments and collect digital assets. Recently, NFTs have been integrated into Metaverse to create a new digital content business \cite{wang2021non}. In March 2021, a breaking NFT event, where a digital work of art created by Beeple was sold for \$69.3 million at Christie's, happened in the UK \cite{Everydays}. This story shows the great potential of NFTs in Metaverse for our future life, which requires secure and reliable NFT management, especially NFT storage. 
	
	Normally, UGC creators record the metadata of NFTs by on-blockchain or off-chain mechanisms~\cite{hong2022cycle,cai2021socialchain}. For the blockchain-based storage, the metadata is permanently stored with their NFT tokens. For example, Decentraland is a virtual platform built on the Ethereum blockchain for UGC storage and monetization as NFTs \cite{goanta2020selling}.  The Decentraland stores ownership of NFTs and other tradable information on the Ethereum blockchain. However, the blockchain storage cost is usually more expensive than that of the off-chain mechanisms. To save storage cost on Ethereum and improve the scalability of Decentraland, the user location information and scene status for real-time interaction is stored on off-chain, such as end devices and edge servers. Similarly, Sandbox is a blockchain-based virtual sandbox game system that can maintain users' ownership of digital land and UGC~\cite{duan2021metaverse}. In Sandbox, the transaction data of digital assets are stored on the Ethereum blockchain, while the media data of digital assets are stored on the InterPlanetary File System (IPFS), and the uncasted digital assets are stored on the Amazon’s S3 cloud servers \cite{duan2021metaverse}. Therefore, off-chain storage solutions are becoming more and more popular for digital asset management, especially NFTs. Currently available off-chain storage solutions mainly include centralized data centers, decentralized cloud/edge storage, and IPFS. Taking one of the representative NFT projects named CryptoPunks as an example, a centralized server is utilized to store the integrated product picture, and smart contracts are used to store the encrypted hash value of the metadata and media data for verification~\cite{dowling2021non}. In this way, with the non-tampering characteristic of the blockchain, the picture can be verified whether there is any modification according to its hash value.  
	However, the media data is stored in the central server instead of the entire blockchain nodes as in NFT ownership storage, which leads to many security risks, including data loss and denial of service.
	
	\subsubsection{\textbf{Game and Social Financial Systems}}
	
	Games and social media are typical applications in the Metaverse, where users can interact with avatars in the Metaverse for entertainment and socializing. Moreover, the game and social finance are also an essential part of the economic systems in the Metaverse. In common usage, Game Financial (GameFi) refers to decentralized applications (dApps) with economic incentives, i.e., play to earn~\cite{why2022olga}. Similarly,  Social Financial (SocialFi) means the social acceptance and effects in the Metaverse that if there are more users in the Metaverse, the valuation of the Metaverse is also higher~\cite{nie2018stackelberg}. The authors provide a comprehensive and detailed overview of blockchain games in~\cite{min2019blockchain}. 
	% They conclude that the blockchain can have epochal significance for the gaming industry. Specifically, the blockchain enables transparency of rules in-game applications, validation of ownership of digital assets, and reuse of assets in games through decentralization and distributed ledger. 
	  Specifically, GameFi and SocialFi in the Metaverse consist of three main components. 
		First, players with resource-constrained edge devices can purchase Metaverse services from edge/cloud servers to access 3D virtual worlds. 
		Second, the blockchain-based economic system in the Metaverse allows players to trade with other players and VSPs for digital or fiat currencies. Third, the value of assets created by players in the virtual worlds might cover the fees charged by PSPs, and thus increasing player satisfaction and enabling players to immerse in the Metaverse with greater enthusiasm.
		
		The authors in~\cite{zhao2020cloudarcade} propose a token-based gaming system called CloudArcade that uses transparent blockchain-backed tokens. In CloudArcade, a silent payment method based on blockchain is proposed to protect the game quality, enhance players' trust in services, and protect their privacy. Cooperative game worlds still lack a suitable system to motivate participants to contribute their trustworthy data. Therefore, the authors in~\cite{xin2020reciprocal} introduce reciprocal crowdsourcing games among wireless users, a new decentralized cooperative game model based on the blockchain to strengthen trust between game participants through transparent and collaborative work. In a campus-based Metaverse, the authors in~\cite{duan2021metaverse} develop a virtual economic system to incentivize academic activities for students on the campus. Based on GPS information of students' smart phones, this financial system is linked to real campus buildings so that students are rewarded with more tokens in library than in dormitories.
	
	\subsection{Decentralized Edge Data Storage and Sharing}
	
	To protect against the risks, IPFS is widely used to achieve redundant backup and stable content addressing. Unlike centralized storage solutions that rely on a central server, IPFS maintains an addressing network running on multiple nodes without the problems of data invalidation and tampering~\cite{kumar2020distributed}. 
	Although off-chain storage has a higher upper bound on storage space, the scalability of storage systems still needs to be improved to meet the increasing demands of NFT storage in the Metaverse. Therefore, sidechain/multiple blockchain-based storage systems can improve the storage capacity of the blockchain for NFTs~\cite{wang2021non}. First, multiple blockchains can cooperate to create a common storage platform with abundant blockchain and off-chain storage resources for NFT storage. However, existing NFT schemes are separated from each other without interoperability due to the limitations of the underlying blockchain platform.
	%To achieve cross-system storage of NFTs, cross-chain communication is essential to improve blockchain interoperability~\cite{tian2021enabling}. Typical cross-chain communication needs external trusted entities, and thus the decentralization nature will inevitably be lost to some extent. Therefore, for NFT storage in the Metaverse, it is meaningful to explore  efficient and scalable NFT storage solutions by using advanced blockchain technologies. 
	
	\subsubsection{\textbf{Blockchain-based Data Storage}}
	Typical decentralized data storage projects include Storj DCS (Decentralized Cloud Storage) and IPFS. Specifically, Storj DCS stores and archives large amounts of data in a decentralized cloud and provides users with encrypted, secure, and affordable blockchain data storage services~\cite{storj}. In contrast, IPFS is a protocol used in peer-to-peer data networks to store and share data in a decentralized file system where each file is uniquely identified to address the content and prevent multiple identical files from being uploaded to the data networks~\cite{IPFS}. 
	 To move IPFS from the cloud to the edge, GoodData File System (GDFS) is a more efficient, domain-aware decentralized file system developed by GoodData based on IPFS and blockchain cloud storage. Compared to IPFS, GDFS can help users store multiple copies and ensure the reliability, availability, and longevity of data storage, especially UGC data in the Metaverse. At the same time, GDFS allocates highly reliable, available, and confidential edge storage resources from PSPs to users nearby, ensuring efficient utilization of storage resources \cite{GDFS}.
	
	In contrast to traditional pricing schemes for blockchain data storage systems based on the number of data accesses, Arweave proposes a permanent pricing scheme for blockchain data storage~\cite{williams2019arweave}. In Arweave's pricing scheme, a data node is rewarded for long-term data storage, i.e., when a node creates a new data block to prove storage, the storage capacity of that proof is used as a tariff for the Arweave network to reward the node. When a node stops storing data blocks, Arweave does not penalize the node but stops rewarding it. In Arweave, users pay a transaction fee for uploading data, a portion of which goes to the miner, while the network retains the rest.
	  To improve sustainability of blockchain-based data storage systems running on edge nodes, the authors in~\cite{liu2020economics} propose a resilient incentive mechanism for blockchain data storage that enables a longer-term, healthier ecology of blockchain data storage systems. Their proposed incentive scheme allows multiple blockchains to balance increasing storage requirements with decreasing mining revenues. Moreover, to improve fairness of the data storage systems running among heterogeneous nodes, the authors in~\cite{daniel2022ipfs} model the incentive scheme for blockchain data storage as a two-stage game analyzing a Nash equilibrium with negative externalities and unfair lagged prices.
	%{IPFS’s Filecoin, Swarm, Storj, and Arweave reward nodes storing data. The reward is either for storing the data over time or for a specific time period. The time period is defined and nodes are preor postpaid, misbehaving storage nodes are then punished or not compensated. In IPFS’s Filecoin, users rent specific storage for a time period. In Swarm, storage guarantees are sold. Swarm, Storj and Arweave reward nodes for storing data over a long time without defined time constraints. In Swarm, storage nodes can participate in a lottery, if they store certain chunks and might be rewarded for the continued storage. In Storj, storage nodes are compensated in time intervals for the data they stored during the interval, in case of storage failures the reward is instead used for file repair compensating the new nodes. In Arweave, the network is paid to store data for a long term. When a node creates a new block, proving storage of data, the node is compensated for its continued provision of storage capacity.}
	
	\subsubsection{\textbf{Blockchain-based Data Sharing}}
	
	Blockchain is intended to be a practical approach to ensure data security and improve efficiency in the Metaverse at mobile edge networks. For example, vehicle-to-vehicle data exchange and data analytics can enrich existing vehicle services in vehicular networks~\cite{liu2022blockchain,xu2022secure}. However, RoadSide Units (RSUs) and vehicles can be malicious and untrusted~\cite{shen2021data}. To solve this problem, the authors in~\cite{kang2018blockchain} propose a consortium blockchain-based system protected by smart contracts for secure data storage and exchange in vehicular networks. This system deploys smart contracts on RSUs and vehicles to achieve secure and efficient data storage and sharing. To improve the trustworthiness of data, they propose a reputation-based data sharing system that uses a three-level subjective logic model to measure the reputation of data sources. To minimize the energy consumption of mobile edge networks and maximize the throughput of the blockchain-based systems, the authors \cite{liu2020blockchain} use an asynchronous learning-based approach for efficient strategy discovery in dynamic network environments. Healthcare services generate a large amount of private medical data about patients. To leverage the data to provide effective and secure medical services to patients while maintaining privacy, the authors in~\cite{abdellatif2021medge} propose a medical edge blockchain framework. In this framework, blockchain technologies are used for fast and secure exchange and storage of a large amount of medical data. They are also developing an optimization model for this framework to improve the latency and computational cost of sharing medical data between different medical institutions.

	The authors in~\cite{li2018blockchain} design a secure and accountable IoT storage system using blockchain. It overcomes the drawbacks of certificate cryptosystems by providing a convenient way to broadcast the public key of any IoT device with the public ledger of the blockchain system. Certificate-free cryptography significantly reduces redundancy while providing efficient ways to authenticate IoT devices.
	Aiming to develop resource-efficient techniques for blockchain-based data exchange in the Metaverse, the authors in~\cite{xu2018making} propose the proof-of-collaboration consensus mechanism by reducing the computational complexity in edge devices. The proposed transaction offloading module and optimization method can save 95\% of storage resources and 30\% of network resources, respectively. By jointly considering data storage and sharing in MEC, the authors in~\cite{zhang2021secure} propose a blockchain-based secure, and efficient data management mechanism for IoT devices with limited resources. In this system, the fault tolerance of this mechanism is improved by key sharing, where the private signature keys are stored in different blocks of the blockchain. Moreover, this mechanism supports anonymous authentication, which enables trusted edge nodes with powerful computing power to implement data proxy signatures, reduce the load on IoT devices with limited computing resources, and ensure the integrity and authenticity of Big Data.
	
	\subsection{Blockchain Scalability and Interoperability}
	
	Due to value isolation among multiple blockchains in the Metaverse, interoperability of multiple virtual worlds is required for data interaction and collaboration between different blockchains~\cite{belchior2021survey}. Therefore, cross-chain is a critical technology for secure data interoperability~\cite{johnson2019sidechains}. Meanwhile, sharding is a flexible way to improve the scalability of the blockchain so that a significant number of connections can be accommodated in the Metaverse~\cite{zamani2018rapidchain}. Therefore, to improve blockchain scalability, the Metaverse will introduce sharding technology to divide the entire Metaverse into multiple connected shards. Each shard of the Metaverse contains its independent state and transaction history, and certain nodes will only process transactions in certain slices, increasing the transaction throughput and scalability of the blockchain \cite{yang2019integrated}. Meanwhile, cross-chain is also an effective solution to scalability issues \cite{jiang2019cross}. Cross-chain is a technology that enables the interconnection between blockchain networks in the Metaverse by allowing the exchange of information and value to improve the scalability and interoperability of blockchains.
	
	\subsubsection{\textbf{Blockchain Scalability}}
	
	Access control without a trusted third-party platform is one of the essential ways to protect privacy in the Metaverse~\cite{zheng2021blockchain}. Given the limited resources for lightweight IoT devices, the authors in~\cite{li2021scaling} propose the Network Sharding Scheme to improve the scalability of the blockchain-based access control systems in IoT. In this scheme, edge nodes are divided into multiple shards that manage their own local blockchains, while cloud nodes manage a global blockchain. In this way, the transaction processing rate can be improved, and the storage pressure of each node can be reduced when multiple blockchains process transactions in parallel. According to their theoretical analysis, the routing cost of querying transactions can be reduced to $O(1)$. To extend ~\cite{li2021scaling} to B5G/6G networks, the authors in~\cite{xie2021resource} propose a blockchain sharding scheme based on a directed acyclic graph (DAG) that combines the efficient data recording of sharding and the secure blockchain update of DAG. In this scheme, DAG can maintain a global state that combines the computational power of different committees without causing security problems. Moreover, they design a resource-efficient consensus mechanism that saves the computational cost of the consensus process and improves the spectrum efficiency in B5G/6G networks. To handle the sub-optimum performance issues caused by blockchain sharding, some intelligence-based blockchain sharding schemes are proposed in~\cite{yuan2021sharding, asheralieva2020reputation}. Specifically, the authors in~\cite{yuan2021sharding} use a DRL-based algorithm to determine the number of partitions, the size of microblocks, and the interval for generating large blocks.
	The authors in~\cite{asheralieva2020reputation} use reputation-based DRL to form shards in a self-organized manner. Each peer selects its shard independently to maximize its payoff, which depends on the peer's throughput. Finally, the authors in~\cite{yuan2021sharding} and~\cite{asheralieva2020reputation} both analyze the security and throughput performance of their proposed schemes.
	
	\subsubsection{\textbf{Blockchain Interoperability}}
	
	In addition to blockchain sharding for scalability, cross-chain is another powerful technology to provide the Metaverse for interoperability that breaks down the siloed nature of the blockchain to create an intertwined distributed Metaverse ecosystem~\cite{9612081}.   To achieve secure and efficient IoT data management, the authors in~\cite{jiang2019cross} design a cross-chain framework based on the notary mechanism to integrate multiple blockchains. The proposed framework can provide fine-grained and interactive decentralized access control for these IoT data in different blockchains. However, the notary mechanism as a cross-chain consensus mechanism may compromise the security of cross-chain transaction storage due to the limited number of notaries.  To secure cross-chain transaction storage among heterogeneous edge nodes, the authors in \cite{wang2020virtual} develop a secure and decentralized cross-chain storage model with their proposed tree structures that has a good performance in tamper-proof data protection and can defend against conspiracy node attacks.
	
	The Metaverse is expected to provide various applications and services that have requirements with multiple QoS parameters~\cite{yu2012building}. These applications need a network fragmentation service level agreement (SLA) to provide customized network services for Metaverse users. To protect the consistency of services on multiple network slices simultaneously using the blockchain, the authors in~\cite{he2021cross} design a multi-chain network slicing framework. The framework deploys a separate blockchain on each edge network slice to reduce the frequency of information exchange between slices and improve service efficiency. In this framework, authorities deploy smart service quality contracts only in the service quality chain to coordinate and formulate the recommended smart contracts for the censorship and the evaluation chains. 
	%In addition, the authors also introduce Cosi protocol~\cite{syta2016keeping} and multi-signcryption algorithm for secure computation of QoS parameters across chains in the scenario of public and private data processing. 
	  However, different types of information about electric vehicles and charging piles need to be stored and queried simultaneously for energy security purposes in shared charging systems. Therefore, the authors in~\cite{he2021charging} propose a multi-chain charging model based on cross-chain trusted smart contracts for electric vehicular networks. In the model, different types of information are stored in different blockchains to ensure the authenticity, real-time, and mutual exclusion for write operations of cross-chain information.
	
	%To extend blockchain's ability in multiple network slices, the authors in~\cite{he2021cross} design a multi-chain 5G network slicing model. This model provides a blockchain for each network slice, reducing the frequency of information exchange and increasing efficiency. They design a cross-chain 5G network slicing service computing smart contract model, where only service quality smart contract are deployed on the service quality chain by authority agency, the consistent smart contract on the censorship chain and the recommended smart contract on the rating chain can be coordinated and formulated as needed through this service quality smart contract. To ensure the cross-chain security calculation of service quality parameters, they introduce Cosi protocol~\cite{syta2016keeping} and multi-signcryption algorithm for data processing of non-private data and private data, respectively. To improve storage and query efficiency of shared charging system, the authors in~\cite{he2021cross} propose a multi-chain charging model that stores different types of information on different blockchains. In this way, different types of information can be stored on different blockchain by deploying a cross-chain trusted smart contract (C2T smart contract) to ensure the authenticity, real-time, and inter-chain write mutual exclusion of cross-chain information.
	
\begin{figure}
		\centering
		\includegraphics[width=0.8\linewidth]{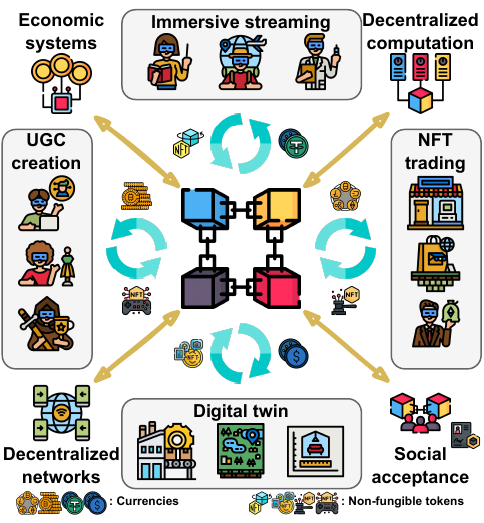}
		\caption{Various scenarios of wireless blockchain for virtual and physical services in the Metaverse. The blue arrows indicate the economic circulation of virtual content generated by users that virtual content is minted as NFTs and traded for cryptocurrency. The brown arrow indicates the interoperation between multiple blockchains for economic systems, avatar society management, and edge resource management.}
		\label{fig:bc_scenarios}
	\end{figure}

	  Decentralized exchanges based on decentralized networks and computation play important roles in the blockchain-based Metaverse economic systems that enable cross-chain token or NFT circulation, as shown in Fig.~\ref{fig:bc_scenarios}. To this end, the authors in~\cite{tian2021enabling} propose a distributed cryptocurrency trading system that employs a decentralized cryptocurrency exchange protocol. In this protocol, two types of consensus mechanisms (i.e., PoW and Proof of Deposit) are used to select trustworthy users for a validation committee. The experimental results show that this cross-chain exchange platform can be suitable for multi-user scenarios. The platform overhead (i.e., provisioning and execution costs) depends only on the number of participants instead of the number of transfers made by a single participant.   With these attractive advantages, the proposed platform is expected to be deployed at edge nodes in mobile edge networks and thus enable transactions to be processed at where they are generated.
	
	% Related work:  sharding/sidechain for scalable blockchains \cite{ksblockchain, kiong2021defi}   

	%  Cross-chain consensus for blockchain networks (Metaverse Networks) \cite{wang2021non}  
	% Related work:
	
	% Cross-chain interaction in blockchain networks \cite{wang2021non}
	
	% sharding/sidechain for scalable blockchains \cite{ksblockchain, kiong2021defi}
	
	\subsection{Blockchain in Edge Resource Management}
	
	The existence of ubiquitous connections in the Metaverse, especially at mobile edge networks, and the mobility of devices pose new challenges for security and privacy in the Metaverse. Therefore, blockchain plays an essential role in communication in the Metaverse. On the one hand, blockchain provides access control~\cite{zyskind2015decentralizing}, process management~\cite{viriyasitavat2020blockchain}, and account auditing~\cite{cannavo2020blockchain} services for AR/VR services in the Metaverse. On the other hand, the blockchain stores the synchronized records of the physical and virtual worlds so that the data is secure and private throughout its lifecycle. 
	
	On the other hand, the computational capabilities provided by the network edge enable users in the Metaverse to process data and draw conclusions with lower latency than in the cloud~\cite{xiong2018mobile}. The computational security of data in the Metaverse is threatened by the widespread presence of heterogeneous edge devices in the mobile edge network, often owned by different stakeholders~\cite{lee2021all}. Fortunately, the blockchain enables these edge devices belonging to various interest parties, such as AR/VR service providers and DT service providers, to process the computation tasks together without a trusted third party. By using blockchain to protect edge computing, the computation tasks, such as AR/VR rendering, in the Metaverse can be transparent and privacy-preserving. On the other hand, the processing of data and knowledge (e.g., AI models) in the Metaverse can be effectively protected by the blockchain~\cite{yang2022fusing}.
	
	\subsubsection{\textbf{Blockchain in Edge Multimedia Networks}}
	
	Multimodal multimedia services require a large amount of user privacy data to optimize the QoS and immerse users in the Metaverse. Therefore, blockchain as a distributed ledger can record private information, such as preferred FoV, for Metaverse users~\cite{shen2021data}. Using fog-based access points as logging nodes of the blockchain, the authors in~\cite{liu2021virtual} propose blockchain-enabled fog radio access networks to provide decentralized logging services for VR services. To reduce the message exchange in the original practical Byzantine Fault Tolerance (PBFT) mechanism~\cite{castro1999practical}, they propose a two-layer PBFT mechanism to improve the scalability and network coverage of multimedia networks. To reduce the energy consumption of multimedia networks, they use the DRL algorithm to learn the optimal resource allocation strategy for blockchain consensus and VR streaming. For a wireless VR-enabled medical treatment system with sensitive health data, the authors in~\cite{lin2021task} propose a blockchain-enabled task offloading and resource allocation scheme. In this system, each edge access point can serve as a mining node to ensure secure storage, processing, and sharing of medical data. Similar to~\cite{liu2021virtual}, a collective RL-based joint resource allocation scheme is proposed in~\cite{lin2021task} for viewport rendering, block consensus, and content transmission. Eye-tracking data helps multimedia service providers assess user attention to improve the quality of multimedia services. However, centralized collection of eye-tracking data is often vulnerable to malicious manipulation and falsification. To ensure the security and privacy of eye-tracking data collected from remote individuals, the authors in~\cite{bozkir2020eye} propose a proof-of-concept (PoC) consensus-based decentralized eye-tracking data collection system. This system implements PoC consensus and smart contracts for eye-tracking data on the Ethernet blockchain. In this way, they do not need a centralized third party to reward and data management, which is optimal from a privacy perspective.
	
	Besides, augmented reality can provide a visual experience for biomedical health informatics applications. In~\cite{djenouri2021secure}, the authors propose a blockchain-based AR framework for collecting, processing, and managing data in health applications. Specifically, they built a system for managing medical models based on Ethereum. In this system, the proof-of-work (PoW) consensus mechanism is used to ensure that data is shared between different agents without compromising privacy. By leveraging the PoW consensus mechanism, this framework can ensure the privacy of biomedical health data across the different agents. To overcome the resource inefficiency of PoW, the authors in~\cite{liao2020information} propose a proof-of-computation offloading (PoCO) consensus mechanism. In this PoCO, the IoT nodes with better caching performance should be given a higher probability of becoming a block verifier and thus receive higher benefits. Compared with the PoW mechanism, the proposed PoCO mechanism is less powerful in terms of cache hit rate, the number of caches hits, and node traffic. To protect the security of military IoT without compromising its performance, the authors in~\cite{islam2020mr} propose a blockchain-based military IoT mixed reality framework. In this framework, inventory management, remote mission control, and battlefield support in military IoT are handled by a blockchain-based data management system. Therefore, smart contracts ensure the security of data storage and sharing in a multi-user environment.
	
	\subsubsection{\textbf{Blockchain in Physical-Virtual Synchronization}}
	
	During the physical-virtual synchronization for constructing and maintaining the Metaverse, the blockchain enhances the data monitoring and command execution in the DT by providing transparency, immutability, decentralized data management, and peer-to-peer collaboration~\cite{yaqoob2020blockchain}. Existing DT networks enable cyber-physical systems (CPS) to predict and respond to changes in real-time by assuming that DT data can be trusted throughout the product life cycle. However, this assumption is vulnerable when there are multiple stakeholders in the system, making data propagation and course correction unreliable. To develop a reliable distributed DT solution, the authors in~\cite{suhail2022towards} propose an AI-assisted blockchain-based DT framework that provides timely and actionable insights for CPS. In this framework, CPS extracts process knowledge from blockchain-protected product lifecycle data. Using AI algorithms, this knowledge is used to predict potential security threats in the system. To preserve data privacy, the authors in~\cite{lu2020low} extend the learning-based prediction scheme of~\cite{suhail2022towards} to FL to improve system reliability and security.
	
	%DT strengthen the security of Cyber Physical System (CPS) throughout the product lifecycle, while assuming that the DT data is trusted, providing agility to predict and respond to real-time changes. However, existing DT solutions in CPS are constrained by untrustworthy data dissemination among multiple stakeholders and timely course correction. Such limitations reinforce the significance of designing trustworthy distributed solutions with the ability to create actionable insights in real-time. To address these limitation, the authors in~\cite{suhail2022towards} propose an AI-aided blockchain-based DT framework to secure CPS. They integrate blockchain to safeguard product lifecycle data. The proposed framework not only acquires process knowledge from the specifications of the CPS, but also relies on AI to learn security threats based on sensor data. Federated learning is also leveraged in~\cite{lu2020low} to improve the reliability and security of the system as well as enhance data pricacy.
	
	In Industry 4.0, actionable insights that the DT obtained from analyzing a massive amount of data are used to help industrial plants make important decisions~\cite{xu2021multi}. However, data inconsistency and unreliability can disrupt the development of data-driven decisions in industrial systems. To address this problem, the authors in \cite{suhail2021trustworthy} propose a blockchain-based DT for the industrial IoT that aggregates heterogeneous data from different data sources. Moreover, this DT can also provide consistency guarantees for data interactions in continuous physical and discrete virtual spaces. In this way, critical events such as device failures and risks in the industrial IoT can be diagnosed and avoided in advance. The authors in~\cite{altun2019liberalization} propose a blockchain-based social DT system for smart home devices and data management. They deploy a private blockchain to support privacy-preserving smart home applications and provide human-centric services. In this private blockchain-based DT framework, a homeowner has shared ownership of physical devices and their DTs at homes.
	
	\subsubsection{\textbf{Blockchain in Edge Computation Offloading}}
	
	As mentioned in the previous section on computation, there are a large number of resource-constrained devices in the Metaverse at mobile edge networks to perform computationally intensive, latency-critical Metaverse services, such as 3D rendering, AI model training, and blockchain mining~\cite{xiong2018mobile,li2020noma}. To reduce computation loads of VR devices in edge medical treatment systems, the authors in~\cite{lin2021task} propose a blockchain-enabled offloading and resource allocation framework. In this framework, edge access points act as the blockchain nodes to execute and reach the consensus on the VR resources management and thus improve the security and privacy of the system. However, the offloading process is vulnerable to data incompleteness, while edge devices are often overloaded or underloaded when processing disproportionate resource requests, resulting in incomplete data at task offloading. The authors in~\cite{xu2019become} propose a blockchain-enabled computation offloading system using genetic algorithms to create a balanced resource allocation policy. Under a fair resource allocation scheme, the system can determine the optimal additive weighting and achieve efficient multi-criteria decision-making. 
	
	To extend the problem of constrained computation resources to edge servers, the authors study peer-offloading scenarios in~\cite{yuan2021coopedge}. To motivate selfish edge servers to participate in collaborative edge computing, they propose a blockchain-based distributed collaborative edge platform called CoopEdge to solve the trust and incentive problem of edge servers. In CoopEdge, the historical performance of edge servers in performing peer-to-peer offloading tasks is recorded in the blockchain to evaluate the reputation of edge servers. The reputation of these edge servers is input into the Proof of Edge Reputation (PoER) consensus mechanism, which allows miners on the blockchain to record their consensus on the performance of peer-offloading tasks on the blockchain. To securely update, validate, and store trust information in a vehicular cloud network, the authors \cite{xu2021blockchain} propose a hierarchical blockchain framework in which long-term reputation and short-term trust variability are jointly considered. In this hierarchical blockchain framework, the vehicle blockchain is used to compute, validate, and store local subjective trust, the RSU blockchain is used for objective trust, and the cloud blockchain is used for social trust, because the heterogeneous resource distribution in this vehicular cloud network places different performance requirements on each blockchain.
	
	% On the other hand, the application of blockchain in wireless mobile networks is hindered by amajor challenge brought by the proof-ofwork puzzle during the mining process, which sets a high demand for the computational capability and storage availability in mobile devices. To this regard, the authors in \cite{liu2018computation} propose a novel mobile edge computing (MEC) enabled wireless blockchain framework where the computation-intensive mining tasks can be offloaded to nearby edge computing nodes and the cryptographic hashes of blocks can be cached in the MEC server. A more comprehensive framework is studied in \cite{chen2019cooperative}, in which Blockchain mining tasks and data processing tasks are studies jointly and edge devices are enabled to connect to the edge servers via relays. The joint offloading problem is formulated as a potential game and then is proved the existence of Nash equilibrium (NE) for this game. To achieve long-term performance, the authors in \cite{qiu2019online} formulate the computation offloading problem into Markov Decision Process (MDP) and accommodate it via Deep Reinforcement Learning.

	\subsubsection{\textbf{Blockchain in Edge Intelligence}}
	  In the Metaverse, the blockchain provides decentralized management of data and knowledge resources that are important for communication and computing services.
		Blockchain-based data and knowledge sharing among edge devices in the physical and virtual worlds can be more secure and reliable~\cite{yaqoob2020blockchain}. Nevertheless, blockchain-based data sharing becomes less efficient and risky as the volume of user data increases and privacy becomes more important. Fortunately, Metaverse can be built on top of many AI models that can extract knowledge and make inferences from vast amounts of edge-aware data. To enable AI models to trade securely in the AI-powered IoT, the authors propose a blockchain-based peer-to-peer knowledge marketplace in~\cite{lin2019making}. In this marketplace, they developed a new cryptocurrency named Knowledge Coin, smart contracts, and a new consensus mechanism called proof-of-transaction (PoT), combining PoW with the proof-of-stake (PoS), for secure and efficient knowledge management and transactions. % One of the more efficient PoT mechanisms, which combines PoW with the proof-of-stake (PoS) mechanism, provides secure knowledge-as-a-service in edge intelligence markets.
	
	 In the Metaverse, multiple physical and virtual entities can leverage FL to collaboratively train a global model without privacy concerns for social-aware avatars and AR/VR recommendation services.
	However, edge devices may upload unreliable local models in FL, thus tempting the global aggregator to degrade the performance of the global model. For example, a mobile device could launch a data poisoning attack on the federation, i.e., intentionally providing unreliable local models, or use only low-quality data for updates, i.e., unintentionally performing unreliable updates. To address this problem, the authors in~\cite{kang2020reliable} propose a secure FL system based on federated chains. First, they capture and compute the performance of historical model updates of edge devices participating in FL with a reputation as a metric. Then, a consortium blockchain is proposed to efficiently and securely manage the reputation of edge devices. To extend reputation-based employee selection in~\cite{kang2020reliable}, the authors propose a reputation-based crowdsourcing incentive mechanism in~\cite{zhao2020privacy} to motivate more customers to participate in FL model training tasks and reduce malicious updates and toxic updates. However, the reliability of the central server is still doubtful. To solve this problem, the authors propose a fully decentralized FL framework with committee consensus in~\cite{li2020blockchain}. In this framework, elected committee members review model updates and aggregate models. In this way, honest FL nodes can reinforce each other and continuously improve the performance of the global model. At the same time, intentionally or unintentionally, unreliable updates are discarded to protect the global model. Finally, the framework is flexible, where FL nodes can join or leave without interrupting the training process.
	
		\subsection{Lessons Learned}
		\subsubsection{Synchronization of Physical-Virtual Economic Systems}
		In the Metaverse economic system, multiple monetary systems co-exist, including the physical monetary system with fiat currencies and virtual monetary systems with in-game tokens, stablecoins, CBDCs, and cryptocurrencies. Under the Metaverse economic system, users can first purchase P2V synchronization services to access the Metaverse. Then, users can create and trade digital assets for digital currencies. Finally, through the V2P synchronization services, users can exchange their digital currencies back into fiat currencies in the physical world. By synchronizing the physical-virtual economic systems in the Metaverse, new liquidity is brought to the global financial system in a secure and robust manner.
		\subsubsection{Off-chain Data Storage}
		Due to the limited storage space in on-chain storage, the enormous volume of data generated by physical and virtual entities in the Metaverse is preferred to be stored in an off-chain manner. In the construction and maintenance of Metaverse, the blockchain-based data storage and sharing system verifies and stores the metadata on-chain while storing the complete data off-chain. This way, the Metaverse is more sustainable while guaranteeing immutability.
		\subsubsection{Intelligent Scalability and Interoperability}
		The Metaverse is expected to support ubiquitous physical and virtual entities for real-time physical-virtual synchronization, which requires the scalability and interoperability of the blockchain. For enhancing the scalability of the Metaverse, AI can provide intelligent solutions for the adaptive sharding of the blockchain.
		In addition, learning-based schemes can be utilized in robust notary selection of cross-chain schemes and thus improving interoperability.
		% The Metaverse is expected to support ubiquitous physical and virtual entities for real-time physical-virtual synchronization, which requires scalability and interoperability of the blockchain. ​Fortunately, AI can provide intelligent solutions for adaptive sharding to enhance scalability of the Metaverse. In addition, learning-based schemes can be utilized in robust notary selection of cross-chain schemes and thus improve interoperability.
		\subsubsection{Anonymous Physical-Virtual Synchronization}
		Based on the blockchain, the Metaverse can provide encrypted addresses and address-based access models for physical and virtual entities to anonymously request immersive streaming and synchronization services in Section~\ref{sec:communication}. This way, physical entities, e.g., mobile devices, can allocate immersive steaming and edge rendering from virtual entities, e.g., AR/VR recommenders, to facilitate the V2P synchronization. In addition, IoT devices and sensors in the physical world can update the DT models anonymously with encrypted addresses.
		\subsubsection{Decentralized Edge Intelligence}
		In the Metaverse, edge intelligence mentioned in Section~\ref{sec:computation} can be empowered by the blockchain to become decentralized during edge training and inference. This way, edge intelligence can avoid a single point of failure during collaborative training between edge devices and servers, and thus improving the reliability of edge networks. In addition, the blockchain-based reputation management scheme and incentive mechanism can enhance the robustness and sustainability of edge networks.

	\section{Future Research Directions}
	\label{sec:open}
	
	This section discusses the future directions that are important when various technologies are implemented in the Metaverse.
	
	\textbf{Advanced Multiple Access for Immersive Streaming:} Heterogeneous services and applications in the Metaverse, such AR/VR, the tactile Internet, and hologram streaming, necessitates the provisioning of unprecedented ubiquitous accessibility, extremely high data rate, and ultra-reliable and low-latency communications of the mobile edge networks. Specifically, the shared Metaverse expects advanced multiple access schemes to accommodate multiple users in the allotted resource blocks, e.g, time, frequency, codes, and power, for sharing the virtual worlds in the most effective manner.
	
	\textbf{Multi-sensory Multimedia Networks: } Unlike the traditional 2D Internet, the Metaverse provides multi-sensory multimedia services to users, including AR/VR, the tactile Internet, and hologram streaming. These multi-sensory multimedia services require mobile edge networks to be capable of providing holographic services (e.g., AR/VR and the tactile Internet) simultaneously. However, it is intractable to design resource allocation schemes for such complex multi-sensory multimedia services, as different types of network resources are required simultaneous. For instance, AR/VR requires eMMB services while the tactile Internet requires URLLC services. In detail, AR services often occupy the network's uplink transmission resources and computing capabilities, while VR services often occupy the network's downlink transmission resources and caching capabilities. Therefore, in these multi-sensory multimedia networks, efficient and satisfactory resource allocation schemes should be proposed to support the immerse experience of users.
	
	\textbf{Multimodal Semantic/Goal-aware Communication:} By automatically interpreting AI models, semantic communication allows mobile edge networks to evolve from data-oriented to semantic-oriented to support a wide variety of context-aware and goal-driven services in the Metaverse. However, existing semantic communication models tend to focus on semantic extraction, encoding, and decoding for a single task, such as speech or images. In the Metaverse, service models tend to be multimodal and include multiple types of immediate interactions, such as audio services and video services. This raises new issues for semantic communication, i.e., i) How to design a multimodal semantic communication model to provide multi-sensory multimedia services in the Metaverse, ii) How to efficiently extract semantics from the data transmitted by users, iii) How to allocate resources in the edge network to support the training and application of semantic communication models.
	
	\textbf{Integrated Sensing and Communication:} Communication and sensing in the edge network are ubiquitous from the construction of the Metaverse (through physical-virtual synchronization) to the user access. However, in future mobile edge networks, the spectrum traditionally reserved for sensing (e.g., mmWave, THz, and visible light) will be used for communication as well. Therefore, the integration of communication and sensing needs to be given the necessary attention to ensure the establishment and sustentation of the Metaverse. For example, the Metaverse is created by a DT for real-world physical entities that contain a large amount of static and dynamic information. Furthermore, the Metaverse needs to instantly distribute such information to desired users and enable Metaverse users to seamlessly communicate and interact with the entities. We envision that the further integration of sensing and communications will enable real-time digital replication of the physical world and empower more diverse services in the Metaverse. %The Metaverse is based on state-of-the-art communication and perception technologies and is the most fantastic testing ground for the role of science in human life.
	
	\textbf{Digital Edge Twin Networks:} DT is the Metaverse's most effective engine to synchronize the physical and virtual worlds in real-time. At mobile edge networks, the DT can digitally replicate, monitor and control real-world entities with the help of a large number of edge nodes and Internet of Everything devices. Moreover, entities in edge networks can operate in the physical world more efficiently through the interconnection of digital entities in the Metaverse. While DT can bring smarter operation and maintenance to mobile edge networks, they also require significant communication, computation, and storage resources, making it hard for edge networks with limited resources to handle. Therefore, more efficient DT solutions and smarter DT-enabled mobile edge network operations and maintenance will be a purposeful research direction.
	
	\textbf{Edge Intelligence and Intelligent Edge:} To realize the Metaverse amid its unique challenges, future research should focus on the edge intelligence driven infrastructure layer, which is a core feature of the future wireless networks. In short, edge intelligence is the convergence between edge computing and AI. The two major features of edge intelligence should be adopted, i.e., i) Edge for AI: which refers to the end-to-end framework of bringing sensing, communication, AI model training, and inference closer to  where data is produced, and ii) AI for Edge: which refers to the use of AI algorithms to improve the orchestration of the aforementioned framework.
	
	% \textbf{Compatibility:} Similar to the traditional Internet, the Metaverse is not owned by any company or anyone. However, each company may release its own cloud/fog/edge computing architecture or product on its own Metaverse platform. Users may not always stay in one Metaverse platform, and they may crossover to other platforms. To boost computing efficiency to minimize waste and energy consumption, it is essential to improve the compatibility between different virtual worlds created by different VSPs. 
	
	\textbf{Sustainable Resource Allocation:} The Metaverse will always be resource hungry. Cloud/fog/edge computing services are always needed so that mobile devices or portable gadgets with low storage and computing capability can provide mobile users with excellent and immersive QoE. Therefore, the amount of energy used to provide the Metaverse with communication and computation power will continue to rise, exacerbating the energy consumption and increasing environmental concerns such as greenhouse gas emissions. Green cloud/fog/edge networking and computing are therefore needed to achieve sustainability. Some possible solutions can be i) new architectures can be designed with sustainability in mind to support green cloud/fog/edge networking and computing in the Metaverse, ii) energy-efficient resource allocation in green cloud/fog/edge networking and computing, and iii) combining green cloud/fog/edge networking and computing with other emerging green technologies.

	\textbf{Avatars (Digital Humans):} The Metaverse users are telepresent and immersed in the virtual worlds as digital humans, a.k.a., avatars. Since human-like avatars may require a large number of resources to create and operate, an important question to ask is: how can we establish dynamic avatar services at mobile edge networks to optimize the QoE of avatar services? Moreover, during the provision of avatar services, each avatar has to collect and store a large amount of private biological data of the user or other users that have mapping relationships with the avatar in the virtual worlds or related physical entities in the physical world. Therefore, it is significant to secure the data and safeguard privacy protection during avatar generation and operation.
	
	% \textbf{Standards for Blockchain Scalability and Interoperability:} Scalability and interoperability of the blockchain are the foundations to support the wide connectivity and interaction of the Metaverse. However, the scalability and interoperability of current blockchains are still in their infancy. Some initial extension schemes such as off-chain, sharding, and cross-chain allow the blockchain technology to provide a vision as the infrastructure of the Metaverse. However, there is still a lack of standardized protocols that allow the scalability and interoperability of blockchains to be efficiently utilized and developed. Standardized scalability and interoperability would allow Metaverse users to share 3D worlds more smoothly without the interruption of teleoperation between different blockchain systems.
	
	\textbf{Intelligent Blockchain:} Intelligent blockchain applies AI algorithms to enable conventional blockchain to be more adaptive to the prevailing network environment in the consensus and the block propagation processes. As AI is the main engine and blockchain is the main infrastructure of the Metaverse, their intersection allows the Metaverse to efficiently protect the security and privacy of users. There are some preliminary studies on the intelligent blockchain, such as adaptive block propagation models and dynamic routing algorithms for payment channel networks. However, the existing research on the intelligent blockchain only focuses on making the blockchain operate more efficiently in digital environments. In the Metaverse, intelligent blockchain will have to be linked to the physical world, changing the manner in which physical entities (e.g., wireless base stations, vehicles, and UAVs) operate in the physical world.

		\textbf{Quality of Experience: } The human perceived QoE of users and their avatars should be properly evaluated and satisfied while immersing in the virtual worlds and interacting with other users. Both subjective and objective QoE assessments based on physiological and psychological studies can be leveraged to score and manage the provisioned services in the Metaverse.
		
		\textbf{Market and Mechanism Design for Metaverse Services: } For interactive and resource-intensive services in the Metaverse, novel market and mechanism design is indispensable to facilitate the allocation and pricing for Metaverse service providers and users. As the Metaverse can blur the boundary between the physical and virtual worlds, the market and mechanism design for Metaverse services should consider local states of the physical and virtual submarkets while taking into account the interplay effects of them.
		
		\textbf{The Industrial/Vehicular Metaverse: } The emerging industrial Metaverse will integrate the physical factories and virtual worlds for next-generation intelligent manufacturing. The industrial Metaverse obtains data from various production and operation lines by the Industrial Internet of Things (IIoT), and thus conducts effective data analysis and decision-making, thereby enhancing the production efficiency of the physical space, reducing operating costs, and maximizing commercial value. Meanwhile, the vehicular Metaverse integrates immersive streaming and real-time synchronization, which is expected to provide better safety, more efficient driving, and immersive experience for drivers and passengers in vehicles.
		
		% \textbf{The Metaverse in the Quantum Internet:} With the continuous development of quantum computing and information technology, the Metaverse, and the quantum Internet will overlap during their development. With the quantum advantage of quantum computers, the large-scale computation problems in the Metaverse will be solved by quantum computers easily. In addition, the quantum Internet can also provide information-theoretic security for communication and networking in the Metaverse. Finally, the blockchain in the Metaverse is expected to be upgraded by the post-quantum cryptography in the Quantum Internet.

	The future research directions for the Metaverse at mobile edge networks are based on three aspects, i.e., embodied user experience, harmonious and sustainable edge networks, and extensive edge intelligence. First, the service delivery with an immersive and human-aware experience of the 3D embodied Internet in the Metaverse will push the development of multi-sensory multimedia networks and human-in-the-loop communication. Second, the emergence of the Metaverse drives mobile edge networks to provide sustainable computing and communication services through real-time physical-virtual synchronization and mutual optimization in digital edge twin networks. Third, empowered by extensive edge intelligence and blockchain services, environments in virtual and physical worlds can be exchanged and mutually affected in the Metaverse without security and privacy threats.
	
	\section{Conclusion}
	\label{sec:conclude}
	In this survey, we first begin by introducing the readers to the architecture of the Metaverse, the current developments, and enabling technologies. With the focus on edge-enabled Metaverse, we next discuss and investigate the importance of solving key communication and networking, computation, and blockchain challenges. Lastly, we discuss the future research directions towards realizing the ultimate vision of the Metaverse. The survey serves as the initial step that precedes a comprehensive investigation of the Metaverse at edge networks, offering insights and guidance for the researchers and practitioners to expand on the edge-enabled Metaverse with continuous efforts.
	
	\bibliographystyle{IEEEtran}
	\bibliography{serverless}

	\newpage

\end{document}